\documentclass[15pt,a4paper]{article}
\usepackage[ ]{natbib} 
\usepackage[english]{babel}
\usepackage[utf8]{inputenc}
\usepackage[T1]{fontenc}
\usepackage{amsmath,amssymb,amsfonts,amsthm}
\usepackage[fleqn,tbtags]{mathtools}
\usepackage{dsfont} 
\usepackage{authblk}
\usepackage{graphicx}
\usepackage{xcolor}
\usepackage{float}
\usepackage{geometry}
\usepackage{booktabs}
\usepackage{multirow}
\usepackage{xr-hyper} 
\usepackage{hyperref}
\usepackage{array}
\usepackage[labelfont=bf]{caption}
\usepackage{subcaption}
\usepackage{floatpag} 
\floatpagestyle{empty} 
\usepackage{setspace} 
\usepackage{mathpazo} 

\usepackage[normalem]{ulem} 

\usepackage[ruled,vlined]{algorithm2e}

\DeclareMathOperator{\Var}{\mathbb{V}\text{ar}}

\DeclareMathOperator{\E}{\mathbb{E}}

\DeclareMathOperator{\diag}{diag}

\newcommand{\norm}[1]{\left\lVert#1\right\rVert}
\newcommand{\argmin}{\operatornamewithlimits{argmin}}
\newcommand{\transpose}{{}^{\text{\sffamily T}}}

\DeclareMathOperator{\ERFE}{\textit{ERFE}}
\DeclareMathOperator{\QRFE}{\textit{QRFE}}
\DeclareMathOperator{\ER}{\textit{ER}}
\DeclareMathOperator{\QR}{\textit{QR}}
\DeclareMathOperator{\SE}{\textit{SE}}
\DeclareMathOperator{\SD}{\textit{SD}}

\DeclareMathOperator{\FE}{\textit{FE}}

\newtheorem{theorem}{Theorem}
\newtheorem{lemma}{Lemma}
\newtheorem{corollary}{Corollary}

\makeatletter
\newcommand*{\addFileDependency}[1]{
  \typeout{(#1)}
  \@addtofilelist{#1}
  \IfFileExists{#1}{}{\typeout{No file #1.}}
}
\makeatother

\title{Weighted asymmetric least squares regression with fixed-effects}

\author[1,2]{Amadou Barry\footnote{Corresponding author: amadou.barry@mail.mcgill.ca.} }
\author[3]{ Karim Oualkacha}
\author[3]{Arthur Charpentier}

\affil[1]{Departments of Epidemiology, Biostatistics and Occupational Health, McGill University, Montréal, Québec, Canada}
\affil[2]{Lady Davis Institute, Jewish General Hospital, Montréal, Québec, Canada}
\affil[3]{Department of Mathematics and Statistics, Université du Québec à Montréal, Montréal, Québec, Canada}

\date{\today}

\begin{document}
\maketitle

\begin{abstract}
\noindent
The fixed-effects model estimates the regressor effects on the mean of the response, which is inadequate to summarize the variable relationships in the presence of heteroscedasticity. In this paper, we adapt the asymmetric least squares (expectile) regression to the fixed-effects model and propose a new model: expectile regression with fixed-effects $(\ERFE).$ The $\ERFE$ model applies the within transformation strategy to concentrate out the incidental parameter and estimates the regressor effects on the expectiles of the response distribution. The $\ERFE$ model captures the data heteroscedasticity and eliminates any bias resulting from the correlation between the regressors and the omitted factors. We derive the asymptotic properties of the $\ERFE$ estimators and suggest robust estimators of its covariance matrix. Our simulations show that the $\ERFE$ estimator is unbiased and outperforms its competitors. Our real data analysis shows its ability to capture data heteroscedasticity (see our R package, \url{github.com/AmBarry/erfe}). 
\end{abstract}

\section{Introduction}\label{intro_erfe}

The fixed-effects $(\FE)$ model is commonly used in econometric to analyze panel data. The $\FE$ model has the ability to account for the correlation between the regressors and the omitted (unmeasured) factors which is common in many applications.  For example, in econometrics, the education level is known to be correlated with the individual unobserved ability \citep{cardEstimatingReturnSchooling2001}. In perinatal studies, the birth weight is influenced by maternal genetic \citep{warringtonMaternalFetalGenetic2019}, which is usually a missing information. Therefore, in such context{\textemdash}where the unmeasured factors are correlated with the regressors, the $\FE$ estimator (within-estimator) is unbiased, consistent and computationally efficient \citep{CornwellRupert1988}.

Several quantile regression $(\QR)$-based methods \citep{koenker_quantile_2004, galvao_penalized_2010, lamarcheRobustPenalizedQuantile2010a} have been proposed to overcome the heteroscedasticity problem in the $\FE$ framework. However, they fail to extend the favorable properties of the within-estimator and suffer from two significant limitations. First, the fixed-effects $\QR$-based methods do not extend the within-transformation strategy to solve the incidental parameter problem. Thus, the fixed-effects $\QR$-based methods simultaneously estimate the parameter of interest and the incidental parameter which results in a computationally demanding algorithm. Additionally, the covariance of the $\QR$-based methods is based on the random error density function which further adds a computational burden and certain numerical issues \citep{Chen2004, YinCai2005a, kocherginskyPracticalConfidenceIntervals2005}. Second, the fixed-effects $\QR$-based method do not control for the correlation between the individual effects and the regressors. Thus, in the presence of such correlations, the fixed-effects $\QR$-based method yields biased and inconsistent estimates.

In this paper, we rely on expectiles to successfully generalize the within-estimator and take into account the heteroscedasticity present in the panel data under the $\FE$ framework. To the best of our knowledge, this is the first approach that estimates the marginal effect of the regressors on the response distribution, and generalizes the within transformation strategy in the $\FE$ framework.

The expectiles are statistics that characterize the distribution function of a random variable \citep{girardFunctionalEstimationExtreme2021}. The expectiles and the expectile regression $(\ER)$ were introduced in the seminal paper by \citet{newey_asymmetric_1987}. The expectiles and quantiles play similar statistical roles, except that expectiles are weighted averages while quantiles are order statistics. This interpretation difference offers significant computational advantages. In other words, quantiles focus on the ordering of the observations while the expectiles target their values. For instance, the mean is a particular expectile as the median is a particular quantile. The research on expectiles is very active and for further details we refer to  \citet{GEEE_Barry2020}.

Typically quantiles are more robust than expectiles, but as mentioned earlier, the proposed $\QR$-based fixed-effects models can not extend the within transformation strategy to solve the computational challenges raised by the incidental parameter problem efficiently. Further, the $\QR$-based fixed-effects models fail to control for the correlation between the individual effects and the regressors. Therefore, the expectile-based approach could be an effective alternative for inference in the $\FE$ framework.

In this paper, we combine the weighted asymmetric least squares regression $(\ER)$ and the $\FE$ model to propose a new panel model that we call: expectile regression for fixed-effects model $(\ERFE).$ The $\ERFE$ model retains the attractive properties of the $\FE$ model, while accounting for the heteroskedasticity present in the panel data. We derive its asymptotic properties and propose a heterogeneous, consistent, and robust estimator of its variance-covariance matrix. We share our code as a free R package available on GitHub (\url{github.com/AmBarry/erfe}) to simplify its implementation.

Our main contributions are: i. Extension of the within-transformation strategy in the $\ER$ framework to solve the incidental parameter problem, offering a significant computational advantage particularly with the advent of high dimensional data, where the sample size can be very large;
ii. Elimination of any bias that might result from the correlation between the individual effects and the regressors;
iii. Derivation of the asymptotic properties of the $\ERFE$ estimators; iv. Proposition of an estimator of its variance-covariance matrix for inference. 

Our $\ERFE$ model accounts for the omitted time-invariant factors and their correlation with the regressors present in the model. It also captures the heteroskedasticity present in the panel data by estimating the effects of the regressors at the conditional expectiles of the response distribution. Indeed, in the presence of heteroskedasticity, the parameters of the model are function of the asymmetric points, and in this case the $\ERFE$ model captures the heteroskedasticity by estimating several regression coefficient vectors at different locations of the conditional response distribution. The $\ERFE$ model provides a detailed overview of the regressor effects on the response distribution without making any assumption about the random error distribution. Our $\ERFE$ model is computationally efficient and easy to implement, with its available R package. We believe that it will be a useful tool for addressing the heteroskedasticity present in the panel data.

In Section \ref{Models_erfe}, we introduce the expectile function and the expectile regression model, and then present the expectile regression with fixed-effects (ERFE) model. In Section \ref{asymp_erfe}, we derive the asymptotic properties of the ERFE estimator, and propose an estimator of its variance-covariance (VC) matrix. We present the sample performance of the ERFE estimator in Section \ref{simulation_erfe} and its application to a real dataset in Section \ref{Application_erfe}. The conclusions is in Section \ref{Conclusion_erfe} and detail of the proofs are in the \textbf{Supplementary} material. 

\section{Models and Methods} \label{Models_erfe}

\subsection{Expectile and expectile regression}

The expectile of level \(\tau\in[0,1]\) of a random variable \(Y\) is defined as the unique solution of

\begin{equation}\label{equ:expect_obj_fun}
\mu_{\tau}(Y)=\argmin_{\theta\;\in\;\mathbb{R}}
\E\{\rho_\tau(Y-\theta)\},   
\end{equation}

where \(\rho_\tau(t)=\lvert \tau-\mathds{1}(t\leq 0)\rvert \cdot t^2\) is the asymmetric square loss function that assigns weights \(\tau\) and \(1-\tau\) to positive and negative deviations, respectively.

The expectiles summarize the cumulative distribution function of a random variable. In this regard, the expectiles play a similar statistical role to the quantiles, except that quantiles are order statistics while expectiles are weighted averages, and this interpretation difference is accompanied by significant computational advantages. The expectiles generalize the mean which corresponds to the expectile of level $\tau=0.5$ and assigns the same weight to positive and negative deviations. The expectiles are location and scale equivariant, that is for \(s>0 \mbox{ and } t\in\mathbb{R}, \ \mu_{\tau}(sY+t)=s\mu_{\tau}(Y)+t.\) A detailed study of the expectiles can be found in \citep{newey_asymmetric_1987}.

Once the optimization problem of equation (\ref{equ:expect_obj_fun}) is solved, for a fixed $\tau,$ the expectile of the random variable $Y$ can be defined as a weighted average:

\begin{equation*}
   \mu_{\tau}(Y)=\mu_{\tau}= \E\Bigg[ \frac{\psi_{\tau}(Y-\mu_{\tau})}{\E\big[\psi_{\tau}(Y-\mu_{\tau})\big]}Y\Bigg],
\end{equation*}

where \(\psi_{\tau}(t)=\lvert \tau-\mathds{1}(t\leq 0)\rvert\) is the check function. The only subtlety is that the weights are random. Given a random sample, \(\lbrace(y_i)\rbrace_{i=1}^{n},\) the corresponding \(\tau\)-th sample expectile

\begin{equation}\label{w_mean_erfe}
    \widehat{\mu}_{\tau}=\sum_{i=1}^{n}\frac{\psi_{\tau}(y_i-\widehat{\mu}_{\tau})}{\sum_{l=1}^{n}\psi_{\tau}(y_l-\widehat{\mu}_{\tau})}y_i
\end{equation}

is the weighted mean, where the weights depend on the sample data. For a fixed \(\theta,\) equation (\ref{w_mean_erfe}) is derived as the solution which minimizes the following empirical risk function:

\begin{equation}\label{exp3_erfe}
\frac{1}{n}\sum_{i=1}^{n}\rho_\tau(y_i-\theta).
\end{equation}

In addition to the expectiles, \citet{newey_asymmetric_1987} introduced the expectile linear regression $(\ER)$ as a tool to study the regressor effects on the response distribution and capture the heteroscedasticity present in the data. Consider the following linear regression model

\begin{equation}\label{lineareg_erfe}
    y_{i}=\boldsymbol{x}_{i}\transpose\boldsymbol{\beta} + \varepsilon_i \; \mbox{ with } \; \mu_{\tau}(\varepsilon_i)=0,
\end{equation}

where \(\boldsymbol{x}_i\) is a \(p\times 1\) vector of regressors, \(y_i\) and \(\varepsilon_i\) are respectively the response variable and the random error with unspecified distribution function. The parameter of interest \(\boldsymbol{\beta} \in \mathbb{R}^p\) is unknown and needs to be estimated. The assumption, \(\mu_{\tau}(\varepsilon_{i})=0,\) ensures that the random error is centered on the \(\tau\)-th expectile. The corresponding $\ER$ model, for a fixed \(\tau \in (0,\ 1),\) is given as:

\begin{equation}\label{explineareg_erfe}
\mu_{\tau}(y_i|\boldsymbol{x}_{i})=\boldsymbol{x}_{i}\transpose\boldsymbol{\beta}_{\tau}.
\end{equation}

Therefore, the $\ER$ estimator \(\widehat{\boldsymbol{\beta}}_{\tau},\) for a fixed \(\tau \in (0,1),\) can be derived by minimizing 
the following objective function:

\begin{equation*}
\sum_{i=1}^{n}\rho_\tau\Big(y_{i}-\boldsymbol{x}_i\transpose\boldsymbol{\beta}_{\tau}\Big)
\end{equation*}

over \(\boldsymbol{\beta}_{\tau}\;\in\;\mathbb{R}^p.\) Since the loss function \(\rho_\tau(t)\) is continuously differentiable, we have through the first order condition:

\begin{equation}\label{er_est_erfe}
 \widehat{\boldsymbol{\beta}}_{\tau}=\Bigg[\sum_{i=1}^{n}\psi_\tau(y_{i}-\boldsymbol{x}_i\transpose \widehat{\boldsymbol{\beta}}_{\tau})\boldsymbol{x}_i\boldsymbol{x}_i\transpose \Bigg]^{-1}
 \sum_{i=1}^{n}\psi_\tau(y_{i}-\boldsymbol{x}_i\transpose \widehat{\boldsymbol{\beta}}_{\tau}) \boldsymbol{x}_i y_i,
\end{equation}

where \(\psi_\tau(t)=\lvert \tau-\mathds{1}(t\leq 0)\rvert\) is the check function. The $\ER$ estimator can be computed as an iterated weighted least squares estimators. For the special case of \(\tau=0.5, \ \widehat{\boldsymbol{\beta}}_{0.5}\)
is the classical ordinary least squares (OLS) estimator and this makes the $\ER$ a natural complement of the OLS regression. 

\subsection{Fixed-effects model for panel data}

Consider the standard linear fixed-effects model for panel data

\begin{equation}\label{longeq_erfe}
y_{ij}=\boldsymbol{x}_{ij}\transpose\boldsymbol{\beta} + \alpha_{i} + \varepsilon_{ij},
\end{equation}

where \(y_{ij}\) is the scalar response variable, the vector \(\boldsymbol{x}_{ij}=(x_{ij}^1,x_{ij}^2, \ldots, x_{ij}^p)\transpose \in \mathbb{R}^p\) is the vector of regressors measured on subject \(i\) at time \(j,\) the parameter \(\alpha_{i}\) is the subject-specific effects parameter, and the variable \(\varepsilon_{ij}\) is a random error with unspecified distribution function. The equation model (\ref{longeq_erfe}) is conveniently represented in individual notation as:

\begin{equation}\label{matlongeq_erfe}
\boldsymbol{y}_i=\boldsymbol{X}_i\boldsymbol{\beta} + \boldsymbol{Z}_i\boldsymbol{\alpha} + \boldsymbol{\varepsilon}_i,
\end{equation}

where \(\boldsymbol{y}_i \mbox{ and } \boldsymbol{\varepsilon}_i\) are \(m \times 1\) vectors, \(\boldsymbol{X}_i\) is an \(m \times p\) design matrix and \(\boldsymbol{Z}_i\) is an \(m \times n\) incidence matrix and \(\boldsymbol{\alpha}\) is an \(n \times 1\) subject-specific effects vector. 
We can also stack all the data and represent the equation model (\ref{matlongeq_erfe}) as:

\begin{equation}\label{bigmatlongeq_erfe} 
    \boldsymbol{y} = \boldsymbol{X}\boldsymbol{\beta} + \boldsymbol{Z}\boldsymbol{\alpha} + \boldsymbol{\varepsilon},
\end{equation}

where \(\boldsymbol{y}\) and \(\boldsymbol{\varepsilon}\) are \(N\times 1\) vectors, \(\boldsymbol{X}\) and \(\boldsymbol{Z}\) are respectively \(N\times p\) and \(N\times n\) matrices with \(N=mn.\) The incidence matrix \(\boldsymbol{Z}\) identifies the \(n\) distinct subjects of the sample.

The fixed-effects $\boldsymbol{Z}\boldsymbol{\alpha}$ of model equation (\ref{bigmatlongeq_erfe}) is infinite in nature and is potentially correlated with the regressors of the model. The traditional estimation method used to overcome this issue is the within-transformation strategy. This technique consists of pre-multiplying both sides of equation model (\ref{bigmatlongeq_erfe}) by the idempotent matrix \(\boldsymbol{M}_{\boldsymbol{Z}}=\mathbb{I}_N-\boldsymbol{Z}(\boldsymbol{Z}\transpose\boldsymbol{Z})^{-1}\boldsymbol{Z}\transpose\) to eliminate the infinite-dimensional parameter, and then applies the ordinary least squares (OLS) regression to the transformed data. The model that results from this transformation is represented as: 

\begin{equation}\label{ols_model_erfe} 
    \boldsymbol{y}^{*} = \boldsymbol{X}^{*}\boldsymbol{\beta} + \boldsymbol{\varepsilon}^{*},
\end{equation}

where \(\boldsymbol{y}^{*}=\boldsymbol{M}_{\boldsymbol{Z}}\boldsymbol{y}\) and \(\boldsymbol{X}^{*} \mbox{ and } \boldsymbol{\varepsilon}^{*}\) are defined similarly. The OLS estimator of the fixed-effects model, known as the within-transformation estimator, is given as:

\begin{equation}\label{ols_est_erfe} 
    \widehat{\boldsymbol{\beta}} = (\boldsymbol{X}^{*}\transpose\boldsymbol{X}^{*})^{-1}\boldsymbol{X}^{*}\transpose\boldsymbol{y}^{*}.
\end{equation}

The within-transformation estimator is consistent and asymptotically normally distributed \citep{Greene2011}. The within-transformation estimator is computationally efficient and is not affected by any bias resulting from the correlation between the individual effects and the regressors. The within-transformation technique does not allow estimation of the time-invariant regressors which could be seen as a limitation. However, this can be a strength when the number of time-invariant confounders is large, and when the collection of some of these variables (genetic factor) is complex and costly \citep{bruderlFixedEffectsPanelRegression2014}. In the following subsection, we present the expectile regression for fixed-effects $(\ERFE)$ model and derive the iterative-within-transformation $\ERFE$ estimator.

\subsection{ERFE model for panel data}

The $\ERFE$ model of the linear fixed-effects model is defined, for fixed \(\tau \in (0,1),\) as:

\begin{equation}\label{erfe_erfe}
\mu_{\tau}(y_{ij}|\alpha_{i},\boldsymbol{x}_{ij})=\boldsymbol{x}_{ij}\transpose\boldsymbol{\beta}_{\tau} + \boldsymbol{z}_{ij}\transpose\boldsymbol{\alpha}.
\end{equation}

In this setting the parameter \(\boldsymbol{\beta}_{\tau} \in \mathbb{R}^p\) captures the influence of the regressors \(\boldsymbol{x}_{ij}\) on the location, scale, and shape of the conditional distribution of the response variable \(y_{ij}.\) The subject-specific effects \(\boldsymbol{\alpha}\) is assumed to be independent of \(\tau\) across the percentiles and to have a pure location-shift effect on the conditional expectile of the response. Assuming a \(\tau\)-dependency of the subject-specific effects implies estimating its distribution with \(m\) number of within-subject observations, which is relatively small in most applications. Take note that no assumption is made about the shape of the response distribution.

The corresponding $\ERFE$ estimator of model equation (\ref{erfe_erfe}) is defined as the vector minimizing the following objective function:

\begin{equation}\label{ObjFunSingFe_erfe}
  \sum_{i=1}^{n}\sum_{j=1}^{m} \rho_{\tau} \big( y_{ij}-\boldsymbol{x}_{ij}\transpose\boldsymbol{\beta}_{\tau} - \boldsymbol{z}_{ij}\transpose\boldsymbol{\alpha} \big).
\end{equation}

Since the loss function \(\rho_{\tau}(\cdot)\) is continuously differentiable, we can apply the first-order condition and derive the resulting $\ERFE$ estimator \(\widehat{\boldsymbol{\beta}}_{\tau}\), which is defined as:

\begin{equation}\label{withinerfebet_erfe}
  \widehat{\boldsymbol{\beta}}_{\tau} 
  = \Big\lbrace\boldsymbol{X}\transpose\boldsymbol{\Psi}_{\tau}\big[\boldsymbol{y}-\boldsymbol{X}\widehat{\boldsymbol{\beta}}_{\tau}- \boldsymbol{Z}\widehat{\boldsymbol{\alpha}}\big]\widehat{\boldsymbol{M}}_{\boldsymbol{Z}}(\tau)
  \boldsymbol{X}\Big\rbrace^{-1}\boldsymbol{X}\transpose\boldsymbol{\Psi}_{\tau}\big[\boldsymbol{y}-\boldsymbol{X}\widehat{\boldsymbol{\beta}}_{\tau}- \boldsymbol{Z}\widehat{\boldsymbol{\alpha}}\big]\widehat{\boldsymbol{M}}_{\boldsymbol{Z}}(\tau)\boldsymbol{y}, 
\end{equation}

where the diagonal check function matrix is:

\begin{equation}\label{check_mat_erfe}
   \boldsymbol{\Psi}_{\tau}\big[\boldsymbol{y}-\boldsymbol{X}\widehat{\boldsymbol{\beta}}_{\tau}- \boldsymbol{Z}\widehat{\boldsymbol{\alpha}}\big]=
  \diag\Big(\psi_{\tau}(y_{11}-\boldsymbol{x}_{11}\transpose\widehat{\boldsymbol{\beta}}_{\tau}- 
   \boldsymbol{z}_{11}\transpose\widehat{\boldsymbol{\alpha}}),\ldots, \psi_{\tau}(y_{nm}-\boldsymbol{x}_{nm}\transpose\widehat{\boldsymbol{\beta}}_{\tau}- 
   \boldsymbol{z}_{nm}\transpose\widehat{\boldsymbol{\alpha}})\Big).
\end{equation}

The projection matrix \(\widehat{\boldsymbol{M}}_{\boldsymbol{Z}}(\tau)\) and its complement \(\widehat{\boldsymbol{P}}_{\boldsymbol{Z}}(\tau)\) are idempotent matrices and are defined as:

\begin{equation*}
\widehat{\boldsymbol{M}}_{\boldsymbol{Z}}(\tau)=\mathbb{I}_{N}-\widehat{\boldsymbol{P}}_{\boldsymbol{Z}}(\tau), \; \
\widehat{\boldsymbol{P}}_{\boldsymbol{Z}}(\tau)=\boldsymbol{Z}(\boldsymbol{Z}\transpose
\boldsymbol{\Psi}_{\tau}\boldsymbol{Z})^{-1}\boldsymbol{Z}\transpose\boldsymbol{\Psi}_{\tau}.
\end{equation*}

The function \(\boldsymbol{\Psi}_{\tau}(\cdot)\) defined in equation (\ref{check_mat_erfe}) depends on the subject-specific parameter estimator \(\widehat{\boldsymbol{\alpha}}\ \) which, by the first-order condition of equation (\ref{ObjFunSingFe_erfe}), verifies the relationship:

\begin{equation}\label{incidence_param_est_erfe}
   \boldsymbol{Z}\widehat{\boldsymbol{\alpha}}=\widehat{\boldsymbol{P}}_{\boldsymbol{Z}}(\tau)(\boldsymbol{y}-\boldsymbol{X}\widehat{\boldsymbol
   {\beta}}_{\tau}).
\end{equation}

Now, using equation (\ref{incidence_param_est_erfe}), the argument of the check function matrix can be written as

\begin{equation}\label{new_u_erfe}
\boldsymbol{y}-\boldsymbol{X}\widehat{\boldsymbol{\beta}}_{\tau}- \boldsymbol{Z}\widehat{\boldsymbol{\alpha}}=
\widehat{\boldsymbol{M}}_{\boldsymbol{Z}}(\tau)(\boldsymbol{y}-\boldsymbol{X}\widehat{\boldsymbol{\beta}}_{\tau}).
\end{equation}

Therefore, the incidental parameter estimate is eliminated from the expression of equation (\ref{withinerfebet_erfe}) of the $\ERFE$ estimator.
Now, using the following relationship:

\begin{equation*}
 \boldsymbol{\Psi}_{\tau}\big[\widehat{\boldsymbol{M}}_{\boldsymbol{Z}}(\tau)(\boldsymbol{y}-\boldsymbol{X}\widehat{\boldsymbol{\beta}}_{\tau})\big]\widehat{\boldsymbol{M}}_{\boldsymbol{Z}}(\tau)=\widehat{\boldsymbol{M}}_{\boldsymbol{Z}}\transpose(\tau)
 \boldsymbol{\Psi}_{\tau}\big[\widehat{\boldsymbol{M}}_{\boldsymbol{Z}}(\tau)(\boldsymbol{y}-\boldsymbol{X}\widehat{\boldsymbol{\beta}}_{\tau})\big],
\end{equation*}

and the idempotent property of the projection matrix \(\widehat{\boldsymbol{M}}_{\boldsymbol{Z}}(\tau),\) we can rewrite the $\ERFE$ estimator as:

\begin{equation}\label{withinerfebet1_erfe}
\begin{split}
  \widehat{\boldsymbol{\beta}}_{\tau} 
  & = \Big\lbrace\boldsymbol{X}\transpose\widehat{\boldsymbol{M}}_{\boldsymbol{Z}}(\tau)\transpose\boldsymbol{\Psi}_{\tau}
  \big[\widehat{\boldsymbol{M}}_{\boldsymbol{Z}}(\tau)(\boldsymbol{y}-\boldsymbol{X}\widehat{\boldsymbol{\beta}}_{\tau})\big]
  \widehat{\boldsymbol{M}}_{\boldsymbol{Z}}(\tau)\boldsymbol{X}\Big\rbrace^{-1}\\
  & \times \boldsymbol{X}\transpose\widehat{\boldsymbol{M}}_{\boldsymbol{Z}}(\tau)\transpose\boldsymbol{\Psi}_{\tau}
  \big[\widehat{\boldsymbol{M}}_{\boldsymbol{Z}}(\tau)(\boldsymbol{y}-\boldsymbol{X}\widehat{\boldsymbol{\beta}}_{\tau})\big]
  \widehat{\boldsymbol{M}}_{\boldsymbol{Z}}(\tau)\boldsymbol{y}.
\end{split}
\end{equation}

In summary, the within-estimator  is extended to the $\ER$ framework. The strategy is derived by applying the projection matrix \(\widehat{\boldsymbol{M}}_{\boldsymbol{Z}}(\tau)\) to the initial data \([\boldsymbol{y},\boldsymbol{X}],\) to eliminate the subject-specific effects parameter. Additionally, like the $\ER$ estimator in equation (\ref{er_est_erfe}), the within $\ERFE$ estimator can be computed iteratively using the iterative weighted least squares algorithm. The detailed algorithm for computing the iterative-within-transformation $\ERFE$ estimator is summarized in the following stepwise procedures.

\begin{algorithm}[H]
\footnotesize
\DontPrintSemicolon
\SetAlgoLined
\SetKwProg{Proc}{Procedure}{}{}
\KwIn{Let, for a fixed \(\tau, \ \widetilde{\boldsymbol{\beta}}_{\tau}^{(0)}=\widehat{\boldsymbol{\beta}}_{\tau},\) the $\ER$ estimator and 
$ \widehat{\varepsilon}_{ij\tau}^{(0)}=\widetilde{y}_{ij}^{(0)}-\widetilde{\boldsymbol{x}}_{ij}^{(0)}\transpose
\widetilde{\boldsymbol{\beta}}_{\tau}^{(0)}.$}
~~\\
 \While{ $\norm{\widehat{\boldsymbol{\beta}}_{\tau}^{(r)}-\widehat{\boldsymbol{\beta}}_{\tau}^{(r-1)}}_{\infty} \leq \; \zeta \quad$ }{
   Given $ \ \widetilde{\boldsymbol{\beta}}_{\tau}^{(r-1)}$ at the $(r-1)$-th step, update: \;   
   \begin{enumerate}  
   \setlength\itemsep{1.3em}
    \item     
        $ \widehat{\boldsymbol{P}}_{\boldsymbol{Z}}^{(r)}(\tau) \leftarrow \boldsymbol{Z}(\boldsymbol{Z}\transpose
\boldsymbol{\Psi}_{\tau}(\widehat{\boldsymbol{\varepsilon}_{\tau}}^{*(r-1)})\boldsymbol{Z})^{-1}\boldsymbol{Z}\transpose\boldsymbol{\Psi}_{\tau}(\widehat{\boldsymbol{\varepsilon}_{\tau}}^{*(r-1)}), \quad \widehat{\boldsymbol{\varepsilon}_{\tau}}^{*(r-1)}=\widetilde{\boldsymbol{y}}^{*(r-1)}-\widetilde{\boldsymbol{X}}^{*(r-1)}\widetilde{\boldsymbol{\beta}}_{\tau}^{(r-1)} $
  \item $ \widetilde{\boldsymbol{y}^{*}}^{(r)} \leftarrow \widehat{\boldsymbol{M}}_{\boldsymbol{Z}}^{(r)}(\tau)\boldsymbol{y} $
    \item $\widetilde{\boldsymbol{X}^{*}}^{(r)} \leftarrow \widehat{\boldsymbol{M}}_{\boldsymbol{Z}}^{(r)}(\tau)\boldsymbol{X}$ 
    \item $  \widehat{\boldsymbol{\beta}}_{\tau}^{(r)} \leftarrow \widehat{\boldsymbol{\beta}}_{\tau}^{(r-1)} + \Big[\widetilde{\boldsymbol{X}}^{*(r-1)}\transpose\boldsymbol{\Psi}_{\tau}(\widehat{\boldsymbol{\varepsilon}_{\tau}}^{*(r-1)})
  \widetilde{\boldsymbol{X}}^{*(r-1)}\Big]^{-1}\widetilde{\boldsymbol{X}}^{*(r-1)}\transpose\boldsymbol{\Psi}_{\tau}
  (\widehat{\boldsymbol{\varepsilon}_{\tau}}^{*(r-1)})\widehat{\boldsymbol{\varepsilon}_{\tau}}^{*(r-1)}$
  \item $\widehat{\boldsymbol{\varepsilon}^{*}_{\tau}}^{(r)} \leftarrow \widetilde{\boldsymbol{y}^{*}}^{(r)}-\widetilde{\boldsymbol{X}^{*}}^{(r)}\widehat{\boldsymbol{\beta}}_{\tau}$
\end{enumerate}        
}
\textbf{Return} $\widehat{\boldsymbol{\beta}}_{\tau}$ \;
\caption{The iterative within-transformation $\ERFE$ algorithm}\label{erfe_algo}
\end{algorithm}

The parameter $\zeta$ is the convergence tolerance and the default value in our code implementation is set to $10^{-7}. \ $ In practice, Algorithm \ref{erfe_algo} is computationally efficient and usually the number of iterations required to achieve convergence is between 3 and 5. Note that when \(\tau=0.5\) we have \(\boldsymbol{\Psi}_{\tau}=0.5\mathbb{I}_{N}\) and the iterative within-transformation $\ERFE$ estimator is nothing else than the OLS within-transformation estimator.

From the above development, the multiplication of a vector (say \(\boldsymbol{y}\)) by the matrix \(\widehat{\boldsymbol{M}}_{\boldsymbol{Z}}(\tau)\) deviates that vector from its expectile as shown by the following expression:

\begin{multline*}
    \widehat{\boldsymbol{M}}_{\boldsymbol{Z}}(\tau)\boldsymbol{y}=\bigg(
      y_{11}-\sum_{j=1}^{m}\frac{\psi_{\tau}(\widehat{\varepsilon}_{1j})}{\sum_{k=1}^{m}\psi_{\tau}(\widehat{\varepsilon}_{1k})}y_{1j}, \ldots,
      y_{1m}-\sum_{j=1}^{m}\frac{\psi_{\tau}(\widehat{\varepsilon}_{1j})}{\sum_{k=1}^{m}\psi_{\tau}(\widehat{\varepsilon}_{1k})}y_{1j}, \ldots,\\
      y_{n1}-\sum_{j=1}^{m}\frac{\psi_{\tau}(\widehat{\varepsilon}_{nj})}{\sum_{k=1}^{m}\psi_{\tau}(\widehat{\varepsilon}_{nk})}y_{nj}, \ldots, y_{nm}-\sum_{j=1}^{m}\frac{\psi_{\tau}(\widehat{\varepsilon}_{nj})}{\sum_{k=1}^{m}\psi_{\tau}(\widehat{\varepsilon}_{nk})}y_{nj}\bigg)\transpose.
\end{multline*}

We can see, from this expression, how the projection matrix \(\widehat{\boldsymbol{M}}_{\boldsymbol{Z}}(\tau)\) eliminates the subject-specific effects parameter and any other time-invariant regressors from the initial model. For a matrix, like the design matrix \(\boldsymbol{X},\) the transformation is applied column-wise.

\textbf{ERFE model for a sequence of expectiles }

The preceding development shows that the classical within-transformation strategy can be generalized in the $\ERFE$ framework. Now, we present the $\ERFE$ estimator for a sequence of expectiles using the transformed data. The sequence of expectiles, for example the mean and a few expectiles above and below it, is necessary in the description of the conditional distribution of the response variable and for capturing the data heteroscedasticity. In addition, the simultaneous estimation allows the multiple expectiles to share strength among each other and to gain better estimation accuracy than individually estimated expectile functions \citep{LiuWu2011}.

The iterative within-transformation $ERFE$ estimator \(\widehat{\boldsymbol{\beta}}_{\boldsymbol{\tau}}=[\widehat{\boldsymbol{\beta}}_{\tau_1}\transpose, \ldots, \widehat{\boldsymbol{\beta}}_{\tau_q}\transpose]\transpose\) for a sequence of asymmetric points \(\boldsymbol{\tau}=(\tau_1,\ldots,\tau_q)\transpose\) is defined as the minimum of the following objective function:

\begin{equation}\label{erfe_obj_erfe}
  \sum_{k=1}^{q}\sum_{i=1}^{n}\sum_{j=1}^{m} v_k \rho_{\tau_k} \Big(  y_{ij}-\boldsymbol{x}_{ij}\transpose\boldsymbol{\beta}_{\tau_k} - \boldsymbol{z}_{ij}\transpose\boldsymbol{\alpha} \Big).  
\end{equation}

The vector \(\boldsymbol{v}=(v_1, \ldots, v_q)\transpose\) is the vector of weights controlling the relative influence of the q asymmetric points \(\lbrace \tau_1, \ldots, \tau_q\rbrace\) and it choice depends on the research question. For a sequence of expectiles, the iterative within-transformation $\ERFE$ estimator is defined as:

\begin{equation}\label{BetMultExp_erfe}
\begin{split}
    \widehat{\boldsymbol{\beta}}_{\boldsymbol{\tau}}=\Big[
    (\mathbb{I}_q\otimes\widehat{\boldsymbol{X}^{*}})\transpose \boldsymbol{\Psi}_{\boldsymbol{\tau}}(\widehat{\boldsymbol{\varepsilon}_{\boldsymbol{\tau}}^{*}})
(\boldsymbol{V}\otimes\widehat{\boldsymbol{X}^{*}})\Big]^{-1}
(\boldsymbol{V}\otimes\widehat{\boldsymbol{X}^{*}})\transpose \boldsymbol{\Psi}_{\boldsymbol{\tau}}(\widehat{\boldsymbol{\varepsilon}_{\boldsymbol{\tau}}^{*}})
(\mathds{1}_q\otimes\widehat{\boldsymbol{y}^{*}})
\end{split}
\end{equation}

where \(\boldsymbol{\Psi}_{\boldsymbol{\tau}}(\widehat{\boldsymbol{\varepsilon}_{\boldsymbol{\tau}}^{*}})=\Big[\diag\big(\boldsymbol{\Psi}_{\tau_k}(\widehat{\boldsymbol{y}^{*}}-\widehat{\boldsymbol{X}^{*}} \widehat{\boldsymbol{\beta}}_{\tau_{k}})\big)\Big]_{k=1}^{q}, \; \boldsymbol{V}=[\diag(v_k)]_{k=1}^{q}\)
and the transformed data \([(\mathds{1}_q\otimes\widehat{\boldsymbol{y}^{*}}),(\mathbb{I}_q\otimes\widehat{\boldsymbol{X}^{*}})]\) is obtained by pre-multiplying the matrix \(\widehat{\boldsymbol{M}}_{\boldsymbol{Z}}(\boldsymbol{\tau})\) to the initial data \([\boldsymbol{y},\boldsymbol{X}].\) The projection matrix is defined as \(\widehat{\boldsymbol{M}}_{\boldsymbol{Z}}(\boldsymbol{\tau})=\mathbb{I}_{nmq}-\widehat{\boldsymbol{P}}_{\boldsymbol{Z}} (\boldsymbol{\tau})\) and

\begin{equation*}
    \widehat{\boldsymbol{P}}_{\boldsymbol{Z}}(\boldsymbol{\tau})=(\boldsymbol{v}\otimes\boldsymbol{Z})
\Big[(\boldsymbol{v}\otimes\boldsymbol{Z})\transpose  \boldsymbol{\Psi}_{\boldsymbol{\tau}}(\widehat{\boldsymbol{\varepsilon}_{\boldsymbol{\tau}}^{*}})
(\mathds{1}_q\otimes\boldsymbol{Z})\Big]^{-1}(\mathds{1}_q\otimes\boldsymbol{Z})\transpose \boldsymbol{\Psi}_{\boldsymbol{\tau}}(\widehat{\boldsymbol{\varepsilon}_{\boldsymbol{\tau}}^{*}}).
\end{equation*}

\section{Asymptotic}\label{asymp_erfe}

In this section, the asymptotic properties of the $\ERFE$ estimator are presented. As stated by \citet{koenker_quantile_2004}, the presence of the incidence parameter, which has an infinite dimension, can raise some challenges. For this reason, we present first the asymptotic results of the $\ERFE$ estimator in the simplest case, namely for a single \(\tau.\) We then present the asymptotic properties of the $\ERFE$ estimator
for a simultaneous sequence of asymmetric points \(\boldsymbol{\tau}=(\tau_1,\ldots,\tau_q).\) The section ends with the suggestion of an estimator of the covariance matrix for the $\ERFE$ estimator. All the proofs are available in the \textbf{Supplementary} file.

\subsection{Asymptotics for the ERFE estimator}

\textbf{Asymptotics for a single expectile }

In the following, the asymmetric square-loss function of the $\ERFE$ model,

\begin{equation*}
\rho_{\tau}\Big(y_{ij}-\boldsymbol{x}_{ij}\transpose\boldsymbol{\beta}_{\tau} - 
\boldsymbol{z}_{ij}\transpose\boldsymbol{\alpha}\Big),
\end{equation*}

is replaced in term of optimization by the equivalent loss function

\begin{equation*}
\rho_{\tau}\Big( y_{ij}-\mu_{ij\tau}-\boldsymbol{x}_{ij}\transpose
\boldsymbol{\delta}_{1\tau}/\sqrt{nm} - \boldsymbol{z}_{ij}\transpose\boldsymbol{\delta}_0/\sqrt{m}\Big)
-\rho_{\tau}(y_{ij}-\mu_{ij\tau}),
\end{equation*}

where \(\mu_{ij\tau}=\boldsymbol{x}_{ij}\transpose \boldsymbol{\beta}_{\tau} + \boldsymbol{z}_{ij}\transpose\boldsymbol{\alpha}.\) Now, observe that the following estimator

\begin{equation*}
\widehat{\boldsymbol{\delta}}=
\begin{pmatrix}
\widehat{\boldsymbol{\delta}}_0\\
\widehat{\boldsymbol{\delta}}_{1\tau}\\
\end{pmatrix}
=
\begin{pmatrix}
\sqrt{m} ( \widehat{\boldsymbol{\alpha}} - \boldsymbol{\alpha})\\
\sqrt{nm}\Big[\widehat{\boldsymbol{\beta}}_{\tau}-\boldsymbol{\beta}_{\tau}\Big]\\
\end{pmatrix}
\end{equation*}

minimizes the new objective function

\begin{equation}\label{Pen1_erfe}
\begin{split}
  {}& R_{nm}(\boldsymbol{\delta})= \sum_{i=1}^{n}\sum_{j=1}^{m} \rho_{\tau} 
  \Big( y_{ij}-\mu_{ij\tau}-\boldsymbol{x}_{ij}\transpose\boldsymbol{\delta}_{1\tau}/\sqrt{nm}-\boldsymbol{z}_{ij}\transpose\boldsymbol{\delta}_0/\sqrt{m}\Big)-\rho_{\tau}(y_{ij}-\mu_{ij\tau}).
\end{split}
\end{equation}

The asymptotic theory of the $\ERFE$ estimator is derived using this new objective function (\ref{Pen1_erfe}) and under the following assumptions.

\textbf{A1}. The data \(\lbrace (\boldsymbol{y}_i,\boldsymbol{X}_i)\rbrace_{i=1}^{n}\) are independent across \(i,\) and,

\begin{equation*}
\Var\Big[\boldsymbol{\Psi}_{\tau}(\boldsymbol{\varepsilon}_{i\tau})\boldsymbol{\varepsilon}_{i\tau}\Big]= \E\Big[\boldsymbol{\Psi}_{\tau}(\boldsymbol{\varepsilon}_{i\tau})\boldsymbol{\varepsilon}_{i\tau}\boldsymbol{\varepsilon}_{i\tau}\transpose
\boldsymbol{\Psi}_{\tau}(\boldsymbol{\varepsilon}_{i\tau})\Big]=\boldsymbol{\Sigma}_{i\tau }, 
\end{equation*}

where $\boldsymbol{\varepsilon}_{i\tau}=(\varepsilon_{i1\tau},\ldots,\varepsilon_{im\tau})\transpose, \ \varepsilon_{ij\tau}=y_{ij}-\boldsymbol{x}_{ij}\transpose\boldsymbol{\beta}_{\tau}\ $ and $ \ \boldsymbol{\Psi}_{\tau }(\boldsymbol{\varepsilon}_{i\tau})=\Big[\diag(\psi_{\tau}(\varepsilon_{ij\tau}))\Big]_{j=1}^{m}.$

\textbf{A2}. The limiting forms of the following matrices are positive definite

\begin{align*}
 \boldsymbol{D}_0 (\tau) &=\;\;
 \lim_{\substack{ \mathllap{m} \rightarrow \mathrlap{\infty} \\  \mathllap{n} \rightarrow \mathrlap{\infty} }} \;\;
 m^{-1} 
 \begin{pmatrix}
  \boldsymbol{Z}\transpose\boldsymbol{\Sigma}_{\tau}\boldsymbol{Z} & \boldsymbol{Z}\transpose\boldsymbol{\Sigma}_{\tau}\boldsymbol{X}/\sqrt{n}  \\
  \boldsymbol{X}\transpose\boldsymbol{\Sigma}_{\tau}\boldsymbol{Z}/\sqrt{n}  & \boldsymbol{X}\transpose\Sigma_{\tau}\boldsymbol{X}/n  
 \end{pmatrix},\\
 \boldsymbol{D}_1 (\tau) &=\;\;
  \lim_{\substack{ \mathllap{m} \rightarrow \mathrlap{\infty} \\ 
 \mathllap{n} \rightarrow \mathrlap{\infty} }} \;\;
 m^{-1} 
 \begin{pmatrix}
  \boldsymbol{Z}\transpose\E[\boldsymbol{\Psi}_{\tau}(\boldsymbol{\varepsilon}_{\tau})]\boldsymbol{Z} & \boldsymbol{Z}\transpose\E[\boldsymbol{\Psi}_{\tau}(\boldsymbol{\varepsilon}_{\tau})]\boldsymbol{X}/\sqrt{n}\\
  \boldsymbol{X}\transpose\E[\boldsymbol{\Psi}_{\tau}(\boldsymbol{\varepsilon}_{\tau})]\boldsymbol{Z}/\sqrt{n}  & \boldsymbol{X}\transpose\E[\boldsymbol{\Psi}_{\tau}(\boldsymbol{\varepsilon}_{\tau})]\boldsymbol{X}/n
 \end{pmatrix},
\end{align*}

where \(\boldsymbol{\Sigma}_{\tau}=\Var[\boldsymbol{\Psi}_{\tau}(\boldsymbol{\varepsilon}_{\tau})\boldsymbol{\varepsilon}_{\tau}]=\diag[\boldsymbol{\Sigma}_{i\tau }]_{i=1}^{n}.\)

\textbf{A3}. The norm of the regressors is bounded by a positive constant $M, \ $ \(\max_{i,j} \norm{\boldsymbol{x}_{ij}}< M.\) 

The stated assumptions \textbf{A1}-\textbf{A3} are standard for panel data models \citep{koenker_quantile_2004}. Condition \textbf{A1} ensures independence across individuals, but allows a within-subject dependency and heterogeneity across individuals. Condition \textbf{A2} is a full rank condition and is used to invoke the Lindeberg-Feller Central Limit Theorem. We observe that, when \(\ \tau =1/2\ \) then \(\boldsymbol{D}_1 (\tau)\) simplifies and Condition \textbf{A2} reduces to a condition on the matrices \( \ \boldsymbol{X}\transpose\boldsymbol{X}/nm \ \) and \( \ \boldsymbol{Z}\transpose\boldsymbol{Z}/m.\ \) Condition \textbf{A3} is useful both for the application of the Lindeberg-Feller Central Limit Theorem and for ensuring the finite dimensional convergence of the objective function.

\begin{theorem}\label{theo1_erfe}
Assume conditions \textbf{A1}-\textbf{A3} are met, with $n,m\rightarrow\infty, \mbox{ and } \E\lvert \psi_{\tau}(\varepsilon_{ij\tau})\rvert^{4+\nu}<\Delta<\infty \mbox{ and }\E\lvert \varepsilon_{ij\tau} \rvert^{4+\nu}<\Delta<\infty$
for some $\nu>0.$ Then $\widehat{\boldsymbol{\delta}}_{1\tau}$ the components of the minimizer, $\widehat{\boldsymbol{\delta}},$ converge in distribution to a Gaussian random vector with mean zero and variance-covariance matrix given by the lower right $p\times p$ block of the matrix $\boldsymbol{D}_1^{-1}(\tau)\boldsymbol{D}_0(\tau)\boldsymbol{D}_1^{-1}(\tau).$ In others words

\begin{equation*}
    \sqrt{nm}\big(\widehat{\boldsymbol{\beta}}_{\tau}-\boldsymbol{\beta}_{\tau}\big)\xrightarrow{d} 
    \mathcal{N}\bigg(0,\Big[\boldsymbol{D}_1^{-1}(\tau)\boldsymbol{D}_0(\tau)\boldsymbol{D}_1^{-1}
    (\tau)\Big]_{22}\bigg).
\end{equation*}
\end{theorem}

To show the closed form of the above matrix \(\Big[\boldsymbol{D}_1^{-1}(\tau)\boldsymbol{D}_0(\tau)\boldsymbol{D}_1^{-1}(\tau)\Big]_{22}\) assume that the limiting forms of the following matrices are positive definite

\begin{equation*}
\begin{split}
  {}&   \widetilde{\boldsymbol{D}}_0(\tau)= \lim_{\substack{ \mathllap{m} \rightarrow \mathrlap{\infty} \\ \mathllap{n} \rightarrow \mathrlap{\infty} }} \; 
    \boldsymbol{X}\transpose \boldsymbol{M}_{\boldsymbol{Z}}(\tau)\transpose\boldsymbol{\Sigma}_{\tau}\boldsymbol{M}_{\boldsymbol{Z}}(\tau)\boldsymbol{X}, \\ 
  {}&  \widetilde{\boldsymbol{D}}_1(\tau)= \lim_{\substack{ \mathllap{m} \rightarrow \mathrlap{\infty} \\  \mathllap{n} \rightarrow \mathrlap{\infty} }} \;
    \boldsymbol{X}\transpose \boldsymbol{M}_{\boldsymbol{Z}}\transpose(\tau)\E[\boldsymbol{\Psi}_{\tau}(\boldsymbol{\varepsilon})]\boldsymbol{M}_{\boldsymbol{Z}}(\tau)\boldsymbol{X}\\
\end{split}
\end{equation*}

where \(\boldsymbol{M}_{\boldsymbol{Z}}(\tau)=\mathbb{I}-\boldsymbol{P}_{\boldsymbol{Z}}(\tau) \mbox{ and } \boldsymbol{P}_{\boldsymbol{Z}}(\tau)=\boldsymbol{Z}\Big[\boldsymbol{Z}\transpose\E[\boldsymbol{\Psi}_{\tau}(\boldsymbol{\varepsilon})] \boldsymbol{Z}\Big]^{-1} \boldsymbol{Z}\transpose\E[\boldsymbol{\Psi}_{\tau}(\boldsymbol{\varepsilon})].\)

Under the above conditions and the conditions of \textbf{Theorem \ref{theo1_erfe}} it follows that:

\begin{lemma}\label{lem1_erfe}
\begin{equation*}
    \Big[\boldsymbol{D}_1^{-1}(\tau)\boldsymbol{D}_0 (\tau)\boldsymbol{D}_1^{-1}(\tau)\Big]_{22}=\widetilde{\boldsymbol{D}}_1^{-1}(\tau)\widetilde{\boldsymbol{D}}_0(\tau)
    \widetilde{\boldsymbol{D}}_1^{-1}(\tau).
\end{equation*}
\end{lemma}

\textbf{ Asymptotics for several expectiles }

The asymptotic properties of the $\ERFE$ estimator for a sequence of asymmetric points \(\boldsymbol{\tau}=(\tau_1,\cdots,\tau_q)\) are derived using the transformed data, \(\ [\boldsymbol{y}^{*};\boldsymbol{X}^{*}],\ \) where \(\boldsymbol{X}^{*}=\boldsymbol{M}_{\boldsymbol{Z}}(\tau)\boldsymbol{X} \ \mbox{ and } \ \boldsymbol{y}^{*} = \boldsymbol{M}_{\boldsymbol{Z}}(\tau)\boldsymbol{y}.\) Both
projection matrices \(\boldsymbol{M}_{\boldsymbol{Z}}(\tau) \mbox{ and } \boldsymbol{P}_{\boldsymbol{Z}}(\tau)\) are idempotent and are defined as:

\begin{equation*}
\boldsymbol{M}_{\boldsymbol{Z}}(\tau)=\mathbb{I}_{N}-\boldsymbol{P}_{\boldsymbol{Z}}(\tau), \quad
\boldsymbol{P}_{\boldsymbol{Z}}(\tau)=\boldsymbol{Z}(\boldsymbol{Z}\transpose
\E[\boldsymbol{\Psi}_{\tau}(\boldsymbol{\varepsilon}^{*}_{\tau})]\boldsymbol{Z})^{-1}\boldsymbol{Z}\transpose
\E[\boldsymbol{\Psi}_{\tau}(\boldsymbol{\varepsilon}^{*}_{\tau})],
\end{equation*}

where \(\boldsymbol{\varepsilon}^{*}_{\tau}=\boldsymbol{y}^{*}- \boldsymbol{X}^{*}\boldsymbol{\beta}_{\tau}.\)

A robust estimator of the covariance matrix is also proposed. Assume the following conditions.

\textbf{B1}. The data \(\lbrace (\boldsymbol{y}_i,\boldsymbol{X}_i) \rbrace_{i=1}^{n}\) are independent across \(i\) and,

\begin{equation*}
\Var\Big[\boldsymbol{\Psi}_{\boldsymbol{\tau}}(\boldsymbol{\varepsilon}_{i\boldsymbol{\tau}}^{*})
\boldsymbol{\varepsilon}_{i\boldsymbol{\tau}}^{*}\Big]= \E\Big[\boldsymbol{\Psi}_{\boldsymbol{\tau}}(\boldsymbol{\varepsilon}_{i\boldsymbol{\tau}}^{*})
\boldsymbol{\varepsilon}_{i\boldsymbol{\tau}}^{*}
\boldsymbol{\varepsilon}_{i\boldsymbol{\tau}}^{*}\transpose
\boldsymbol{\Psi}_{\boldsymbol{\tau}}(\boldsymbol{\varepsilon}_{i\boldsymbol{\tau}}^{*}) \Big]=\boldsymbol{\Sigma}_{i\boldsymbol{\tau}}^{*}, 
\end{equation*}

where  $\boldsymbol{\varepsilon}_{i\boldsymbol{\tau}}^{*}=\Big(\boldsymbol{\varepsilon}_{i\tau_1}^{*}\transpose,\ldots,\boldsymbol{\varepsilon}_{i\tau_q}^{*}\transpose\Big)\transpose, \; \boldsymbol{\varepsilon}_{i\tau_k}^{*}=(\varepsilon_{i1\tau_k}^{*},\ldots,\varepsilon_{im\tau_k}^{*})\transpose, \; \varepsilon_{ij\tau_k}^{*}=y_{ij}^{*}-\boldsymbol{x}_{ij}^{*}\transpose\boldsymbol{\beta}_{\tau_k} \ $ and
$ \ \boldsymbol{\Psi}_{\boldsymbol{\tau}}(\boldsymbol{\varepsilon}_{i\boldsymbol{\tau}}^{*})=
[\diag(\boldsymbol{\Psi}_{\tau_k}(\boldsymbol{\varepsilon}_{i{\tau_k}}^{*}))]_{k=1}^{q}.$

\textbf{B2}. The limiting forms of the following matrices are positive definite

\begin{equation*}
\begin{split}
    {}& \boldsymbol{D}_0 (\boldsymbol{\tau})= \lim_{\substack{ \mathllap{n} \rightarrow \mathrlap{\infty} }} \quad
     (\boldsymbol{V}\otimes\boldsymbol{X}^{*})\transpose\E[\boldsymbol{\Psi}_{\boldsymbol{\tau}}(\boldsymbol{\varepsilon}_{\boldsymbol{\tau}}^{*})
     \boldsymbol{\varepsilon}_{\boldsymbol{\tau}}^{*}\boldsymbol{\varepsilon}_{\boldsymbol{\tau}}^{*}\transpose
    \boldsymbol{\Psi}_{\boldsymbol{\tau}}(\boldsymbol{\varepsilon}_{\boldsymbol{\tau}}^{*})](\boldsymbol{V}\otimes\boldsymbol{X}^{*})/nm. \\
    {}& \boldsymbol{D}_1 (\boldsymbol{\tau})= \lim_{\substack{ \mathllap{n} \rightarrow \mathrlap{\infty} }} \quad
     (\mathbf{I}_q\otimes\boldsymbol{X}^{*})\transpose\E[\boldsymbol{\Psi}_{\boldsymbol{\tau}}(\boldsymbol{\varepsilon}_{\boldsymbol{\tau}}^{*})](\boldsymbol{V}\otimes\boldsymbol{X}^{*})/nm \\
\end{split}
\end{equation*}

\textbf{B3}. The norm of the regressors is bounded by a positive constant $M, \ $ \(\max_{\substack{ 1\leq i \leq n \\ 1 \leq j \leq m} }\norm{x_{ij}^{*}} < M.\) 

\begin{theorem}\label{theo2_erfe}
Suppose conditions \textbf{B1}-\textbf{B3} are satisfied, and that $n,m\rightarrow\infty.$ If $\; \E\lvert \psi_{\tau}(\varepsilon_{ij\tau}^{*})\rvert^{4+\nu}<\Delta<\infty \mbox{ and }\E\lvert \varepsilon_{ij\tau}^{*} \rvert^{4+\nu}<\Delta<\infty$  then 
\begin{equation*}
 \sqrt{nm}\big(\widehat{\boldsymbol{\beta}}_{\boldsymbol{\tau}}-
    \boldsymbol{\beta}_{\boldsymbol{\tau}}\big)\xrightarrow{d} 
    \mathcal{N}\bigg(0,    \boldsymbol{D}_1^{-1}(\boldsymbol{\tau})
    \boldsymbol{D}_0(\boldsymbol{\tau})\boldsymbol{D}_1^{-1}
    (\boldsymbol{\tau})\bigg).
\end{equation*}
\end{theorem}

In order to use the $\ERFE$ estimator to make inference, an estimator of its covariance matrix is presented in \textbf{Theorem \ref{theo3_erfe}}. This will make it possible to construct large sample confidence intervals or hypothesis tests. The proposed covariance matrix estimator is robust and consistent, and is a generalization of the commonly advocated covariance matrix estimator proposed by \citet{White1980}.

\begin{theorem}\label{theo3_erfe}

Let the matrices $\widehat{\boldsymbol{D}}_0(\boldsymbol{\tau})$ and $\widehat{\boldsymbol{D}}_1(\boldsymbol{\tau})$ defined as:
\begin{align*}
    \begin{split}
        {}& \widehat{\boldsymbol{D}}_0(\boldsymbol{\tau})=\frac{1}{nm}(\boldsymbol{V}\otimes\widehat{ \boldsymbol{X}^{*}})\transpose\boldsymbol{\Psi}_{\boldsymbol{\tau}}(\widehat{\boldsymbol{\varepsilon}_{\boldsymbol{\tau}}^{*}})
        \widehat{\boldsymbol{\varepsilon}_{\boldsymbol{\tau}}^{*}}\widehat{\boldsymbol{\varepsilon}_{\boldsymbol{\tau}}^{*}}\transpose
        \boldsymbol{\Psi}_{\boldsymbol{\tau}}(\widehat{\boldsymbol{\varepsilon}_{\boldsymbol{\tau}}^{*}}) (\boldsymbol{V}\otimes\widehat{\boldsymbol{X}^{*}}), \\
        {}& \widehat{\boldsymbol{D}}_1(\boldsymbol{\tau})=\frac{1}{nm}(\mathbf{I}_q\otimes\widehat{\boldsymbol{X}^{*}})\transpose\boldsymbol{\Psi}_{\boldsymbol{\tau}}(\widehat{\boldsymbol{\varepsilon}_{\boldsymbol{\tau}}^{*}})(\boldsymbol{V}
        \otimes\widehat{\boldsymbol{X}^{*}}),
    \end{split}
\end{align*}

where the transformed data is obtained by pre-multiplying the initial data with the projection matrix $\widehat{\boldsymbol{M}}_{\boldsymbol{Z}}(\boldsymbol{\tau})=\mathbb{I}_{nmq}-\widehat{\boldsymbol{P}}_{\boldsymbol{Z}}
(\boldsymbol{\tau})$ and

\begin{equation*}
    \widehat{\boldsymbol{P}}_{\boldsymbol{Z}}(\boldsymbol{\tau})=(\boldsymbol{v}\otimes\boldsymbol{Z})
\Big[(\boldsymbol{v}\otimes\boldsymbol{Z})\transpose  \boldsymbol{\Psi}_{\boldsymbol{\tau}}(\widehat{\boldsymbol{\varepsilon}_{\boldsymbol{\tau}}^{*}})
(\mathds{1}_q\otimes\boldsymbol{Z})\Big]^{-1}(\mathds{1}_q\otimes\boldsymbol{Z})\transpose \boldsymbol{\Psi}_{\boldsymbol{\tau}}(\widehat{\boldsymbol{\varepsilon}_{\boldsymbol{\tau}}^{*}}).
\end{equation*}

Then, for every fixed, $\boldsymbol{\tau}$ we have:

\begin{equation*}
  \widehat{\boldsymbol{D}}_1^{-1}(\boldsymbol{\tau}) \widehat{\boldsymbol{D}}_0(\boldsymbol{\tau}) \widehat{\boldsymbol{D}}_1^{-1}(\boldsymbol{\tau})
  \xrightarrow{p} 
  \boldsymbol{D}_1^{-1}(\boldsymbol{\tau}) \boldsymbol{D}_0(\boldsymbol{\tau})\boldsymbol{D}_1^{-1}(\boldsymbol{\tau}).
\end{equation*}
\end{theorem}

We end this section with the result for a single \(\tau.\)

\begin{corollary}\label{cor1_erfe}
Let the matrices $\widehat{\boldsymbol{D}}_0(\tau)$ and $\widehat{\boldsymbol{D}}_1(\tau)$ defined as:

\begin{align*}
    \begin{split}
{}& \widehat{\boldsymbol{D}}_0(\tau)= \frac{1}{nm}\sum_{i=1}^{n}\widehat{\boldsymbol{X}_{i}^{*}}\transpose\boldsymbol{\Psi}_{\tau}(\widehat{\boldsymbol{\varepsilon}_{i\tau}^{*}})\widehat{\boldsymbol{\varepsilon}_{i\tau}^{*}}\widehat{\boldsymbol{\varepsilon}_{i\tau}^{*}}\transpose
\boldsymbol{\Psi}_{\tau}(\widehat{\boldsymbol{\varepsilon}_{i\tau}^{*}})\widehat{\boldsymbol{X}_{i}^{*}},\\
{}& \widehat{\boldsymbol{D}}_1(\tau)=\frac{1}{nm}\sum_{i=1}^{n}\widehat{\boldsymbol{X}_{i}^{*}}\transpose\boldsymbol{\Psi}_{\tau}
(\widehat{\boldsymbol{\varepsilon}_{i\tau}^{*}})\widehat{\boldsymbol{X}_{i}^{*}}
    \end{split}
\end{align*}

with the corresponding projection matrices

\begin{equation*}
\widehat{\boldsymbol{M}}_{\boldsymbol{Z}}(\tau)=\mathbb{I}_{N}-\widehat{\boldsymbol{P}}_{\boldsymbol{Z}}(\tau) \mbox{ and }
\widehat{\boldsymbol{P}}_{\boldsymbol{Z}}(\tau)=\boldsymbol{Z}(\boldsymbol{Z}\transpose
\boldsymbol{\Psi}_{\tau}(\widehat{\boldsymbol{\varepsilon}_{\tau}^{*}})\boldsymbol{Z})^{-1}\boldsymbol{Z}\transpose\boldsymbol{\Psi}_{\tau}(
\widehat{\boldsymbol{\varepsilon}_{\tau}^{*}}).
\end{equation*}

Then, under the above conditions and for every fixed $\tau,$ we have

\begin{equation*}
  \widehat{\boldsymbol{D}}_1^{-1}(\tau) \widehat{\boldsymbol{D}}_0(\tau) \widehat{\boldsymbol{D}}_1^{-1}(\tau)
  \xrightarrow{p} 
  \boldsymbol{D}_1^{-1}(\tau) \boldsymbol{D}_0(\tau)\boldsymbol{D}_1^{-1}(\tau).
\end{equation*}

\end{corollary}

\section{Simulations}\label{simulation_erfe}
In this section we conducted a simulation study to evaluate the performance of the $\ERFE$ estimator. We started by presenting the simulation design, then the metrics to evaluate the estimators and the results.

\subsection{Design}

The random samples were generated from the following linear model:
\begin{equation}\label{sim_mod_erfe}
    y_{ij} = x_{ij1}\beta_1 + x_{ij2}\beta_2 + \alpha_i + (1 + \gamma x_{ij2})\varepsilon_{ij}, \qquad i\in\lbrace 1,\ \ldots, \ n \rbrace \; \mbox{ and } \; j\in\lbrace 1,\ \ldots, \ m \rbrace. 
\end{equation}

We considered two versions of model equation (\ref{sim_mod_erfe}) according to the heteroscedastic parameter \(\gamma\in\lbrace 0, \ 3/10 \rbrace.\)
The value of \(\gamma=0\) corresponds to a location shift model \((M_0)\) where the regressors are uncorrelated to the random error. The model \((M_0)\) is used to assess the performance of the estimators for a homoscedastic scenario. In contrast, when the value of \(\gamma = 3/10,\) then there is a correlation between the predictor $x_{2}$ and the random error. In that case, the model is a location-scale shift model \((M_{3/10})\) and is set to assess the performance of the estimators in the presence of heteroscedasticity.

In the location shift scenario, the $\ERFE$ model corresponds to \(\mu_{\tau}(y_{ij})=x_{ij1}\beta_1 + x_{ij2}\beta_2 + \alpha_i + \mu_{\tau}(\varepsilon_{ij})\) where only the intercept term, \(\beta_{0\tau}=\alpha_i +\mu_{\tau}(\varepsilon_{ij}),\) varies with \(\tau\) and the expectile functions are parallel lines. In the location-scale shift scenario, the related $\ERFE$ model is defined as: \(\mu_{\tau}(y_{ij})= x_{ij1}\beta_1 + x_{ij2}\beta_{2\tau} + \alpha_i + \mu_{\tau}(\varepsilon_{ij})\) where the intercept $\beta_{0\tau}= \alpha_i + \mu_{\tau}(\varepsilon_{ij})$ and $\beta_{2\tau}= \beta_{2} + \gamma\mu_{\tau}(\varepsilon_{ij}).$ 
Therefore, in the presence of heteroscedasticity both the intercept and the slope of the predictor $x_2$ vary with \(\tau.\)

The parameters are set to $\beta_1=0.6$ and $\beta_2=1,$ and the corresponding regressors are generated from a non-central student distribution with 3 degree of freedom $(\mathcal{T}_2(1.3))$ and a normal distribution \((\mathcal{N}(2,\ 1.5)),\) respectively. The individual-specific effects parameter \(\alpha\) is generated from a normal distribution \((\mathcal{N}(1,\ 1)),\) and is correlated $(\rho=0.5)$ to the predictor $x_2.$ Indeed, in real data applications it is more likely that omitted factors are correlated with regressors in the model. The random error \(\varepsilon\) of the model equation (\ref{sim_mod_erfe}) is generated from three different distributions: normal distribution \((\mathcal{N}(0,1)),\) Student distribution $(\mathcal{T}_3)$ with 3 degrees of freedom, and chi-squared distribution \((\chi^2_3)\) with \(3\) degrees of freedom. We have set the sample size and the repeated measurements to \(n\times m\in \ \lbrace 100,\ 250,\ 500 \rbrace \times \lbrace 5,\ 15,\ 30 \rbrace.\)  The extensive simulation was carried out with 400 replications. In each case the focus is on the regressor effects at the asymmetric points $\tau\in\lbrace 0.1, \ 0.3, 0.5, \ 0.8, \ 0.9 \rbrace.$

All simulations were conducted using high performance computing clusters provided by Calcul Quebec and Compute Canada. All computations were performed with the R (v3.6.0) statistical programming language \citet{rcran}. The implemented R package \textbf{erfe} that comes with this manuscript is publicly available on GitHub at \url{https://github.com/AmBarry/erfe}.

\subsection{Performance measures}
We compared our $\ERFE$ model to the quantile regression with fixed-effects $(\QRFE)$ model proposed by \citet{koenker_quantile_2004}. The $\QRFE$ model estimated the parameter of interest and the nuisance parameter of the model which could be computationally demanding as the sample size increased. We also considered the expectile regression model $(\ER)$ and the quantile regression model $(\QR),$ which ignored the individual fixed-effects parameter. Given that expectile and quantile of the same level $\tau$ were generally different, we carried out the appropriate conversions between the asymmetric points and the percentiles to ensure that the expectile-based regressions and the quantile-based regressions estimated the same statistics (that is quantiles and expectiles are identical). For example, the Gaussian quantiles of level \(\tau=(0.33, \ 0.5, \ 0.67)\) are identical to the Gaussian expectiles of level \(\tau=(0.25,\ 0.5,\ 0.75).\)  In other words, the $\ER$ based-model and the $\QR$ based-model estimate the same locations of the response distribution.

We evaluated the quality of the estimators by reporting the distribution of their coefficient estimate as box-plots. We also evaluated the performance of the asymptotic standard error $(\SE)$ presented in Theorem \ref{theo3_erfe} by reporting the distribution of the ratio between the asymptotic standard error $(\SE)$ and the Monte Carlo standard deviation $(\SD)$ defined as: 

\begin{equation*}
\SD^{2}(\beta_{k\tau}) = \frac{1}{400}  \sum_{j=1}^{400} \Big(\widehat{\beta}^{(j)}_{k\tau} - \overline{\widehat{\beta}}_{k\tau}\Big)^2, \, \quad k \in \lbrace 1,\ 2\rbrace, 
\end{equation*}

where $\overline{\widehat{\beta}}_{k\tau} = \frac{1}{400}  \sum_{j=1}^{400} \widehat{\beta}^{(j)}_{k\tau}.$ 

We estimated the $\ERFE$ model with the \textbf{erfe} package and the $\QRFE$ model with the rqpd package \citep{rqpd}. The $\ER$ model and the $\QR$ model was obtained from the well-known packages: expectreg \citep{expectreg} and quantreg \citep{quantreg}, respectively. 

\subsection{Results}

\begin{figure}[H]
\centering
\includegraphics[width=0.8\linewidth]{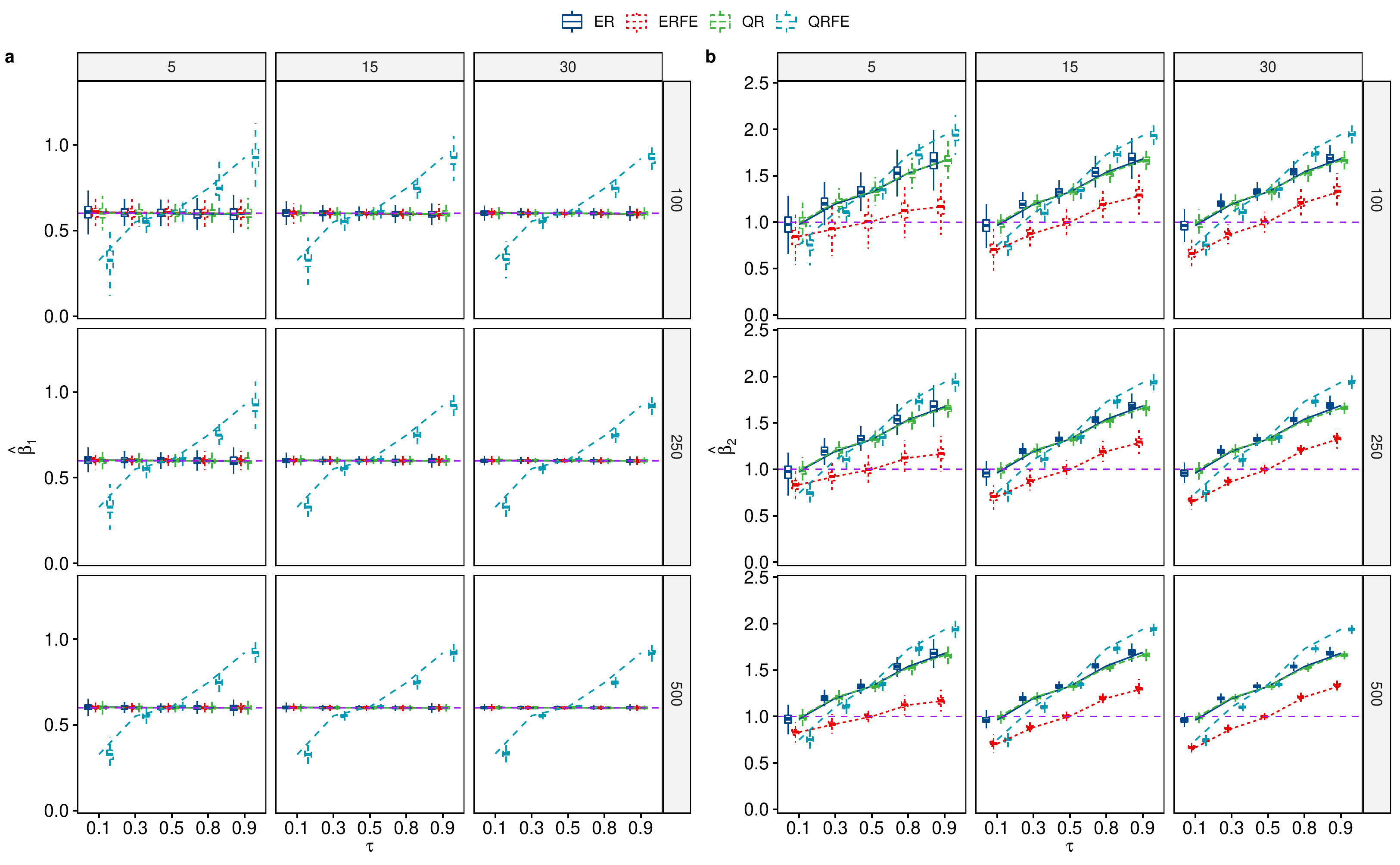}
 \caption{Distribution of the coefficient estimates of the parameter $\beta_1$ (Figure \ref{fig:gnorm2_erfe}\textbf{a}) and the parameter $\beta_2$ (Figure \ref{fig:gnorm2_erfe}\textbf{b}) represented as boxplot according to the sample size $n\in(100,  \ 250,  \ 500),$ the repeated measurements $m=(5,\ 15,\ 30),$ the asymmetric points $\tau\in (0.1,  \ 0.3,  \  0.5, \  0.8,\ 0.9)$ and the error term $\varepsilon\sim\mathcal{N}(0, \ 1)$ in the location-scale-shift scenario.}\label{fig:gnorm2_erfe}
\end{figure}

We present here the results related to the Gaussian random error and we brought the results for the Student and Chi-square random errors in the \textbf{Supplementary} file. \textbf{Figure \ref{fig:gnorm1_erfe}} and \textbf{Figure \ref{fig:gnorm2_erfe}} report the distribution of the coefficient estimates in the location-shift and location-scale-shift scenarios, respectively. 

In the location-shift scenario, we observe that the coefficient estimates of our $\ERFE$ model are centered around the true value of the parameters with a small interquartile range. We also observe that the coefficient estimate of the $\ER$ and $\QR$ models are centered around the true value for the parameter $\beta_1$ only. We notice that the coefficient estimates of the $\QRFE$ model are not close to the true value of the parameters except when $\tau=0.5$ for the parameter $\beta_1$ only. In other words our $\ERFE$ model performs well in estimating the parameter coefficients of the model in the location-shift scenario. The $\ER$ and $\QR$ models perform similarly in the location-shift scenario, with an unbiased estimator for the parameter $\beta_1$ and a biased estimator for the parameter $\beta_2.$ The $\ER$ and $\QR$ models do not account for the individual fixed-effects which are correlated to the regressor $x_2,$ which could explain the bias for the parameter $\beta_2.$ In contrast, the $\QRFE$ model performed poorly in estimating the parameter coefficients of the model in the location-shift scenario. The $\QRFE$ model includes the individual fixed-effects in its specification but, similarly to the random-effect model, it did not account for the dependence between the individual fixed-effects and the regressors of the model. This could explained the poor performance of the $\QRFE.$ 

Indeed, similar to the within-estimator, the $\ERFE$ model transforms the data by subtracting the person-specific expectile of level $\tau$ from the observed values of each variable and then applied the $\ER$ method to the de-expectilized model given by:

\begin{equation}\label{eq:de_expect}
y_{ij}^{*} = x_{ij1}^{*}\beta_1 + x_{ij2}^{*}\beta_2 + \varepsilon_{ij}^{*}, 
\end{equation}

where $y_{ij}^{*} = y_{ij} - \widehat{\mu}_{\tau}(y_{ij}), \ x_{ij}^{*}$ and $\varepsilon_{ij}^{*}$ are defined similarly. This transformation concentrated out the individual fixed-effects and any bias that could result from its association with the regressors.

The $\ER$ and $\QR$ models do not take into account the individual fixed-effects parameter, which is included in the random error component. Since, the individual fixed-effects parameter is correlated to the predictor $x_2,$ then the random error of the model equation (\ref{sim_mod_erfe}) is also correlated to the predictor $x_2$ of the model. Hence, the coefficient estimate of the $\ER$ and $\QR$ methods for the parameter $\beta_2$ is biased.

Consider, the reformulation of model equation (\ref{sim_mod_erfe}) in the location-shift scenario: 

\begin{equation*}
    \begin{split}
        y_{ij} & = x_{ij1}\beta_1 + x_{ij2}\beta_2 + Q_{\tau}(\alpha_i| x_{ij1}, x_{ij2}) + (\alpha_i - Q_{\tau}(\alpha_i| x_{ij1}, x_{ij2})) +
        \varepsilon_{ij} \\
               & = x_{ij1}\beta_1 + x_{ij2}\beta_2 + Q_{\tau}(\alpha_i| x_{ij1}, x_{ij2}) + \eta_{ij}, \qquad \eta_{ij} = (\alpha_i - Q_{\tau}(\alpha_i| x_{ij1}, x_{ij2}) ) + \varepsilon_{ij}, \\
    \end{split}
\end{equation*}

where $Q_{\tau}(\alpha_i| x_{ij1}, x_{ij2})$ is the quantile of the individual fixed-effects $\alpha_i$ of level $\tau$  and $\eta_{ij}$ the new random variable. The corresponding $\QRFE$ model, for a fixed $\tau,$ can be specified as: $ Q_{\tau}(y_{ij}| x_{ij1}, x_{ij2})=x_{ij1}\beta_1 + x_{ij2}\beta_2 + f(x_{ij1}, x_{ij2}), \ $ where $ \ f(x_{ij1}, x_{ij2})=Q_{\tau}(\alpha_i| x_{ij1}, x_{ij2})$ (since the individual fixed-effects are correlated to the regressors). Thus, in this context, the coefficient estimates of the $\QRFE$ model would be biased.

\textbf{Figure \ref{fig:gnorm2_erfe}} report the distribution of the coefficient estimates in the location-scale-shift scenario. Again, we observe that the $\ERFE$ model performs well in estimating the parameter coefficients of the model and outperformed its competitors. 
The apparent bias of the $\ERFE$ estimator for the parameter $\beta_2$ is due to the effect of the heteroscedasticity in this formulation of the model and is not surprising. Indeed, in the location-scale-shift scenario, because of the correlation between the predictor $x_2$ and the error term, the parameter of the predictor $x_2$ is function of the asymmetric point and then different to $\beta_2$ except when $\tau=0.5$ for the symmetric distributions (Normal and Student), where the expectile of level $\tau=0.5$ is zero. The same remark could be applied to the other methods which in addition did not account for the individual fixed-effects $(\ER \ \mbox{ and } \ \QR)$ or its correlation with the regressors $(\QRFE).$
We observed similar results for the Student and Chi-square random errors (results are available in the \textbf{Supplementary} file). 

Overall, the $\ERFE$ model outperforms its competitor and extends the favorable properties of the fixed-effects model. The $\ERFE$ model accounts for the time-invariant omitted variables and for the heteroscedasticity present in the data.

\begin{figure}[hbt!]
\centering
\includegraphics[width=0.75\linewidth]{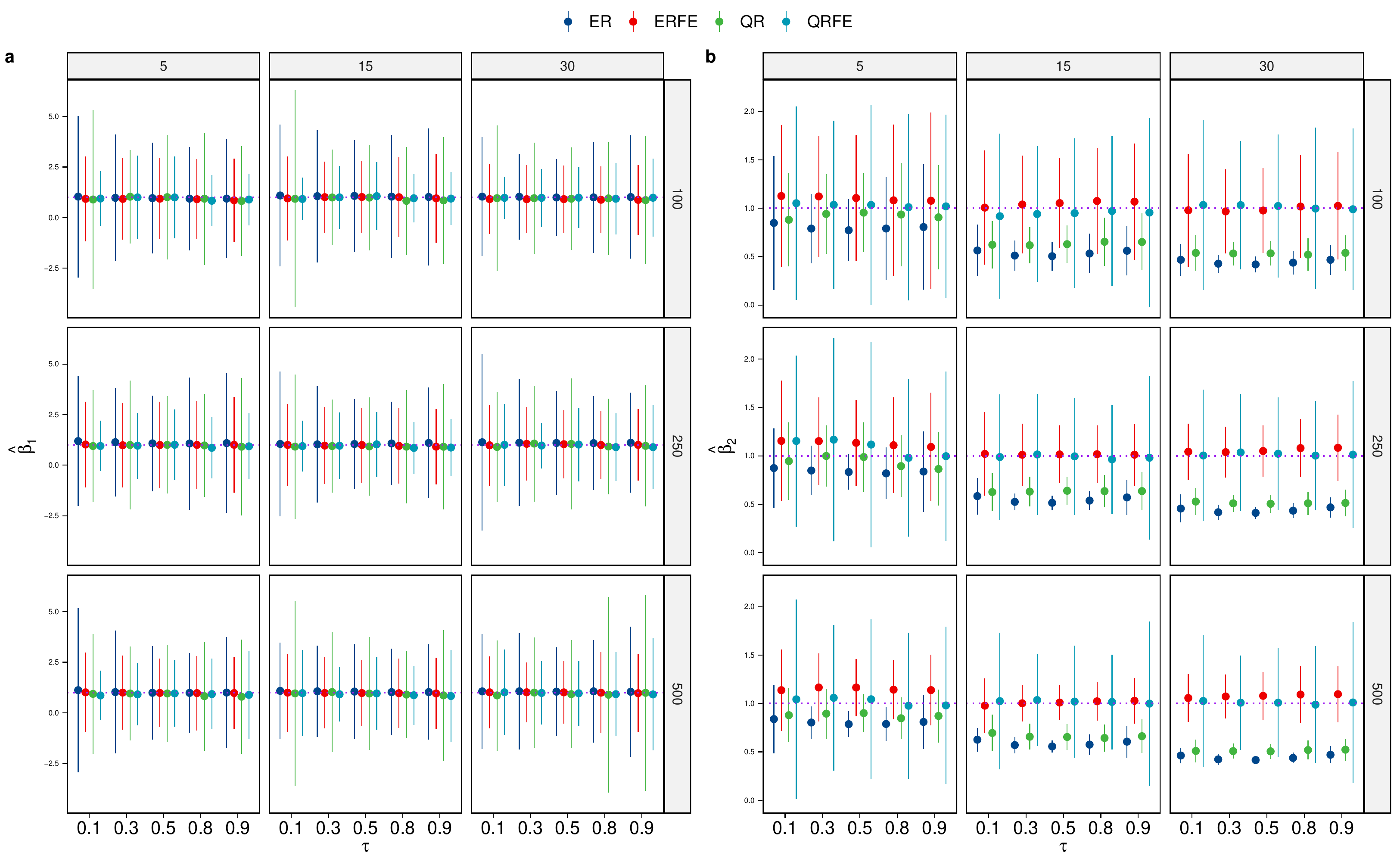}
    \caption{Distribution of the ratio $\frac{\SE}{\SD}$ for the parameter estimate $\widehat{\beta}_1$ (Figure \ref{fig:SdSe_norm_homo}\textbf{a}) and the parameter estimate $\widehat{\beta}_2$ (Figure \ref{fig:SdSe_norm_homo}\textbf{b}) represented as an error plot with respect to the sample size $n\in(100,  \ 250,  \ 500),$ the repeated measurements $m=(5,\ 15,\ 30),$ the asymmetric points $\tau\in (0.1,  \ 0.3,  \  0.5, \  0.8,\ 0.9)$ and the error term $\varepsilon\sim\mathcal{N}(0, \ 1)$ in the location-shift scenario.} \label{fig:SdSe_norm_homo}
\end{figure}

\begin{figure}[hbt!]
\centering
\includegraphics[width=0.75\linewidth]{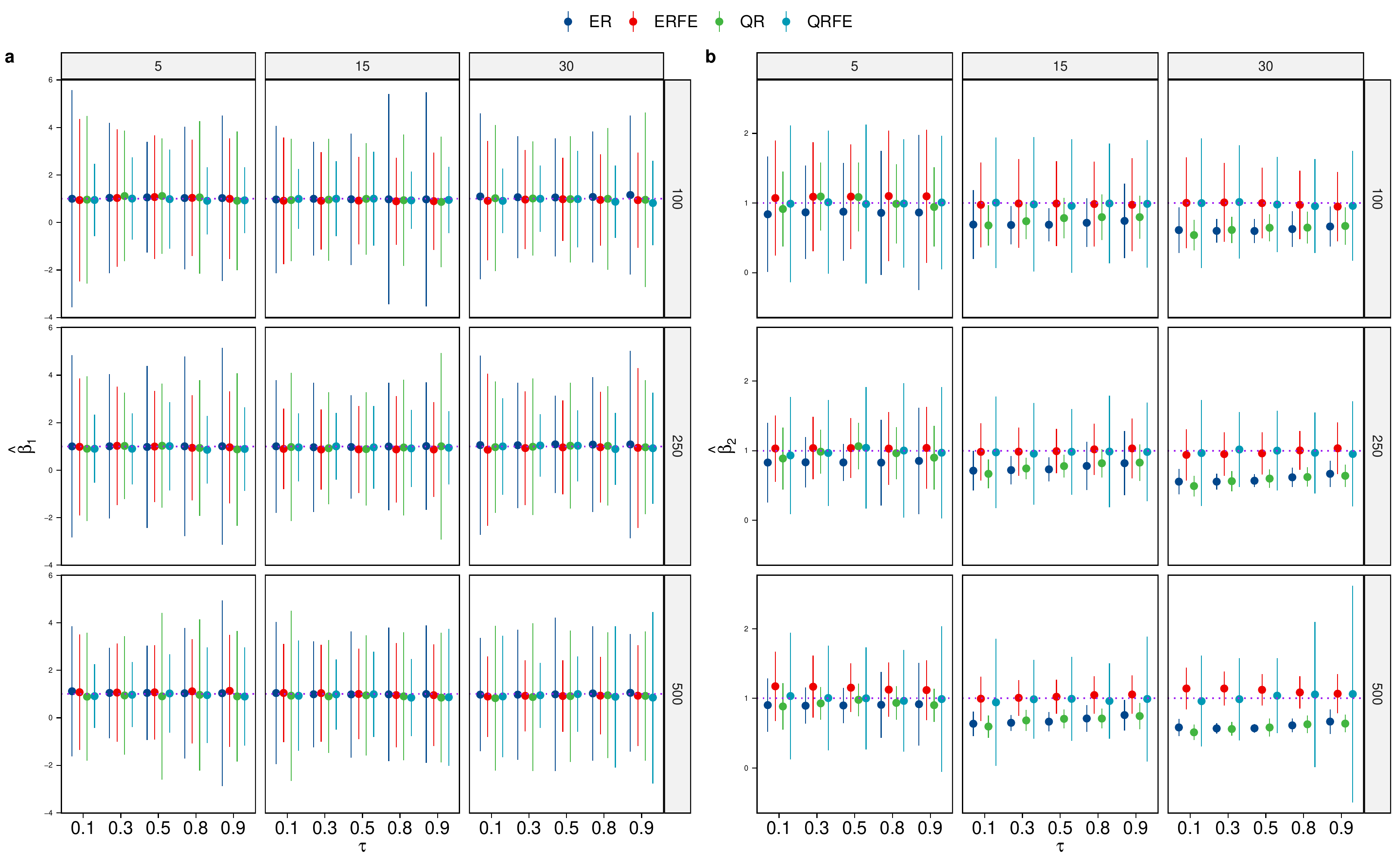}
    \caption{Distribution of the ratio $\frac{\SE}{\SD}$ for the parameter estimate $\widehat{\beta}_1$ (Figure \ref{fig:SdSe_norm_hetero}\textbf{a}) and the parameter estimate $\widehat{\beta}_2$ (Figure \ref{fig:SdSe_norm_hetero}\textbf{b}) represented as an error plot with respect to the sample size $n\in(100,  \ 250,  \ 500),$ the repeated measurements $m=(5,\ 15,\ 30),$ the asymmetric points $\tau\in (0.1,  \ 0.3,  \  0.5, \  0.8,\ 0.9)$ and the error term $\varepsilon\sim\mathcal{N}(0, \ 1)$ in the location-scale-shift scenario.} \label{fig:SdSe_norm_hetero}
\end{figure}

To evaluate the asymptotic standard error $(\SE)$ of the $\ERFE$ parameter estimates, we use the Monte Carlo standard deviation $(\SD)$ as a benchmark and present the distribution of the ratio $\frac{\SE}{\SD}$ as an error plot centered at the mean, \textbf{Figure \ref{fig:SdSe_norm_homo}} and \textbf{Figure \ref{fig:SdSe_norm_hetero}}. In general the error plots of the $\ERFE$ model and the $\QRFE$ model are centered around 1, which means that on average the asymptotic standard error $\SE$ and the Monte Carlo standard deviation $\SD$ are identical. However, we observe that the error plots of the $\ER$ model and the $\QR$ model are not centered around the mean for the $\beta_2$ parameter and the range of their error plot is generally larger. Similar performances were observed for the Student and Chi-Squared random error which results can be found in the \textbf{Supplementary} file.

\begin{figure}[hbt!]
\centering
\includegraphics[width=0.5\linewidth]{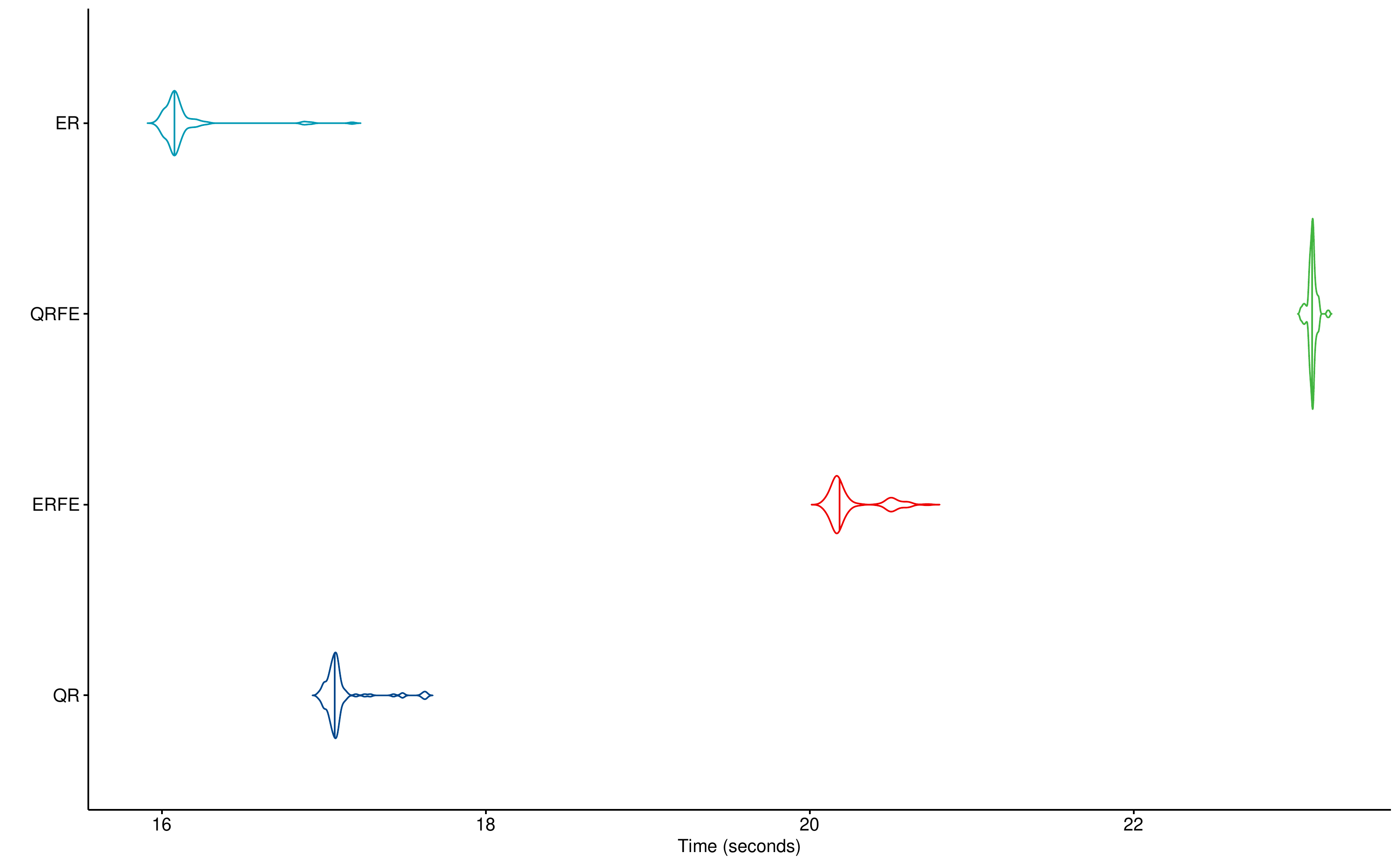}
    \caption{Distribution of the computation time (in seconds) of the $\ER$-based algorithms and the $\QR$-based algorithms. The algorithms are fitted to a dataset $(n=300, \ m=10)$ generated by a location-shift model with a Gaussian random error. All other settings are identical to those used in the simulation. } \label{fig:bench}
\end{figure}

We end this section by comparing the run-times of the $\ER$-based algorithms and the $\QR$-based algorithms. We fitted the methods to a dataset $(n=300, \ m=10)$ generated by a location-shift model with a Gaussian random error. We used the microbenchmark package \citep{microbenchmark2019} with 100 replications to evaluate the computation time of the different algorithm. The results in \textbf{Figure \ref{fig:bench}} show that the cross-sectional algorithms $\ER$ and $\QR$ are the fastest algorithms, and our $\ERFE$ algorithm is faster than the $\QRFE$ algorithm.
We also performed the comparison for a larger sample size $(n>2500),$ but the algorithm stopped due to a shortage of memory for the $\QRFE$ algorithm. This problem has also been reported by \citet{canaySimpleApproachQuantile2011a}.

\begin{figure}[H]
\centering
\includegraphics[width=0.8\linewidth]{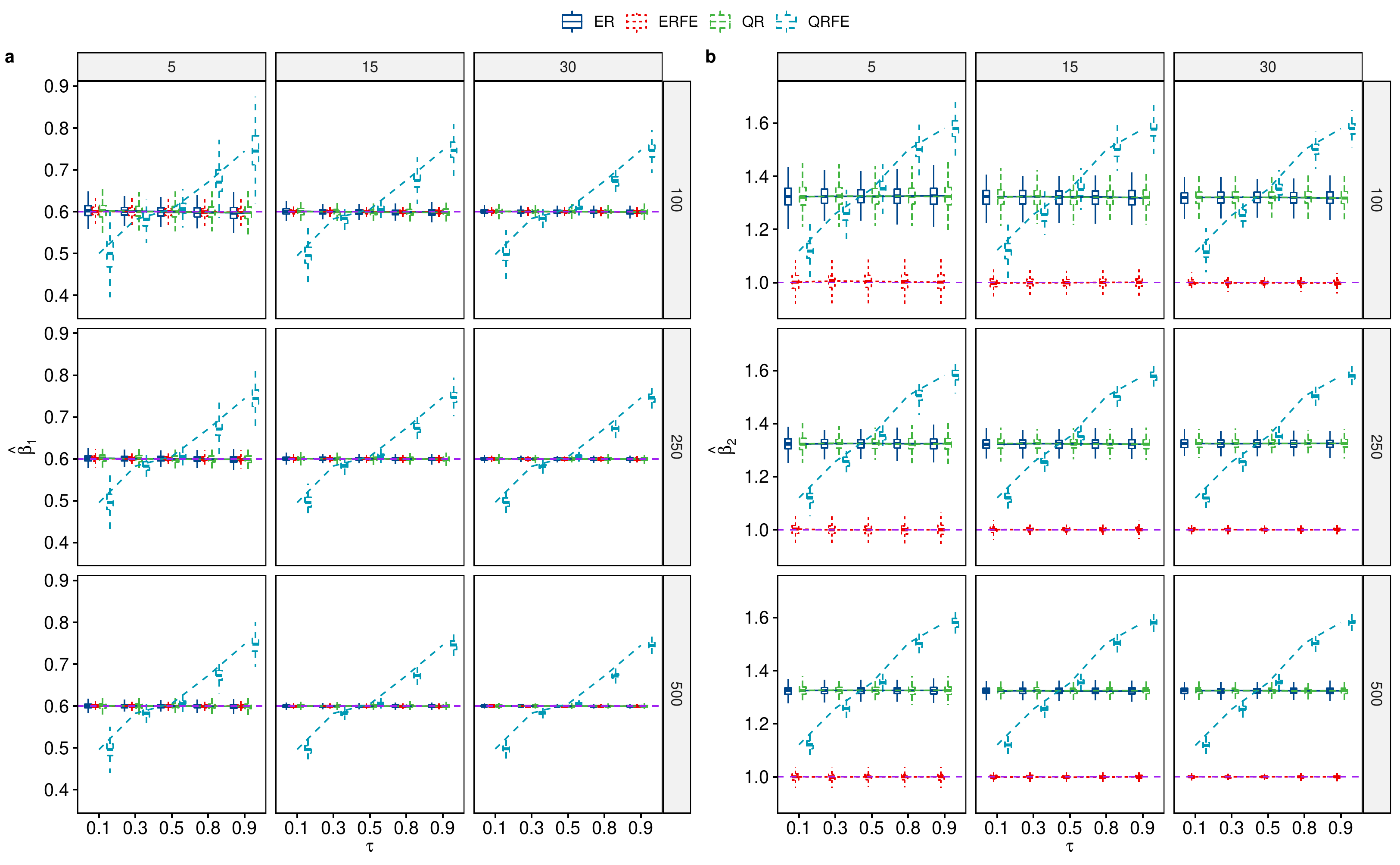}
 \caption{Distribution of the coefficient estimates of the parameter $\beta_1$ (Figure \ref{fig:gnorm1_erfe}\textbf{a}) and the parameter $\beta_2$ (Figure \ref{fig:gnorm1_erfe}\textbf{b}) represented as boxplot according to the sample size $n\in(100,  \ 250,  \ 500),$ the repeated measurements $m=(5,\ 15,\ 30),$ the asymmetric points $\tau\in (0.1,  \ 0.3,  \  0.5, \  0.8,\ 0.9)$ and the error term $\varepsilon\sim\mathcal{N}(0, \ 1)$ in the location-shift scenario.}\label{fig:gnorm1_erfe}
\end{figure}

\section{Application}\label{Application_erfe}

\textit{Returns to schooling} also known as returns to education is a topic widely studied in empirical economics. It is often presented in standard econometric textbooks \citep{baltagi2008, Greene2011, CameronTrivedi2005} as an example of an endogeneity model. Indeed, there is a potential correlation between individual's ability and the other regressors such as education. In the presence of endogeneity, the $\FE$ model is often preferred than other mean regression models for panel data. Despite the fact that it does not estimate the effect of the time-invariant regressors, the $\FE$ estimator is consistent even if the individual effects are correlated with the regressors of the model \citep{baltagi2008}. 

In this section, we replicated \citet{BaltagiKhantiAkom1990}'s study using the Panel Study of Income Dynamics (PSID) dataset. The dataset is a cohort of 595 individuals observed over the period 1976--1982. The respondents, aged between 18 and 65 in 1976, are those who reported a positive wage in private non-farm employment for all 7 years, \citep{CornwellRupert1988}.

The log wage is the dependent variable and is regressed on weeks worked (WKS), years of full-time work experience (EXP), occupation (OCC=1, if the individual is in a blue-collar occupation), residence (SOUTH = 1, SMSA = 1, if the individual resides in the South, or in a standard metropolitan statistical area), marital status (MS = 1, if the individual is married), industry (IND = 1, if the individual works in a manufacturing industry), and union coverage (UNION = 1, if the individual's wage is set by a union contract).

We fitted the $\ERFE$ model to the PSID dataset. In addition to the regressor effects on the average salary \citep{baltagi2008, BaltagiKhantiAkom1990,  CornwellRupert1988}, the $\ERFE$ model captures the regressor effects on the entire wage distribution. Consequently, the $\ERFE$ model controls for the endogeneity resulting from unmeasured factors and captures the heterogeneity present in the data. The corresponding Mincer equation of the $\ERFE$ model, for a fixed $\tau\in (0,1),$ is specified as:

\begin{equation*}
    \begin{split}
   \mu_{\tau}(\log(\text{Wage}_{ij})^*)  
    & = \beta_{1\tau} \text{WKS}_{ij}^* + \beta_{2\tau} \text{EXP}_{ij}^* + \beta_{3\tau} \text{EXP}_{ij}^{2*} \\
    & + \beta_{4\tau} \text{UNION}_{ij}^* + \beta_{5\tau} \text{IND}_{ij}^* + \beta_{6\tau} \text{MS}_{ij}^* + \beta_{7\tau} \text{OCC}_{ij}^* \\
    & + \beta_{8\tau} \text{SOUTH}_{ij}^* + \beta_{9\tau} \text{SMSA}_{ij}^*,\\
   \end{split}
\end{equation*}

where the initial model is transformed to eliminate the individual effects.

We estimated the conditional expectiles of the log wage distribution using 91 asymmetric points \( (\tau \in (0.05,\ 0.06,\ 0.07,\ \ldots,\ 0.95)).\) We generated the confidence intervals using the asymptotic standard error of the $\ERFE$ model. For comparison, we also fitted the $\ER$ model, the $\QR$ model and the $\QRFE$ model. Notice that the covariance matrix of the $\QR$-based method depends on the random error density function which add a computational burden and some numerical issues \citep{Chen2004, YinCai2005a, kocherginskyPracticalConfidenceIntervals2005}. We used a kernel estimate of the sandwich as proposed by \citet{barnettNonparametricSemiparametricMethods1991} to compute the standard error of the $\QR$ estimates and the generalized bootstrap of \citet{boseGeneralizedBootstrapEstimators2003}  to compute the standard error of the $\QRFE$ estimates.
Moreover, since an expectile of level \(\tau\) is not necessarily equal to a quantile of the same level, the comparison between the $\ER$-based results and the $\QR$-based results must be done globally. 

\begin{figure}
     \centering
     \begin{subfigure}[b]{0.49\textwidth}
         \centering
         \includegraphics[width=\textwidth]{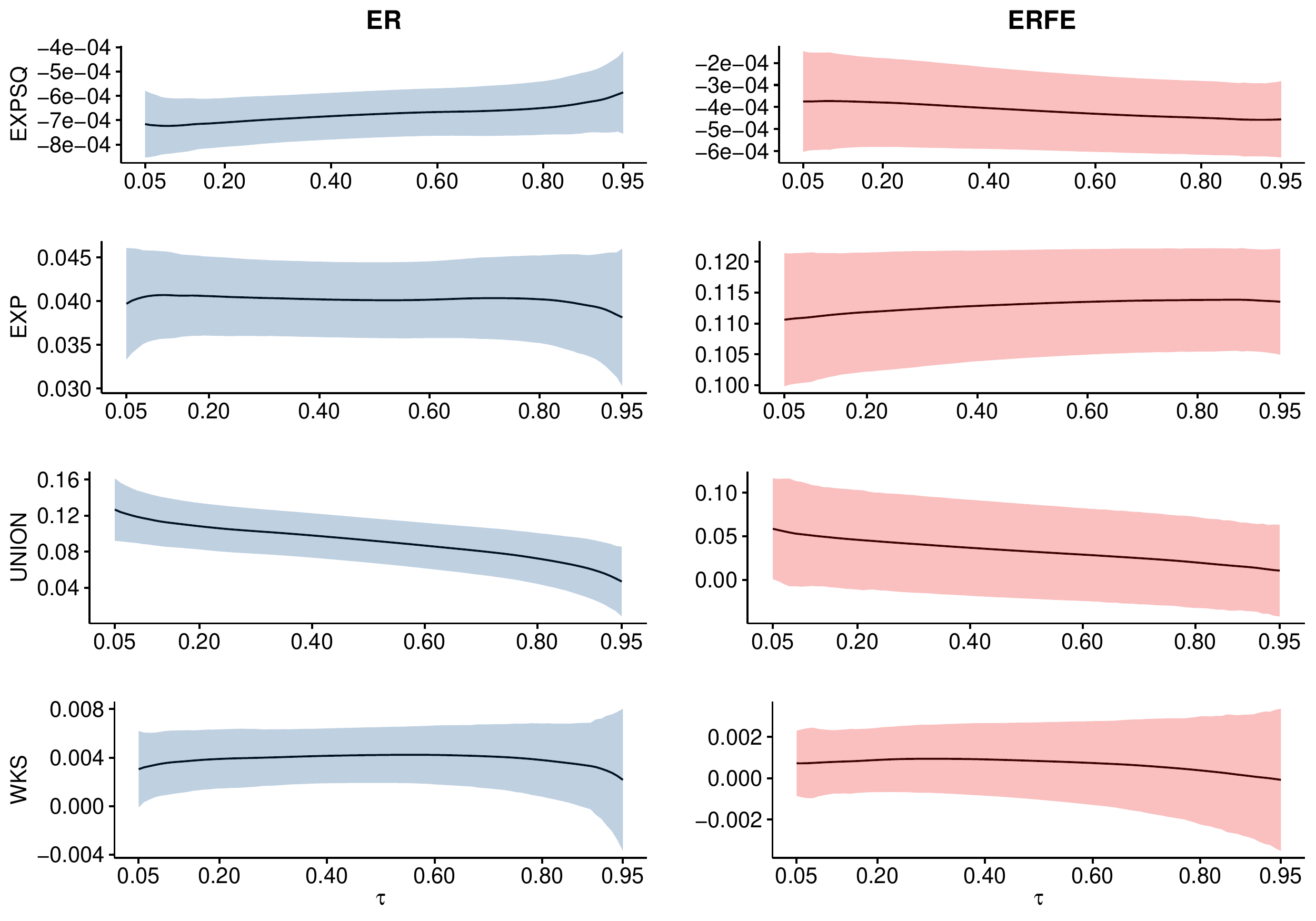}
         \caption{}
         \label{fig:cross1_er_erfe1}
     \end{subfigure}
     \hfill
     \begin{subfigure}[b]{0.49\textwidth}
         \centering
         \includegraphics[width=\textwidth]{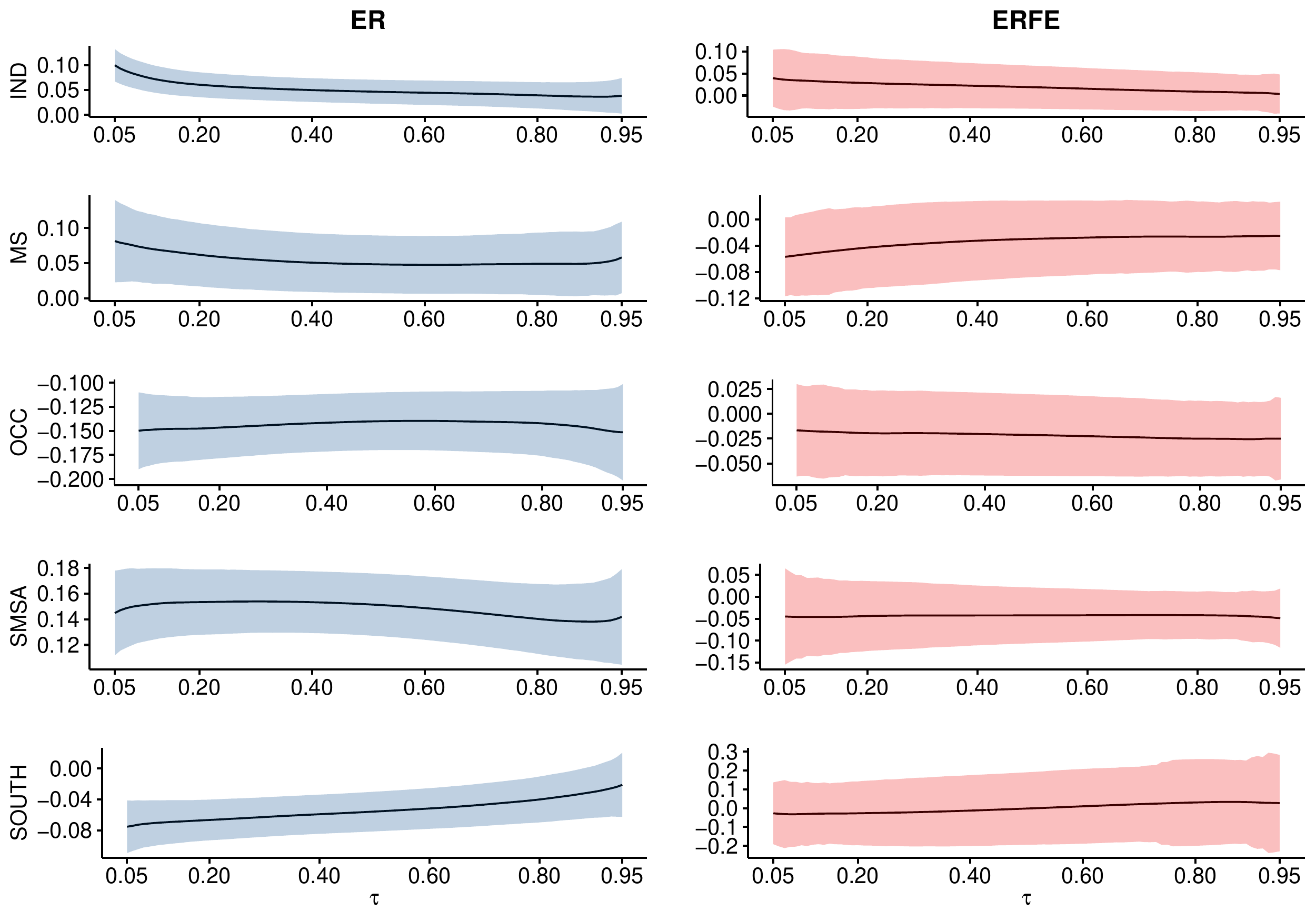}
         \caption{}
         \label{fig:cross1_er_erfe2}
     \end{subfigure}
       \caption{ $\ER$ and $\ERFE$ coefficient estimates. Fig. \textbf{(a)} displays coefficient estimates of the regressors: EXPSQ, EXP, UNION and WKS, with their estimated confidence intervals. Fig. \textbf{(b)} displays coefficient estimates of the regressors: IND, MS, OCC, SMSA and SOUTH, with their estimated confidence intervals. The EXPSQ variable is the square of the experience variable (EXP) and the log of wage is the dependent variable.}
        \label{fig:cross1_er_erfe}
\end{figure}

\begin{figure}
     \centering
     \begin{subfigure}[b]{0.49\textwidth}
         \centering
         \includegraphics[width=\textwidth]{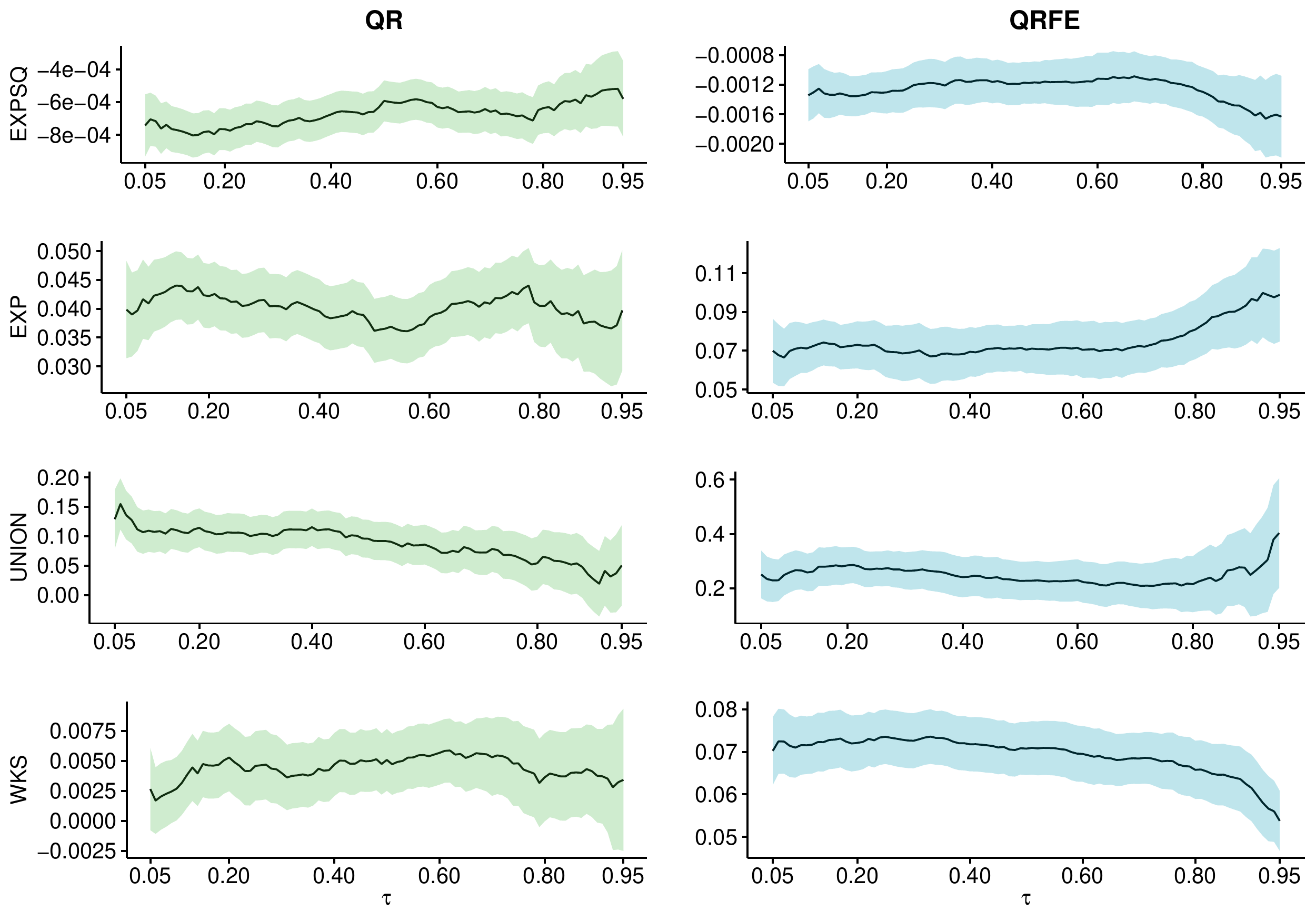}
         \caption{}
         \label{fig:cross1_qr_qrfe1}
     \end{subfigure}
     \hfill
     \begin{subfigure}[b]{0.49\textwidth}
         \centering
         \includegraphics[width=\textwidth]{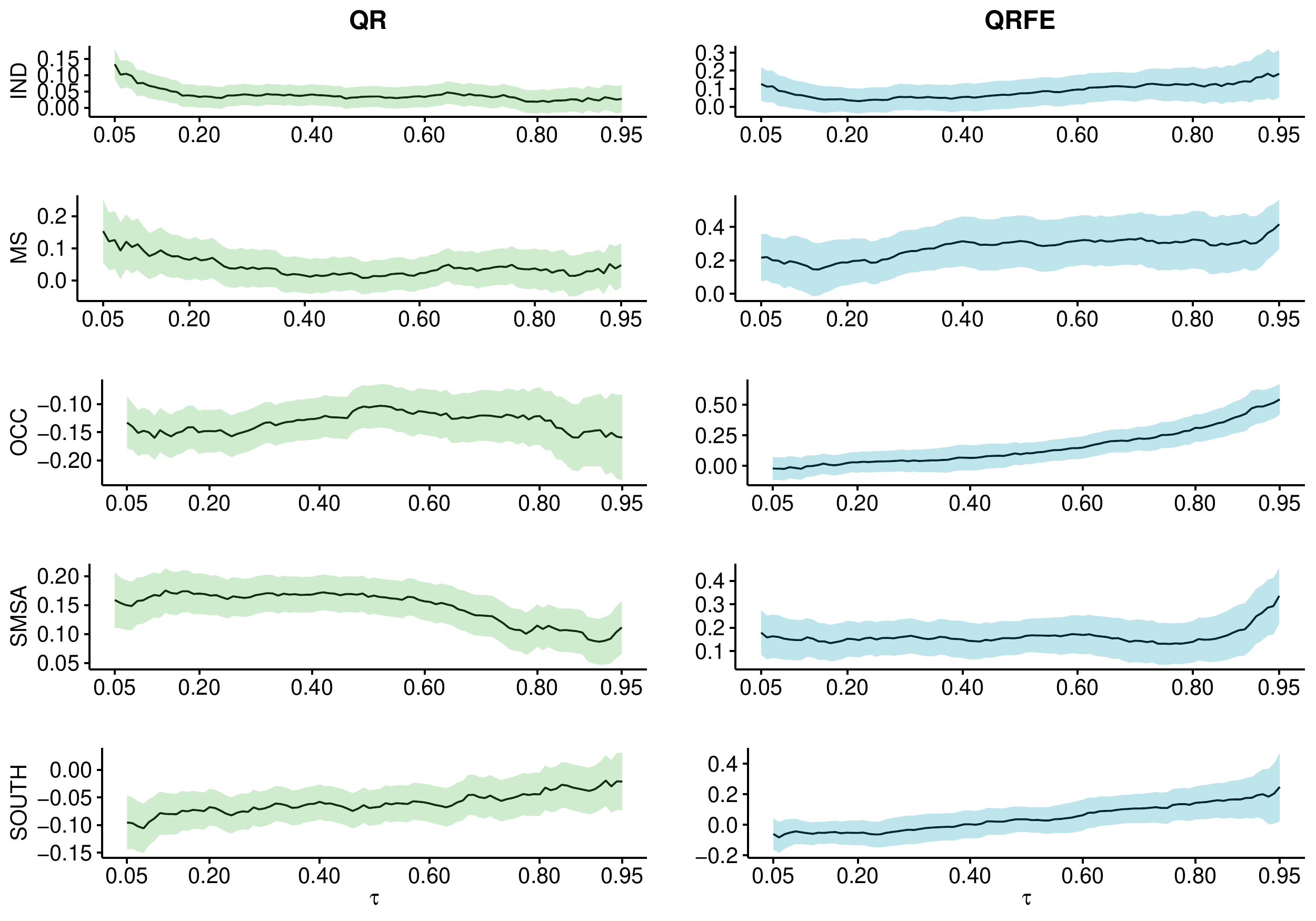}
         \caption{}
         \label{fig:cross1_qr_qrfe2}
     \end{subfigure}
       \caption{ $\QR$ and $\QRFE$ coefficient estimates. Fig. \textbf{(a)} displays coefficient estimates of the regressors: EXPSQ, EXP, UNION and WKS, with their estimated confidence intervals. Fig. \textbf{(b)} displays coefficient estimates of the regressors: IND, MS, OCC, SMSA and SOUTH, with their estimated confidence intervals. The EXPSQ variable is the square of the experience variable (EXP) and the log of wage is the dependent variable.}
        \label{fig:cross1_qr_qrfe}
\end{figure}

\textbf{Figure \ref{fig:cross1_er_erfe1}} and \textbf{Figure \ref{fig:cross1_er_erfe2}} display the coefficient estimates of the regressors
obtained by fitting the $\ER$-based methods while \textbf{Figure \ref{fig:cross1_qr_qrfe1}} and \textbf{Figure \ref{fig:cross1_qr_qrfe2}} display the coefficient estimates of the regressors obtained by fitting the $\QR$-based methods. The overall results show the potential of both $\ER$-based and $\QR$-based methods to reveal the  heterogeneous regressor effects on the response distribution and therefore to capture the heteroscedasticity present in the data. We observe that the parameter estimates of some regressors (UNION, IND and SOUTH, for example) vary with respect to the asymmetric points or percentiles suggesting the presence of heteroscedasticity in the data. For example, we observe that the parameter estimates of the UNION variable decrease with respect to the asymmetric points or percentiles suggesting that individuals with low salary have more advantage of being unionized than individuals with high salary. We also observe that the parameter estimates of some regressors may vary a little or not at all with respect to the asymmetric points, suggesting that the mean effect of these regressors would be enough to summarize their relationship with the response variable.

We also observe that the curves of the $\ER$-based results are smoother than those from the $\QR$-based method which seem to be more wiggly and unstable. Indeed, the $\QR$-based results is more volatile and it is more difficult to identify an overall trend of the heterogeneity of the regressor effects. For example, the $\QRFE$ parameter estimates of the IND variable is decreasing between the percentiles 0.1  and 0.25, and then increasing between the percentiles 0.25 and  0.9.

Despite the similar trend, the parameter estimates of the different methods have different statistical properties. The coefficient estimates of the $\ER$ and $\QR$ have similar range and are biased upward. For example, the $\ER$ and $\QR$ coefficient estimates of the WKS variable fluctuate between 0.0025 and 0.005. While the $\QRFE$ coefficient estimates of the WKS variable vary between 0.06 and 0.07, 10 times higher than that of the $\ERFE$ parameter estimates. Therefore, the $\QRFE$ coefficient estimates is severely biased because of its inability to account for the correlation between the individual fixed-effects and some regressors in the model.   

This results are in line with the simulation results, where we observed that the $\ER$ and the $\QR$ estimates have similar and lower bias than the $\QRFE$ estimates which have higher bias particularly when the individual fixed-effects is correlated to the regressors in the model.

In summary, the data analysis shows that some parameter estimates vary according to the asymmetric points or the percentiles. Therefore, we need to consider beyond the mean or median regression in order to capture the heterogeneity present in the data. The $\FE$ model, like other methods that estimate the mean effect, is not sufficient to analyze the returns to schooling because the impact of most of the regressors vary across the wage distribution.

\section{Conclusion}\label{Conclusion_erfe}

We introduced the $\ERFE$ model which inherits the attractive properties of the weighted asymmetric least squares regression $(\ER)$ and the $\FE$ model. As with the $\FE$ model, the ERFE model is an endogenous model that takes into account the possible correlation between the omitted time-invariant variables and the regressors included in the model. In addition, the $\ERFE$ model estimates the regressor effects on the conditional expectiles of the response distribution allowing to study the influence of the regressors on the location, scale, and shape of the conditional response distribution.

We derived the asymptotic properties of the $\ERFE$ estimator and suggest an estimator of its variance covariance matrix. We showed that the $\ERFE$ estimator is an iterative-within-transformation estimator. That is, the $\ERFE$ estimator can be derived by using iteratively the within-transformation strategy to concentrate out the incidental parameter from the model. The $\ERFE$ model is computationally efficient and easy to implement. See our GitHub for a free R package that simplifies the implementation (\url{github.com/AmBarry/erfe}). 

The exhaustive simulations showed that the $\ERFE$ estimator outperformed its competitors, including the $\QRFE$ estimator in the location-shift and location-scale-shift scenarios. These results are not surprising because our $\ERFE$ estimator inherits the properties of the within-estimator which is simply an $\ERFE$ estimator of level $\tau=0.5.$ The real data application showed that some parameter estimates vary according to the asymmetric points signaling the presence of heteroscedasticity in the data. Therefore we need to go beyond the mean regression to capture unobserved heterogeneity of the data and provide an overview of the relationship between the regressors and the dependent variables for a better decision making.

Our $\ERFE$ model suffers from the same limitations as the $\FE$ model which corresponds to the $\ERFE$ model of level $\tau = 0.5.$ 
The $\ERFE$ model estimates only the effects of the time-variant regressors. The $\ERFE$ model ignores also the between-subject variations which can affect the efficiency of its standard error. The $\ERFE$ model is a weighted mean regression and as such it is sensitive to aberrant values. Fortunately, there is a large number of regression diagnostic tools available to mitigate their influence.

There are alternatives in the literature that have been proposed to circumvent the lack of inference for the time-invariant regressors
\citep{CornwellRupert1988, BaltagiKhantiAkom1990}, while keeping the favorable properties of the $\FE$ model. Future research should investigate the possibility of adapting these methods to the $\ERFE$ framework. 

In addition to the research avenues mentioned above we are currently exploring different alternatives such as penalizing the individual fixed-effects parameter to solve the incidental parameter problem while allowing inference on the time-invariant regressors.

\clearpage
\bibliographystyle{apalike}
\bibliography{bib_fix_effect.bib}
\end{document}


\maketitle

\section{Proof}

\begin{proof}[\textbf{Proof of Theorem \ref{theo1_erfe}}]$ $ ~~\\~~\\
The proof of \textbf{Theorem \ref{theo1_erfe}} is adapted from a proof of \citet{koenker_quantile_2004}. First, we present a philosophical approach and, secondly, a thorough one. We adopt a purely heuristic approach and ignore the complications introduced by the infinite dimensional nature of the incidental parameter. In the completed version, we explicitly concentrate out the incidental parameter before showing the asymptotic property of the parameter estimator of interest. The Lemma \ref{lem1_erfe} establishes the equivalence between the two approaches.
~~\\ 
~~\\
\textbf{Part 1.} Let $\mu_{ij\tau}= \boldsymbol{x}_{ij}\transpose\boldsymbol{\beta}_{\tau} + \boldsymbol{z}_{ij}\transpose\boldsymbol{\alpha}$ and consider the following objective function
\begin{equation} \label{risk_function_erfe}
    R_{nm}(\boldsymbol{\delta})=\sum_{i=1}^{n}\sum_{j=1}^{m}\rho_\tau \bigg\lbrace y_{ij}-\mu_{ij\tau}-\boldsymbol{z}_{ij}\transpose\boldsymbol{\delta}_0/\sqrt{m}-\boldsymbol{x}_{ij}
    \transpose\boldsymbol{\delta}_1/\sqrt{nm}\bigg\rbrace-\rho_\tau\lbrace y_{ij}-\mu_{ij\tau}\rbrace.
\end{equation}
The above objective function is a convex function of $\boldsymbol{\delta}$ that is minimized by 

\begin{equation}\label{delta_erfe}
\widehat{\boldsymbol{\delta}}=
\begin{pmatrix}
\widehat{\boldsymbol{\delta}}_0\\
\widehat{\boldsymbol{\delta}}_1\\
\end{pmatrix}
=
\begin{pmatrix}
\sqrt{m}(\widehat{\boldsymbol{\alpha}}-\boldsymbol{\alpha})\\
\sqrt{nm}(\widehat{\boldsymbol{\beta}}_{\tau}-\boldsymbol{\beta}_{\tau})\\
\end{pmatrix}.
\end{equation}
Our goal is to approximate $R_{nm}$ by a quadratic function with a unique minimizing value, and use results from \citet{hjort_asymptotics_2011} to show that $\widehat{\boldsymbol{\delta}}$ has the same asymptotic distribution of that minimizing value. This quadratic approximation is mainly composed by the Taylor expansion of the expected value and by a linear approximation function.
~~\\ 
~~\\
Let $\quad \widetilde{\boldsymbol{x}_{ij}}=(\boldsymbol{z}_{ij}\transpose,\boldsymbol{x}_{ij}\transpose)
\transpose, \qquad \widetilde{\boldsymbol{\delta}}=(\boldsymbol{\delta}_0\transpose/\sqrt{m},
\boldsymbol{\delta}_1\transpose/\sqrt{nm})\transpose\quad \mbox{ and } \quad \varepsilon_{ij\tau}=y_{ij}-\mu_{ij\tau}.$ The function $\E(\rho_\tau(\varepsilon_{ij\tau}-\widetilde{\boldsymbol{x}_{ij}}\transpose\widetilde{\boldsymbol{\delta}})-
\rho_\tau(\varepsilon_{ij\tau}))$ is convex, twice continuously differentiable and reaches its minimum at $\widetilde{\boldsymbol{\delta}}=\boldsymbol{0}.$ It can be represented in the neighbourhood of $\widetilde{\boldsymbol{\delta}}=\boldsymbol{0}$ as

\begin{equation}\label{Exploss1_erfe}
\begin{split}
    \E\big[\rho_\tau(\varepsilon_{ij\tau}-\widetilde{\boldsymbol{x}_{ij}}\transpose\widetilde{\boldsymbol{\delta}})-
    \rho_\tau(\varepsilon_{ij\tau})\big]&=\widetilde{\boldsymbol{\delta}}\transpose\widetilde{\boldsymbol{x}_{ij}}\E[\psi_\tau(\varepsilon_{ij\tau})]\widetilde{\boldsymbol{x}_{ij}}\transpose\widetilde{\boldsymbol{\delta}}  \\
    & -2\widetilde{\boldsymbol{\delta}}\transpose\widetilde{\boldsymbol{x}_{ij}}\E[\psi_\tau(\varepsilon_{ij\tau}).\varepsilon_{ij\tau}]   + \smallO{\norm{\widetilde{\boldsymbol{\delta}}}^2},
\end{split}
\end{equation}
where $\psi_\tau(\lambda)=\tau-\mathds{1}(\lambda<0).$ Since

\begin{equation*}
   \argmin_{\boldsymbol{\delta}\;\in\;\mathbb{R}^{n+p}} \E\big[\rho_\tau(\varepsilon_{ij\tau}-\widetilde{\boldsymbol{x}_{ij}}\transpose\widetilde{\boldsymbol{\delta}})-\rho_\tau(\varepsilon_{ij\tau})\big]=\boldsymbol{0}
\end{equation*}
we have by the first order condition

\begin{equation}\label{FirstCondition_erfe}
    \E[\psi_\tau(\varepsilon_{ij\tau}).\varepsilon_{ij\tau}]=0,
\end{equation}
and equation (\ref{Exploss1_erfe}) can be reduced to:

\begin{equation}\label{Exploss2_erfe}
    \E\big[\rho_\tau(\varepsilon_{ij\tau}-\widetilde{\boldsymbol{x}_{ij}}\transpose\widetilde{\boldsymbol{\delta}})-
    \rho_\tau(\varepsilon_{ij\tau})
    \big]=\widetilde{\boldsymbol{\delta}}\transpose\widetilde{\boldsymbol{x}_{ij}}\E[\psi_\tau(\varepsilon_{ij\tau})]
    \widetilde{\boldsymbol{x}_{ij}}\transpose \widetilde{\boldsymbol{\delta}} + \smallO{\norm{\widetilde{\boldsymbol{\delta}}}^2}.
\end{equation}
The linear approximation function can be seen as a sort of Taylor expansion of $R_{nm}(\boldsymbol{\delta})$ around  $\boldsymbol{\delta}=0,$ see \citep{pollard_asymptotics_1991}. Define

\begin{equation}\label{Approx1_erfe}
D_{ij}(\varepsilon_{ij\tau})= -2\psi_\tau(\varepsilon_{ij\tau}).\varepsilon_{ij\tau}.
\end{equation}
Notice that by (\ref{FirstCondition_erfe}), \;$\E(D_{ij}(\varepsilon_{ij\tau}))=0.$ Define

\begin{equation*}
    r_{ij}(\widetilde{\boldsymbol{\delta}})=\rho_\tau(\varepsilon_{ij\tau}-\widetilde{\boldsymbol{x}_{ij}}\transpose
    \widetilde{\boldsymbol{\delta}})-\rho_\tau(\varepsilon_{ij\tau}) - \widetilde{\boldsymbol{\delta}}\transpose\widetilde{\boldsymbol{x}_{ij}} D_{ij}(\varepsilon_{ij\tau}).
\end{equation*}
Then
\begin{equation*}
\begin{split}
    R_{nm}(\widetilde{\boldsymbol{\delta}}) &=\sum_{i=1}^{n}\sum_{j=1}^{m}\bigg(\E\big[\rho_\tau(\varepsilon_{ij\tau}-
\widetilde{\boldsymbol{x}_{ij}}\transpose\widetilde{\boldsymbol{\delta}})-\rho_\tau(\varepsilon_{ij\tau})\big]\bigg) \\
    & + \sum_{i=1}^{n}\sum_{j=1}^{m}\widetilde{\boldsymbol{\delta}}\transpose\widetilde{\boldsymbol{x}_{ij}} D_{ij}(\varepsilon_{ij\tau}) + \sum_{i=1}^{n}\sum_{j=1}^{m} \bigg(r_{ij}(\widetilde{\boldsymbol{\delta}})-\E\big[r_{ij}(\widetilde{\boldsymbol{\delta}})\big]\bigg).
\end{split}
\end{equation*}
Using Lemma \ref{lemma1}, the objective function $R_{nm}(\widetilde{\boldsymbol{\delta}})$ reduce to 
\begin{equation}\label{risk_function1_erfe}
\begin{split}
    R_{nm}(\widetilde{\boldsymbol{\delta}}) &= \widetilde{\boldsymbol{\delta}}\transpose\sum_{i=1}^{n}\sum_{j=1}^{m}\bigg(\widetilde{\boldsymbol{x}_{ij}}
    \E[\psi_\tau(\varepsilon_{ij\tau})]\widetilde{\boldsymbol{x}_{ij}}\transpose\bigg)\widetilde{\boldsymbol{\delta}}+
    \widetilde{\boldsymbol{\delta}}\transpose\sum_{i=1}^{n}\sum_{j=1}^{m}
    \widetilde{\boldsymbol{x}_{ij}}D_{ij}(\varepsilon_{ij\tau}) + \bigO{\norm{\widetilde{\boldsymbol{\delta}}}^2} + \smallO{\norm{\widetilde{\boldsymbol{\delta}}}^2} \\
    &=    \widetilde{\boldsymbol{\delta}}\transpose\sum_{i=1}^{n}\sum_{j=1}^{m}\bigg(\widetilde{\boldsymbol{x}_{ij}}
    \E[\psi_\tau(\varepsilon_{ij\tau})]\widetilde{\boldsymbol{x}_{ij}}\transpose\bigg)\widetilde{\boldsymbol{\delta}}+
    \widetilde{\boldsymbol{\delta}}\transpose\sum_{i=1}^{n}\sum_{j=1}^{m}
    \widetilde{\boldsymbol{x}_{ij}}D_{ij}(\varepsilon_{ij\tau})
    + o_p(1)\\
    & \simeq \widetilde{\boldsymbol{\delta}}\transpose\sum_{i=1}^{n}\sum_{j=1}^{m}\bigg(\widetilde{\boldsymbol{x}_{ij}}
    \E[\psi_\tau(\varepsilon_{ij\tau})]\widetilde{\boldsymbol{x}_{ij}}\transpose\bigg)\widetilde{\boldsymbol{\delta}}+
    \widetilde{\boldsymbol{\delta}}\transpose\sum_{i=1}^{n}\sum_{j=1}^{m}
    \widetilde{\boldsymbol{x}_{ij}}D_{ij}(\varepsilon_{ij\tau}). \\
\end{split}
\end{equation}
By replacing $\; \widetilde{\boldsymbol{x}_{ij}}=(\boldsymbol{z}_{ij}\transpose,\boldsymbol{x}_{ij}\transpose)\transpose \; \mbox{ and } \; \widetilde{\boldsymbol{\delta}}=(\boldsymbol{\delta}_0\transpose/\sqrt{m},
\boldsymbol{\delta}_1\transpose/\sqrt{nm})\transpose$ by their initial value, we have
\begin{align*}
\begin{split}
   {}& R_{nm}(\boldsymbol{\delta}) = \\
    & -2\frac{1}{\sqrt{m}}\sum_{i=1}^{n}\sum_{j=1}^{m}(\boldsymbol{z}_{ij}\transpose\boldsymbol{\delta}_0 + \boldsymbol{x}_{ij}\transpose\boldsymbol{\delta}_1/\sqrt{n})
     \psi_\tau(y_{ij}-\mu_{ij\tau}).(y_{ij}-\mu_{ij\tau})\\ 
    & +\frac{1}{m}\sum_{i=1}^{n}\sum_{j=1}^{m}\E[\psi_\tau(y_{ij}-\mu_{ij\tau})](\boldsymbol{z}_{ij}\transpose\boldsymbol{\delta}_0+\boldsymbol{x}_{ij} \transpose\boldsymbol{\delta}_1/\sqrt{n})^2
\end{split}\\
\begin{split}
    {}&= -2\frac{1}{\sqrt{m}}\bigg[\boldsymbol{\delta}_0\transpose\boldsymbol{Z}\transpose\boldsymbol{\Psi}_\tau(\boldsymbol{\varepsilon}_{\tau})\boldsymbol{\varepsilon}_{\tau} +\boldsymbol{\delta}_1\transpose/\sqrt{n}\boldsymbol{X}\transpose\boldsymbol{\Psi}_\tau(\boldsymbol{\varepsilon}_{\tau})\boldsymbol{\varepsilon}_{\tau}\bigg]\\
    & + \frac{1}{m}\bigg[ \boldsymbol{\delta}_0\transpose\boldsymbol{Z}\transpose\E[\boldsymbol{\Psi}_\tau(\boldsymbol{\varepsilon}_{\tau})]\boldsymbol{Z}
    \boldsymbol{\delta}_0 + 2\boldsymbol{\delta}_0\transpose\boldsymbol{Z}\transpose\E[\boldsymbol{\Psi}_\tau(\boldsymbol{\varepsilon}_{\tau})]
    \boldsymbol{X}\boldsymbol{\delta}_1/\sqrt{n} +\boldsymbol{\delta}_1\transpose\boldsymbol{X}\transpose\E[\boldsymbol{\Psi}_\tau(\boldsymbol{\varepsilon}_{\tau})]\boldsymbol{X}
    \boldsymbol{\delta}_1/n\bigg]\\
    & = R_{nm}^{(1)}(\boldsymbol{\delta}) + R_{nm}^{(2)}(\boldsymbol{\delta})
\end{split}\\
\end{align*}
~~\\
The condition \textbf{A2} and \textbf{A3} imply a Lindberg condition and we have
\begin{equation*}
    R_{nm}^{(1)}(\boldsymbol{\delta})=-2\frac{1}{\sqrt{m}}\Big[\boldsymbol{\delta}_0\transpose\boldsymbol{Z}
    \transpose+\boldsymbol{\delta}_1\transpose/\sqrt{n}\boldsymbol{X}\transpose\Big]\boldsymbol{\Psi}_\tau(\boldsymbol{\varepsilon}_{\tau})\boldsymbol{\varepsilon}_{\tau}
    \xrightarrow{d}-2\boldsymbol{\delta}\transpose\boldsymbol{B}.
\end{equation*}
While by condition \textbf{A2}

\begin{equation*}
\begin{split}
    R_{nm}^{(2)}(\boldsymbol{\delta}) 
    & =  \frac{1}{m}\bigg[ \boldsymbol{\delta}_0\transpose\boldsymbol{Z}\transpose\E[\boldsymbol{\Psi}_\tau(\boldsymbol{\varepsilon}_{\tau})]\boldsymbol{Z}
    \boldsymbol{\delta}_0 + 2\boldsymbol{\delta}_0\transpose\boldsymbol{Z}\transpose\E[\boldsymbol{\Psi}_\tau(\boldsymbol{\varepsilon}_{\tau})]
    \boldsymbol{X}\boldsymbol{\delta}_1/\sqrt{n} +\boldsymbol{\delta}_1\transpose\boldsymbol{X}\transpose\E[\boldsymbol{\Psi}_\tau(\boldsymbol{\varepsilon}_{\tau})]\boldsymbol{X}
    \boldsymbol{\delta}_1/n\bigg]\\
    & \rightarrow \boldsymbol{\delta}\transpose \boldsymbol{D}_1 \boldsymbol{\delta}.
\end{split}
\end{equation*}
Thus the limiting form of the objective function is
\begin{equation*}
    R_0(\boldsymbol{\delta})=-2\boldsymbol{\delta}\transpose\boldsymbol{B}+\boldsymbol{\delta}\transpose \boldsymbol{D}_1 \boldsymbol{\delta}
\end{equation*}
where $\boldsymbol{B}$ is a zero mean Gaussian vector with covariance matrix $\boldsymbol{D}_0.$ The objective function $R_{nm}$ is convex and its limiting form $R_0$ has a unique minimum. Therefore, by \textbf{Theorem}\ref{hjort2}, $\widehat{\boldsymbol{\delta}}$ converge to the argmin of $R_0.$
~~\\
~~\\
\textbf{Part 2.} In the precedent proof we overlook the complications related to the infinite dimensional nature of $\boldsymbol{\alpha}.$ \citet{koenker_quantile_2004}, following \citep{Bickel1975,ruppert_trimmed_1980}, used the Bahadur-Kiefer representation of the incidental parameter in order to concentrate out its effect and express the objective function solely in terms of the finite dimensional parameter $\boldsymbol{\beta}.$ With the smoothness of the asymmetric least square loss, we can use the first order condition to represent the incidental parameter as a function of the structural parameter. 
\newline
\newline
Consider the following objective function
\begin{equation*}
\begin{split}
   R_{nm}(\boldsymbol{\delta})&= 
   -2\frac{1}{\sqrt{m}}\sum_{i=1}^{n}\sum_{j=1}^{m}(\boldsymbol{z}_{ij}\transpose\boldsymbol{\delta}_0
        + \boldsymbol{x}_{ij}\transpose\boldsymbol{\delta}_1/\sqrt{n})\psi_\tau(y_{ij}-\mu_{ij\tau}).
        (y_{ij}-\mu_{ij\tau})\\
        & +\frac{1}{m}\sum_{i=1}^{n}\sum_{j=1}^{m}\E[\psi_\tau(y_{ij}-\mu_{ij\tau})]
   (\boldsymbol{z}_{ij}\transpose\boldsymbol{\delta}_0+\boldsymbol{x}_{ij} \transpose\boldsymbol{\delta}_1/\sqrt{n})^2.
\end{split}
\end{equation*}
This function is twice derivable and the first order condition gives us the exact representation of the incidental parameter
\begin{equation*}
   \frac{\widehat{\delta}_{0i}}{\sqrt{m}}=\frac{1}{m\overline{\boldsymbol{\psi}}_i}\sum_{k=1}^{m}\psi_{\tau}(y_{ik}-\mu_{ik\tau})(y_{ik}-\mu_{ik\tau})-\frac{1}{m\overline{\boldsymbol{\psi}}_i}
    \sum_{k=1}^{m}\E\big[\psi_{\tau}(y_{ik}-\mu_{ik\tau})\big]\boldsymbol{x}_{ik}\transpose\frac{\boldsymbol{\delta}_1}{\sqrt{nm}},
\end{equation*}
where $\overline{\boldsymbol{\psi}}_i=m^{-1}\sum_{k=1}^{m} \E\Big[\psi_{\tau}\big(y_{ik}-\mu_{ik\tau}\big)\Big].$
Substituting $\frac{\widehat{\delta}_{0i}}{\sqrt{m}}$ we have
\begin{equation*}
\begin{split}
   {}&\frac{\widehat{\boldsymbol{\delta}}_{1}}{\sqrt{mn}}=\bigg(\sum_{i=1}^{n}\sum_{j=1}^{m}\boldsymbol{x}_{ij}\E[\psi_\tau(y_{ij}-\mu_{ij\tau})]\Big[\boldsymbol{x}_{ij}\transpose-\frac{1}{m
   \overline{\boldsymbol{\psi}}_i}
   \sum_{k=1}^{m}\E\big[\psi_{\tau}(y_{ik}-\mu_{ik\tau})\big]\boldsymbol{x}_{ik}\transpose\Big]\bigg)^{-1}\\
   {}&\times \bigg(\sum_{i=1}^{n}\sum_{j=1}^{m}\boldsymbol{x}_{ij}\psi_\tau(y_{ij}-\mu_{ij\tau})
   (y_{ij}-\mu_{ij\tau})\\
   {}& -\sum_{i=1}^{n}\sum_{j=1}^{m}\boldsymbol{x}_{ij}\E[\psi_\tau(y_{ij}-
   \mu_{ij\tau})]\Big[\frac{1}{m\overline{\boldsymbol{\psi}}_i}\sum_{k=1}^{m}
   \psi_{\tau}(y_{ik}-\mu_{ik})(y_{ik}-\mu_{ik})\Big]\bigg).
\end{split}
\end{equation*}
Note that,
\begin{equation*}
\begin{split}
   {}& \sum_{i=1}^{n}\sum_{j=1}^{m}\boldsymbol{x}_{ij}\E[\psi_\tau(y_{ij}-\mu_{ij\tau})]\Big[\boldsymbol{x}_{ij}\transpose-\frac{1}{m\overline{\boldsymbol{\psi}}_i}  \sum_{k=1}^{m}\E\big[\psi_{\tau}(y_{ik}-\mu_{ik})\big]\boldsymbol{x}_{ik}\transpose\Big]\\
   {}&=\boldsymbol{X}\transpose\E[\boldsymbol{\Psi}_\tau(\boldsymbol{\varepsilon}_{\tau})]\boldsymbol{X}-\boldsymbol{X}\transpose\boldsymbol{P}_{\boldsymbol{Z}}\transpose\E[\boldsymbol{\Psi}_\tau(\boldsymbol{\varepsilon}_{\tau})]\boldsymbol{X}\\
   {}&=\boldsymbol{X}\transpose[\mathbb{I}-\boldsymbol{P}_{\boldsymbol{Z}}]\E[\boldsymbol{\Psi}_\tau(\boldsymbol{\varepsilon}_{\tau})]\boldsymbol{X}\\   
   {}&=\boldsymbol{X}\transpose\boldsymbol{M}_{\boldsymbol{Z}}\transpose\E[\boldsymbol{\Psi}_\tau(\boldsymbol{\varepsilon}_{\tau})]\boldsymbol{X}\\
   {}&=\boldsymbol{X}\transpose\boldsymbol{M}_{\boldsymbol{Z}}\transpose\boldsymbol{M}_{\boldsymbol{Z}}\transpose\E[\boldsymbol{\Psi}_\tau(\boldsymbol{\varepsilon}_{\tau})]\boldsymbol{X}\\
   {}&=\boldsymbol{X}\transpose\boldsymbol{M}_{\boldsymbol{Z}}\transpose\E[\boldsymbol{\Psi}_\tau(\boldsymbol{\varepsilon}_{\tau})]\boldsymbol{M}_{\boldsymbol{Z}}\boldsymbol{X}
\end{split}
\end{equation*}
and
\begin{multline*}
\sum_{i=1}^{n}\sum_{j=1}^{m}\boldsymbol{x}_{ij}\psi_\tau(y_{ij}-\mu_{ij\tau})(y_{ij}-\mu_{ij\tau})-\sum_{i=1}^{n}\sum_{j=1}^{m}\boldsymbol{x}_{ij}\E[\psi_\tau(y_{ij}-\mu_{ij\tau})] \times \\ \frac{1}{m\overline{\boldsymbol{\psi}}_i}\sum_{k=1}^{m}\psi_{\tau}(y_{ik}-\mu_{ik\tau})(y_{ik}-\mu_{ik\tau})
 =\boldsymbol{X}\transpose\boldsymbol{\Psi}_\tau(\boldsymbol{\varepsilon}_{\tau})\boldsymbol{\varepsilon}_{\tau}-\boldsymbol{X}\transpose\boldsymbol{P}_{\boldsymbol{Z}}\transpose\boldsymbol{\Psi}_\tau(\boldsymbol{\varepsilon}_{\tau})\boldsymbol{\varepsilon}_{\tau}
 =\boldsymbol{X}\transpose\boldsymbol{M}_{\boldsymbol{Z}}\transpose\boldsymbol{\Psi}_\tau(\boldsymbol{\varepsilon}_{\tau})
 \boldsymbol{\varepsilon}_{\tau}.
\end{multline*}
Consequently,
\begin{equation}\label{delta1_erfe}
\begin{split}
\frac{\widehat{\boldsymbol{\delta}}_{1}}{\sqrt{mn}} &= \bigg(\boldsymbol{X}\transpose\boldsymbol{M}_{\boldsymbol{Z}}\transpose \E[\boldsymbol{\Psi}_\tau(\boldsymbol{\varepsilon}_{\tau})]\boldsymbol{M}_{\boldsymbol{Z}}\boldsymbol{X}\bigg)^{-1}
   \boldsymbol{X}\transpose\boldsymbol{M}_{\boldsymbol{Z}}\transpose\boldsymbol{\Psi}_\tau(\boldsymbol{\varepsilon}_{\tau})\boldsymbol{\varepsilon}_{\tau} \\
   &= \bigg(\boldsymbol{X}\transpose\boldsymbol{M}_{\boldsymbol{Z}}\transpose \E[\boldsymbol{\Psi}_\tau(\boldsymbol{\varepsilon}_{\tau})]\boldsymbol{M}_{\boldsymbol{Z}}\boldsymbol{X}/m\bigg)^{-1}
   m^{-1}\boldsymbol{X}\transpose\boldsymbol{M}_{\boldsymbol{Z}}\transpose\boldsymbol{\Psi}_\tau(\boldsymbol{\varepsilon}_{\tau})\boldsymbol{\varepsilon}_{\tau}. \\
\end{split}
\end{equation}
Let $\boldsymbol{X}_{\boldsymbol{Z}}=\boldsymbol{M}_{\boldsymbol{Z}}\boldsymbol{X}$ then we have $ m^{-1}\boldsymbol{X}_{\boldsymbol{Z}}\transpose\boldsymbol{\Psi}_\tau(\boldsymbol{\varepsilon}_{\tau})\boldsymbol{\varepsilon}_{\tau}= \sum_{i=1}^{n}\sum_{j=1}^{m}\boldsymbol{x}_{ij\boldsymbol{Z}}\psi_\tau(\varepsilon_{ij\tau})\varepsilon_{ij\tau}/m.$ Let $T_{ni}=\sum_{j=1}^{m}m^{-1}\boldsymbol{\lambda}\transpose\boldsymbol{x}_{ij\boldsymbol{Z}}\psi_\tau(\varepsilon_{ij\tau})\varepsilon_{ij\tau}$
and consider $n^{-1/2}\sum_{i=1}^{n}T_{ni},$ where $\boldsymbol{\lambda}$ is a $p\times 1$ unit vector, $\boldsymbol{\lambda}\transpose\boldsymbol{\lambda}=1.$
The summands $T_{ni}$ are independent with $\E[T_{ni}]=0$ and $\Var\Big[n^{-1/2}\sum_{i=1}^{n}T_{ni}\Big]>\nu\prime>0,$ by condition \textbf{A2}. By application of the Minkowski's inequality, we have
\begin{align*}
\begin{split}
 \E\lvert T_{ni} \rvert^{2+\nu} & = \E\bigg\lvert \sum_{j=1}^{m}\sum_{k=1}^{p} m^{-1}\lambda_{k} x_{ij\boldsymbol{Z}}^{k} \psi_{\tau}(\varepsilon_{ij\tau})\varepsilon_{ij\tau} \bigg\rvert^{2+\nu} \\
& \leq \bigg[\sum_{j=1}^{m}\sum_{k=1}^{p}\bigg(\E\Big\lvert m^{-1} \lambda_{k}x_{ij\boldsymbol{Z}}^{k}\psi_{\tau}(\varepsilon_{ij\tau})\varepsilon_{ij\tau}\Big\rvert^{2+\nu}\bigg)^{\frac{1}{2+\nu}} \bigg]^{2+\nu}\\
& \leq \bigg[\sum_{j=1}^{m}\sum_{k=1}^{p} m^{-1} \lvert \lambda_{k}x_{ij\boldsymbol{Z}}^{k} \rvert
\bigg(\E\Big\lvert\psi_{\tau}(\varepsilon_{ij\tau})\varepsilon_{ij\tau}\Big\rvert^{2+\nu}\bigg)^{\frac{1}{2+\nu}} \bigg]^{2+\nu}\\
& \leq M \Delta p^{1+\nu}, \\
\end{split}
\end{align*}
where the last inequality follows by $\E\lvert \psi_{\tau}(\varepsilon_{ij\tau})\rvert^{4+\nu}<\Delta$ and $\E\lvert \varepsilon_{ij\tau} \rvert^{4+\nu}<\Delta.$
~~\\
~~\\
Then by the Liapounov CLT $\ \boldsymbol{X}\transpose\boldsymbol{M}_{\boldsymbol{Z}}\transpose\boldsymbol{\Psi}_\tau(\boldsymbol{\varepsilon}_{\tau})\boldsymbol{\varepsilon}_{\tau}\ $ is Gaussian and by condition \textbf{A2}, $\widehat{\boldsymbol{\delta}}_{1}$ is zero mean Gaussian vector with covariance matrix $\Var(\widehat{\boldsymbol{\beta}}_{\tau})=\widetilde{\boldsymbol{D}}_1^{-1}\widetilde{\boldsymbol{D}}_0\widetilde{\boldsymbol{D}}_1^{-1}.$
\end{proof}

\begin{proof}[\textbf{Proof of Lemma \ref{lem1_erfe}}]$ $ ~~\\~~\\
we intend to show that this new covariance matrix $\widetilde{\boldsymbol{D}}_1^{-1}\widetilde{\boldsymbol{D}}_0\widetilde{\boldsymbol{D}}_1^{-1}$ is identical to the lower right diagonal block matrix $\boldsymbol{D}_1^{-1}\boldsymbol{D}_0\boldsymbol{D}_1^{-1}.$
 ~~\\
 ~~\\
We have $(\boldsymbol{D}_1^{-1}\boldsymbol{D}_0\boldsymbol{D}_1^{-1})_{22}=(\boldsymbol{D}_1^{-1})_{2.}\boldsymbol{D}_0(\boldsymbol{D}_1^{-1})_{2.}\transpose.$ Using the standard partitioned inverse formula for a general $2\times 2$ partitioned matrix, we have
\begin{equation*} 
\begin{split}
mn(\boldsymbol{D}_1^{-1}\boldsymbol{D}_0\boldsymbol{D}_1^{-1})_{22} & = 
\begin{pmatrix}
-\boldsymbol{F}\boldsymbol{E}^{-1} & \boldsymbol{E}^{-1}\\
\end{pmatrix} 
 \begin{pmatrix}
  \boldsymbol{Z}\transpose\boldsymbol{\Sigma}_{\tau}\boldsymbol{Z} & \boldsymbol{Z}\transpose\boldsymbol{\Sigma}_{\tau}\boldsymbol{X}  \\
  \boldsymbol{X}\transpose\boldsymbol{\Sigma}_{\tau}\boldsymbol{Z}  & \boldsymbol{X}\transpose\boldsymbol{\Sigma}_{\tau}\boldsymbol{X}  
 \end{pmatrix}
 \begin{pmatrix}
-\boldsymbol{F}\boldsymbol{E}^{-1} \\
\boldsymbol{E}^{-1}
\end{pmatrix}\\
 & = \boldsymbol{E}^{-1}[\boldsymbol{F}\transpose\boldsymbol{Z}\transpose\boldsymbol{\Sigma}_{\tau}\boldsymbol{Z}\boldsymbol{F}-
 \boldsymbol{X}\transpose\boldsymbol{\Sigma}_{\tau}\boldsymbol{Z}\boldsymbol{F}-\boldsymbol{F}\transpose\boldsymbol{Z}\transpose\boldsymbol{\Sigma}_{\tau}
 \boldsymbol{X}+\boldsymbol{X}\transpose\boldsymbol{\Sigma}_{\tau}\boldsymbol{X}]\boldsymbol{E}^{-1}
\end{split}
\end{equation*}
where $\boldsymbol{E}=\boldsymbol{X}\transpose\E[\boldsymbol{\Psi}_{\tau}(\boldsymbol{\varepsilon}_{\tau})]\boldsymbol{X}-\boldsymbol{X}\transpose\E[\boldsymbol{\Psi}_{\tau}(\boldsymbol{\varepsilon}_{\tau})]\boldsymbol{Z}(\boldsymbol{Z}\transpose\E[\boldsymbol{\Psi}_{\tau}(\boldsymbol{\varepsilon}_{\tau})]\boldsymbol{Z})^{-1}\boldsymbol{Z}\transpose\E[\boldsymbol{\Psi}_{\tau}(\boldsymbol{\varepsilon}_{\tau})]
\boldsymbol{X}=mn\widetilde{\boldsymbol{D}}_1, \mbox{ and } \boldsymbol{Z}\boldsymbol{F}=\boldsymbol{Z}(\boldsymbol{Z}\transpose\E[\boldsymbol{\Psi}_{\tau}(\boldsymbol{\varepsilon}_{\tau})]\boldsymbol{Z})^{-1}\boldsymbol{Z}\transpose\E[\boldsymbol{\Psi}_{\tau}(\boldsymbol{\varepsilon}_{\tau})]\boldsymbol{X}=\boldsymbol{P}_{\boldsymbol{Z}}\boldsymbol{X}.$ The term in square brackets is

\begin{align*}
\begin{split}
   {}& \boldsymbol{F}\transpose\boldsymbol{Z}\transpose\boldsymbol{\Sigma}_{\tau}\boldsymbol{Z}\boldsymbol{F}-
 \boldsymbol{X}\transpose\boldsymbol{\Sigma}_{\tau}\boldsymbol{Z}\boldsymbol{F}-\boldsymbol{F}\transpose\boldsymbol{Z}\transpose\boldsymbol{\Sigma}_{\tau}
 \boldsymbol{X}+\boldsymbol{X}\transpose\boldsymbol{\Sigma}_{\tau}\boldsymbol{X}
\end{split}\\
\begin{split}
    {}&= \boldsymbol{X}\transpose\boldsymbol{P}_{\boldsymbol{Z}}\transpose\boldsymbol{\Sigma}_{\tau}\boldsymbol{P}_{\boldsymbol{Z}}
    \boldsymbol{X}- \boldsymbol{X}\transpose\boldsymbol{\Sigma}_{\tau}\boldsymbol{P}_{\boldsymbol{Z}}\boldsymbol{X}
    -\boldsymbol{X}\transpose\boldsymbol{P}_{\boldsymbol{Z}}\transpose\boldsymbol{\Sigma}_{\tau}\boldsymbol{X}  +\boldsymbol{X}\transpose\boldsymbol{\Sigma}_{\tau}\boldsymbol{X}\\     
    & =\boldsymbol{X}\transpose[\boldsymbol{P}_{\boldsymbol{Z}}\transpose\boldsymbol{\Sigma}_{\tau}\boldsymbol{P}_{\boldsymbol{Z}}
     -\boldsymbol{\Sigma}_{\tau}\boldsymbol{P}_{\boldsymbol{Z}} -\boldsymbol{P}_{\boldsymbol{Z}}\transpose\boldsymbol{\Sigma}_{\tau} +\boldsymbol{\Sigma}_{\tau}]\boldsymbol{X}
\end{split}\\
\begin{split}
    {}&= \boldsymbol{X}\transpose[\mathbf{I}-\boldsymbol{P}_{\boldsymbol{Z}}\transpose]\boldsymbol{\Sigma}_{\tau}[\mathbf{I}-
    \boldsymbol{P}_{\boldsymbol{Z}}]\boldsymbol{X}\\    
    & =\boldsymbol{X}\transpose\boldsymbol{M}_{\boldsymbol{Z}}\transpose\boldsymbol{\Sigma}_{\tau}\boldsymbol{M}_{\boldsymbol{Z}}\boldsymbol{X}\\
    &=\widetilde{\boldsymbol{D}}_0.
\end{split}\\
\end{align*}
Finally, the result follows.
\end{proof}
\begin{proof}[\textbf{Proof of Theorem \ref{theo2_erfe}}]$ $
~~\\
In the following, we remove the asterix as an exponent to lighten the notation. Using the same approach to the proof of \textbf{Theorem} \ref{theo1_erfe}, the objective function $R_{mnq}(\boldsymbol{\delta})$ can be decomposed into two parts
\begin{equation*}
    R_{mnq}(\boldsymbol{\delta})=R_{mnq}^{(1)}(\boldsymbol{\delta})+R_{mnq}^{(2)}
    (\boldsymbol{\delta}).
\end{equation*}

\begin{align*}
\begin{split}
R_{nmq}^{(1)}(\boldsymbol{\delta})&= \frac{-2}{\sqrt{nm}}\sum_{k=1}^{q}\sum_{i=1}^{n}\sum_{j=1}^{m} v_k
        \boldsymbol{\delta}_{\tau_k}\transpose\boldsymbol{x}_{ij}\psi_{\tau_k}(\varepsilon_{ij\tau_k}).\varepsilon_{ij\tau_k}\\ 
    &=\frac{-2}{\sqrt{nm}}\sum_{k=1}^{q}v_k\boldsymbol{\delta}_{\tau_k}\transpose\boldsymbol{X}\transpose\boldsymbol{\Psi}_{\tau_k}(\boldsymbol{\varepsilon}_{\tau_k})\boldsymbol{\varepsilon}_{\tau_k}\\
    &=\frac{-2}{\sqrt{nm}}\boldsymbol{\delta}\transpose (\boldsymbol{V}\otimes\boldsymbol{X})\transpose
   \boldsymbol{\Psi}_{\boldsymbol{\tau}}(\boldsymbol{\varepsilon}_{\boldsymbol{\tau}})\boldsymbol{\varepsilon}_{\boldsymbol{\tau}}\\
    &=\frac{-2}{\sqrt{nm}}\boldsymbol{\delta}\transpose
        \begin{pmatrix}
            v_1\boldsymbol{X}\transpose & \cdots & \boldsymbol{0} \\
            \vdots    & \ddots & \vdots  \\
            \boldsymbol{0} & \cdots & v_q\boldsymbol{X}\transpose
        \end{pmatrix}
       \begin{pmatrix}
            \boldsymbol{\Psi}_{\tau_1}(\boldsymbol{\varepsilon}_{\tau_1})  & \cdots & \boldsymbol{0} \\
            \vdots    & \ddots & \vdots  \\
            \boldsymbol{0} & \cdots & \boldsymbol{\Psi}_{\tau_q}(\boldsymbol{\varepsilon}_{\tau_q})
        \end{pmatrix}
       \begin{pmatrix}
            \boldsymbol{\varepsilon}_{\tau_1} \\
            \vdots  \\
            \boldsymbol{\varepsilon}_{\tau_q} 
        \end{pmatrix}\\
    &=\frac{-2}{\sqrt{nm}}\boldsymbol{\delta}\transpose\sum_{i=1}^{n} (\boldsymbol{V}\otimes\boldsymbol{X}_i)
	\transpose\boldsymbol{\Psi}_{\boldsymbol{\tau}}(\boldsymbol{\varepsilon}_{i\boldsymbol{\tau}})\boldsymbol{\varepsilon}_{i\boldsymbol{\tau}}\\
        &=\frac{-2}{\sqrt{nm}}\boldsymbol{\delta}\transpose\sum_{i=1}^{n}
        \begin{bmatrix}
            v_1\sum_{j=1}^{m}x_{ij}^{1}\psi_{\tau_1}(\varepsilon_{it\tau_{1}})\varepsilon_{it\tau_{1}} \\
            \vdots \\
            v_1\sum_{j=1}^{m}x_{ij}^{p}\psi_{\tau_1}(\varepsilon_{it\tau_{1}})\varepsilon_{it\tau_{1}} \\
            \vdots \\
            v_q\sum_{j=1}^{m}x_{ij}^{1}\psi_{\tau_q}(\varepsilon_{it\tau_{q}})\varepsilon_{it\tau_{q}} \\
            \vdots \\
            v_q\sum_{j=1}^{m}x_{ij}^{p}\psi_{\tau_q}(\varepsilon_{it\tau_{q}})\varepsilon_{it\tau_{q}} \\
        \end{bmatrix}\\
    & \xrightarrow{d}-2\boldsymbol{\delta}\transpose\boldsymbol{B}.
\end{split}
\end{align*}
To show the asymptotic normality of $\boldsymbol{B},$ we apply the Cramér-Wold device and verify the Lyapunov's condition.
~~\\
Let $T_{ni}=m^{-1}\boldsymbol{\lambda}\transpose(\boldsymbol{V}\otimes\boldsymbol{X}_i)\transpose\boldsymbol{\Psi}_{\boldsymbol{\tau}}(
\boldsymbol{\varepsilon}_{i\boldsymbol{\tau}})\boldsymbol{\varepsilon}_{i\boldsymbol{\tau}}$
and consider $n^{-1/2}\sum_{i=1}^{n}T_{ni},$ where $\boldsymbol{\lambda}$ is a $pq \times 1$ unit vector, $\boldsymbol{\lambda}\transpose\boldsymbol{\lambda}=1.$ 
The summands $T_{ni}$ are independent with $\E[T_{ni}]=0$ and $\Var\Big[n^{-1/2}\sum_{i=1}^{n}T_{ni}\Big]>\nu\prime>0,$ by condition \textbf{A2}. By the Minkowski's inequality, we have
\begin{align*}
\begin{split}
 \E\lvert T_{ni} \rvert^{2+\nu} & = \E\bigg\lvert \sum_{k=1}^{q}\sum_{l=1}^{p}v_k\lambda_{kl}\sum_{j=1}^{m}m^{-1}x_{ij}^{l} \psi_{\tau_k}(\varepsilon_{ij\tau_{k}})\varepsilon_{ij\tau_{k}} \bigg\rvert^{2+\nu} \\
& = \E\bigg\lvert \sum_{k=1}^{q}\sum_{l=1}^{p}\sum_{j=1}^{m}m^{-1}v_k\lambda_{kl}x_{ij}^{l} \psi_{\tau_k}(\varepsilon_{ij\tau_{k}})\varepsilon_{ij\tau_{k}} \bigg\rvert^{2+\nu} \\
& \leq \bigg[ \sum_{k=1}^{q}\sum_{l=1}^{p}\sum_{j=1}^{m}\bigg(\E\Big\lvert m^{-1} v_k \lambda_{kl} x_{ij}^{l} \psi_{\tau_k}(\varepsilon_{ij\tau_{k}})\varepsilon_{ij\tau_{k}}\Big\rvert^{2+\nu}\bigg)^{\frac{1}{2+\nu}} \bigg]^{2+\nu}.\\
\end{split}
\end{align*}
By the Cauchy-Schwarz inequality, we have
\begin{equation*}
    \begin{split}
        \E\Big\lvert m^{-1} v_k \lambda_{kl} x_{ij}^{l} \psi_{\tau_k}(\varepsilon_{ij\tau_{k}})\varepsilon_{ij\tau_{k}}\Big\rvert^{2+\nu} 
            &=\lvert m^{-1} v_k\lambda_{kl}x_{ij}^{l}\rvert^{2+\nu}\E\Big\lvert \psi_{\tau_k}(\varepsilon_{ij\tau_{k}})\varepsilon_{ij\tau_{k}}\Big\rvert^{2+\nu} \\
            &\leq (m^{-1}M\lvert\lambda_{kl}\rvert)^{2+\nu} \Big[\E\lvert \psi_{\tau_k}(\varepsilon_{ij\tau_{k}})\rvert^{4+\nu}\Big]^{1/2} 
            \Big[\E\lvert \varepsilon_{ij\tau_{k}}\rvert^{4+\nu}\Big]^{1/2} \\
            & \leq (m^{-1}M\lvert\lambda_{kl}\rvert)^{2+\nu} \Delta,
    \end{split}
\end{equation*}
where the last inequality follows by $\E\lvert \psi_{\tau_k}(\varepsilon_{ij\tau_{k}})\rvert^{4+\nu}<\Delta$ and $\E\lvert \varepsilon_{ij\tau_{k}} \rvert^{4+\nu}<\Delta.$ Therefore,
\begin{align*}
\begin{split}
 \E\lvert T_{ni} \rvert^{2+\nu} & \leq \bigg[ \sum_{k=1}^{q}\sum_{l=1}^{p}\sum_{j=1}^{m} m^{-1} M \lvert\lambda_{kl}\rvert \Delta^{\frac{1}{2+\nu}} \bigg]^{2+\nu}\\
                                & \leq \Delta M^{2+\nu} \lvert\boldsymbol{\lambda}\transpose\mathds{1}_{pq}\rvert^{2+\nu} \\
                                & \leq \Delta (pq)^{1+\nu}. \\
\end{split}
\end{align*}
Then by the Liapounov CLT $\boldsymbol{B}$ is a zero mean Gaussian vector with covariance matrix $\boldsymbol{D}_{0}(\boldsymbol{\tau}).$

\begin{align*}
\begin{split}
   R_{nmq}^{(2)}(\boldsymbol{\delta})&= \frac{1}{nm}\sum_{k=1}^{q}\sum_{i=1}^{n}\sum_{j=1}^{m}v_k\boldsymbol{\delta}_{\tau_k}\transpose\boldsymbol{x}_{ij}\E[\psi_{\tau_k}
   (\varepsilon_{ij{\tau_k}})]
        \boldsymbol{x}_{ij}\transpose\boldsymbol{\delta}_{\tau_k}\\
    &=\frac{1}{nm}\sum_{k=1}^{q}v_k \boldsymbol{\delta}_{\tau_k}\transpose\boldsymbol{X}\transpose\E[\boldsymbol{\Psi}_{\tau_k}(\boldsymbol{\varepsilon}_{\tau_k})]\boldsymbol{X}\boldsymbol{\delta}_{\tau_k}\\
    & \rightarrow\boldsymbol{\delta}\transpose\boldsymbol{D}_1(\boldsymbol{\tau})\boldsymbol{\delta}.
\end{split}
\end{align*}
Thus the limiting form of the objective function is
\begin{equation*}
    R_{0q}(\boldsymbol{\delta})=-2\boldsymbol{\delta}\transpose\boldsymbol{B}+\boldsymbol{\delta}\transpose \boldsymbol{D}_1(\boldsymbol{\tau}) \boldsymbol{\delta}
\end{equation*}
where $\boldsymbol{B}$ is a zero mean Gaussian vector with covariance matrix $\boldsymbol{D}_0(\boldsymbol{\tau}).$ Application of Theorem 2.2 of \citet{hjort_asymptotics_2011} gives the result of Theorem \ref{theo2_erfe}.
\end{proof}
\begin{proof}[\textbf{Proof of Theorem \ref{theo3_erfe}}]$ $
~~\\
We have 
\begin{equation*}
\begin{split}
\widehat{\boldsymbol{D}}_1(\boldsymbol{\tau})&=\frac{1}{nm}(\mathbf{I}_q\otimes\widehat{\boldsymbol{X}^{*}})\transpose\boldsymbol{\Psi}_{\boldsymbol{\tau}}(
\widehat{\boldsymbol{\varepsilon}_{\boldsymbol{\tau}}^{*}})
(\boldsymbol{V}\otimes \widehat{\boldsymbol{X}^{*}})\\
&=\frac{1}{nm}\sum_{i=1}^{n}\diag\Big(v_1\widehat{\boldsymbol{X}^{*}_i}\transpose\boldsymbol{\Psi}_{\tau_1}(\widehat{\boldsymbol{\varepsilon}^{*}_{i\tau_1}})\widehat{\boldsymbol{X}^{*}_i},\ldots,v_q\widehat{\boldsymbol{X}^{*}_i}\transpose
\boldsymbol{\Psi}_{\tau_q}(\widehat{\boldsymbol{\varepsilon}^{*}_{i\tau_q}})\widehat{\boldsymbol{X}^{*}_i} \Big). \\
\end{split}
\end{equation*}
The convergence of $\widehat{\boldsymbol{D}}_1(\boldsymbol{\tau})$ is obtained by showing the convergence of the general term $\frac{1}{nm}\sum_{i=1}^{n}v_k\widehat{\boldsymbol{X}^{*}_i}\transpose\boldsymbol{\Psi}_{\tau_k}(\widehat{\boldsymbol{\varepsilon}^{*}_{i\tau_k}})\widehat{\boldsymbol{X}^{*}_i}.$  This general term breaks down as follows:
\begin{equation}\label{conv_d1_erfe}
\begin{split}
    \sum_{i=1}^{n}\widehat{\boldsymbol{X}^{*}_i}\transpose\boldsymbol{\Psi}_{\tau}(\widehat{\boldsymbol{\varepsilon}^{*}_{i\tau}})\widehat{\boldsymbol{X}^{*}_i} 
    &= \widehat{\boldsymbol{X}^{*}}\transpose\boldsymbol{\Psi}_{\tau}(\widehat{\boldsymbol{\varepsilon}_{\tau}^{*}}) \widehat{\boldsymbol{X}^{*}} 
     = \boldsymbol{X}\transpose\widehat{\boldsymbol{M}}_{\boldsymbol{Z}}(\tau)\transpose\boldsymbol{\Psi}_{\tau}(\widehat{\boldsymbol{\varepsilon}_{\tau}^{*}})\widehat{\boldsymbol{M}}_{\boldsymbol{Z}}(\tau)\boldsymbol{X}\\
    &= \boldsymbol{X}\transpose\boldsymbol{\Psi}_{\tau}(\widehat{\boldsymbol{\varepsilon}_{\tau}^{*}})\widehat{\boldsymbol{M}}_{\boldsymbol{Z}}(\tau)\boldsymbol{X}
    = \boldsymbol{X}\transpose\boldsymbol{\Psi}_{\tau}(\widehat{\boldsymbol{\varepsilon}_{\tau}^{*}})[\mathbf{I}_{nm}-\widehat{\boldsymbol{P}}_{\boldsymbol{Z}}(\tau)]\boldsymbol{X} \\
    &=\boldsymbol{X}\transpose\boldsymbol{\Psi}_{\tau}(\widehat{\boldsymbol{\varepsilon}_{\tau}^{*}})\boldsymbol{X} - \boldsymbol{X}\transpose\boldsymbol{\Psi}_{\tau}(\widehat{\boldsymbol{\varepsilon}_{\tau}^{*}})\widehat{\boldsymbol{P}}_{\boldsymbol{Z}}(\tau)\boldsymbol{X}.
\end{split}
\end{equation}
We will show the convergence of each of the terms separately. First, consider the following expression: $\lvert\psi_{\tau}(\widehat{\varepsilon}^{*}_{ij\tau})-\psi_{\tau}(\varepsilon^{*}_{ij\tau})\rvert.$ This expression is 0 except when $\boldsymbol{x}_{ij}^{*}\transpose\widehat{\boldsymbol{\beta}}_{\tau} \leq y_{ij}^{*} \leq \boldsymbol{x}_{ij}^{*}\transpose\boldsymbol{\beta}_{\tau}$ or $\boldsymbol{x}_{ij}^{*}\transpose \boldsymbol{\beta}_{\tau} \leq y_{ij}^{*} \leq \boldsymbol{x}_{ij}^{*}\transpose\widehat{\boldsymbol{\beta}}_{\tau}.$ It can be bounded as follows:
\begin{equation}\label{diff_phi_erfe}
    \begin{split}
        \lvert\psi_{\tau}(\widehat{\varepsilon}^{*}_{ij\tau})-\psi_{\tau}(\varepsilon^{*}_{ij\tau})\rvert 
        &\leq \lvert 1- 2\tau \rvert \mathds{1}\big(\lvert \varepsilon^{*}_{ij\tau}
            \rvert\leq\lvert\boldsymbol{x}_{ij}^{*}\transpose(\widehat{\boldsymbol{\beta}}_{\tau}-\boldsymbol{\beta}_{\tau})\rvert\big) \\ 
        &\leq \lvert 1- 2\tau \rvert \mathds{1}\big(\lvert \varepsilon^{*}_{ij\tau} 
            \rvert\leq \norm{\boldsymbol{x}_{ij}^{*}}\norm{\widehat{\boldsymbol{\beta}}_{\tau}-\boldsymbol{\beta}_{\tau}}\big) \\
        &\leq \lvert 1- 2\tau \rvert \mathds{1}\big(\lvert \varepsilon^{*}_{ij\tau} 
            \rvert\leq pM\norm{\widehat{\boldsymbol{\beta}}_{\tau}-\boldsymbol{\beta}_{\tau}}\big). \\   
    \end{split}
\end{equation}
As $\text{plim} \; \widehat{\boldsymbol{\beta}}_{\tau}=\boldsymbol{\beta}_{\tau},$ we have, by equation (\ref{diff_phi_erfe}) and Markov's inequality:
\begin{equation*}
\begin{split}
    (nm)^{-1}\boldsymbol{X}\transpose[\boldsymbol{\Psi}_{\tau}(\widehat{\boldsymbol{\varepsilon}^{*}_{\tau}})-\boldsymbol{\Psi}_{\tau}(\boldsymbol{\varepsilon}^{*}_{\tau})]\boldsymbol{X} =
    (nm)^{-1}\sum_{i=1}^{n}\boldsymbol{X}_i\transpose[\boldsymbol{\Psi}_{\tau}(\widehat{\boldsymbol{\varepsilon}^{*}_{i\tau}})-\boldsymbol{\Psi}_{\tau}
    (\boldsymbol{\varepsilon}^{*}_{i\tau})]\boldsymbol{X}_i \xrightarrow{p} \boldsymbol{0}.
\end{split}
\end{equation*}
In other words:
\begin{equation}\label{conv_diff_phi_erfe}
(nm)^{-1}\boldsymbol{X}\transpose\boldsymbol{\Psi}_{\tau}(\widehat{\boldsymbol{\varepsilon}^{*}_{\tau}})\boldsymbol{X} = (nm)^{-1}\boldsymbol{X}\transpose\boldsymbol{\Psi}_{\tau}(\boldsymbol{\varepsilon}^{*}_{\tau})\boldsymbol{X} + \smallO_{p}(1). 
\end{equation}
This result (\ref{conv_diff_phi_erfe}) is important and will be used frequently below. Another useful result is the convergence of the function $\widehat{w}^{*}_{i\tau}=(\sum_{j=1}^{m}\psi_{\tau}(\widehat{\varepsilon}^{*}_{ij\tau}))^{-1}$ which appears in the expression of $ \widehat{\boldsymbol{P}}_{\boldsymbol{Z}}(\tau).$ This function is bounded by:
\begin{equation*}
 m \min(\tau,1-\tau)\leq\widehat{w}^{*{-1}}_{i\tau}\leq m \max(\tau,1-\tau).
\end{equation*}
We have 
\begin{equation}\label{conv_poids_erfe} 
    \begin{split}
        \Big\lvert \frac{1}{nm}\sum_{i=1}^{n} \widehat{w}^{*}_{i\tau} - \frac{1}{nm}\sum_{i=1}^{n} \E[w^{*}_{i\tau}] \Big\rvert & \leq 
        \Big\lvert \frac{1}{nm}\sum_{i=1}^{n} w^{*}_{i\tau} - \frac{1}{nm}\sum_{i=1}^{n} \E[w^{*}_{i\tau}] \Big\rvert \\
        & + \Big\lvert \frac{1}{nm}\sum_{i=1}^{n} \widehat{w}^{*}_{i\tau} - \frac{1}{nm}\sum_{i=1}^{n} w^{*}_{i\tau}\Big\rvert. 
    \end{split}
\end{equation}
The first term converges by Markov's Law of Large Numbers (LLN). The second term is bounded by:
\begin{equation*}
    \Big\lvert \frac{1}{nm}\sum_{i=1}^{n} \widehat{w}^{*}_{i\tau} - \frac{1}{nm}\sum_{i=1}^{n} w^{*}_{i\tau}\Big\rvert \leq \frac{1}{nm^3}\sum_{i=1}^{n}\sum_{j=1}^{m}\lvert\psi_{\tau}(\widehat{\varepsilon}^{*}_{ij\tau})-\psi_{\tau}(\varepsilon^{*}_{ij\tau})\rvert.
\end{equation*}
Thus, convergence is achieved by the application of (\ref{diff_phi_erfe}).
 ~~\\
 ~~\\
Now, let's show the convergence of the first term on the last line of equation (\ref{conv_d1_erfe}). We have:
\begin{align}\label{first_d1_term_conv_erfe}
\begin{split}
& \Big\lvert \frac{1}{nm}\sum_{i=1}^{n}\boldsymbol{X}_i\transpose\boldsymbol{\Psi}_{\tau}(\widehat{\boldsymbol{\varepsilon}^{*}_{i\tau}}) \boldsymbol{X}_i - \frac{1}{nm}\sum_{i=1}^{n}\boldsymbol{X}_i\transpose
\E[\boldsymbol{\Psi}_{\tau}(\boldsymbol{\varepsilon}^{*}_{i\tau})]\boldsymbol{X}_i \Big\rvert \\
& \quad \leq    \bigg\lvert \frac{1}{nm}\sum_{i=1}^{n}\boldsymbol{X}_i\transpose[\boldsymbol{\Psi}_{\tau}(\widehat{\boldsymbol{\varepsilon}^{*}_{i\tau}})-\boldsymbol{\Psi}_{\tau}(\boldsymbol{\varepsilon}^{*}_{i\tau})] \boldsymbol{X}_i\bigg\rvert \\
& \quad  + \bigg\lvert\frac{1}{nm} \sum_{i=1}^{n}\boldsymbol{X}_i\transpose\boldsymbol{\Psi}_{\tau}(\boldsymbol{\varepsilon}^{*}_{i\tau})\boldsymbol{X}_i-\frac{1}{nm}\sum_{i=1}^{n}\boldsymbol{X}_i\transpose
\E[\boldsymbol{\Psi}_{\tau}(\boldsymbol{\varepsilon}^{*}_{i\tau})]\boldsymbol{X}_i\bigg \rvert.\\
\end{split}
\end{align}
The result (\ref{conv_diff_phi_erfe}) gives the convergence of the first term on the right of the inequality and the application of Markov's Law of large-number (LLN) gives the convergence of the second term.
 ~~\\
 ~~\\
The second term on the last line of equation (\ref{conv_d1_erfe}) is expressed by:
 \begin{equation}\label{second_d1_erfe}
     \boldsymbol{X}\transpose\boldsymbol{\Psi}_{\tau}(\widehat{\boldsymbol{\varepsilon}_{\tau}^{*}})\widehat{\boldsymbol{P}}_{\boldsymbol{Z}}(\tau)\boldsymbol{X} =
     \sum_{i=1}^{n}\sum_{j=1}^{m}\sum_{k=1}^{m}\widehat{w}^{*}_{i\tau}\psi_{\tau}(\widehat{\varepsilon}^{*}_{ij\tau})\psi_{\tau}(\widehat{\varepsilon}^{*}_{ik\tau})\boldsymbol{x}_{ij}\boldsymbol{x}_{ik}\transpose.
\end{equation}
We will use the following relation to show its convergence
\begin{equation}\label{second_d1_term_conv_inter_erfe}
\begin{split}
\widehat{w}^{*}_{i\tau}\psi_{\tau}(\widehat{\varepsilon}^{*}_{ij\tau})\psi_{\tau}(\widehat{\varepsilon}^{*}_{ik\tau}) & -w^{*}_{i\tau}\psi_{\tau}(\varepsilon^{*}_{ij\tau})\psi_{\tau}(\varepsilon^{*}_{ik\tau}) =
(\widehat{w}^{*}_{i\tau}-w^{*}_{i\tau})\psi_{\tau}(\widehat{\varepsilon}^{*}_{ij\tau})\psi_{\tau}(\widehat{\varepsilon}^{*}_{ik\tau})  \\
& + w^{*}_{i\tau}\big\lbrace \psi_{\tau}(\widehat{\varepsilon}^{*}_{ij\tau})[\psi_{\tau}(\widehat{\varepsilon}^{*}_{ik\tau})-\psi_{\tau}(\varepsilon^{*}_{ik\tau})] + \psi_{\tau}(\varepsilon^{*}_{ik\tau})[\psi_{\tau}(\widehat{\varepsilon}^{*}_{ij\tau})-\psi_{\tau}(\varepsilon^{*}_{ij\tau})]\big\rbrace. 
\end{split}
\end{equation}
To show the convergence of the expression (\ref{second_d1_erfe}), consider:
\begin{align}\label{second_d1_term_conv_erfe}
\begin{split}
& \Big\lvert\frac{1}{nm}\sum_{i=1}^{n}\sum_{jk}^{m}\boldsymbol{x}_{ij}\boldsymbol{x}_{ik}\transpose\widehat{w}^{*}_{i\tau}\psi_{\tau}(\widehat{\varepsilon}^{*}_{ij\tau})\psi_{\tau}(\widehat{\varepsilon}^{*}_{ik\tau}) 
        - \frac{1}{nm}\sum_{i=1}^{n}\sum_{jk}^{m}\boldsymbol{x}_{ij}\boldsymbol{x}_{ik}\transpose\E[w^{*}_{i\tau}\psi_{\tau}(\varepsilon^{*}_{ij\tau})\psi_{\tau}(\varepsilon^{*}_{ik\tau})]\Big\rvert  \\
& \quad \leq \Big\lvert\frac{1}{nm}\sum_{i=1}^{n}\sum_{jk}^{m}\boldsymbol{x}_{ij}\boldsymbol{x}_{ik}\transpose
        [\widehat{w}^{*}_{i\tau}\psi_{\tau}(\widehat{\varepsilon}^{*}_{ij\tau})\psi_{\tau}(\widehat{\varepsilon}^{*}_{ik\tau})- w^{*}_{i\tau}\psi_{\tau}(\varepsilon^{*}_{ij\tau})\psi_{\tau}(\varepsilon^{*}_{ik\tau})]\Big\rvert \\
& \quad  + \Big\lvert\frac{1}{nm}\sum_{i=1}^{n}\sum_{jk}^{m}\boldsymbol{x}_{ij}\boldsymbol{x}_{ik}\transpose w^{*}_{i\tau}\psi_{\tau}(\varepsilon^{*}_{ij\tau})\psi_{\tau}(\varepsilon^{*}_{ik\tau})
        -\frac{1}{nm}\sum_{i=1}^{n}\sum_{jk}^{m}\boldsymbol{x}_{ij}\boldsymbol{x}_{ik}\transpose\E[w^{*}_{i\tau}\psi_{\tau}(\varepsilon^{*}_{ij\tau})\psi_{\tau}(\varepsilon^{*}_{ik\tau})]\Big\rvert  \\
& \quad \leq \Big\lvert\frac{1}{nm}\sum_{i=1}^{n}\sum_{jk}^{m}\boldsymbol{x}_{ij}\boldsymbol{x}_{ik}\transpose (\widehat{w}^{*}_{i\tau}-w^{*}_{i\tau})\psi_{\tau}(\widehat{\varepsilon}^{*}_{ij\tau})\psi_{\tau}(\widehat{\varepsilon}^{*}_{ik\tau})
        \Big\rvert \\
& \quad +    \Big\lvert\frac{1}{nm}\sum_{i=1}^{n}\sum_{jk}^{m}\boldsymbol{x}_{ij}\boldsymbol{x}_{ik}\transpose w^{*}_{i\tau} \big\lbrace \psi_{\tau}(\widehat{\varepsilon}^{*}_{ij\tau})  
        [\psi_{\tau}(\widehat{\varepsilon}^{*}_{ik\tau})-\psi_{\tau}(\varepsilon^{*}_{ik\tau})] + \psi_{\tau}(\varepsilon^{*}_{ik\tau}) [\psi_{\tau}(\widehat{\varepsilon}^{*}_{ij\tau})-\psi_{\tau}(\varepsilon^{*}_{ij\tau})]\big\rbrace\Big\rvert \\
& \quad  + \Big\lvert\frac{1}{nm}\sum_{i=1}^{n}\sum_{jk}^{m}\boldsymbol{x}_{ij}\boldsymbol{x}_{ik}\transpose w^{*}_{i\tau}\psi_{\tau}(\varepsilon^{*}_{ij\tau})\psi_{\tau}(\varepsilon^{*}_{ik\tau}) -  
        \frac{1}{nm}\sum_{i=1}^{n}\sum_{jk}^{m}\boldsymbol{x}_{ij}\boldsymbol{x}_{ik}\transpose\E[w^{*}_{i\tau}\psi_{\tau}(\varepsilon^{*}_{ij\tau})\psi_{\tau}(\varepsilon^{*}_{ik\tau})]\Big\rvert  \\
& \quad \leq \Big\lvert\frac{1}{nm}\sum_{i=1}^{n}\sum_{jk}^{m}\boldsymbol{x}_{ij}\boldsymbol{x}_{ik}\transpose (\widehat{w}^{*}_{i\tau}-w^{*}_{i\tau})\Big\rvert + 
        \Big\lvert\frac{1}{nm^2}\sum_{i=1}^{n}\sum_{jk}^{m}\boldsymbol{x}_{ij}\boldsymbol{x}_{ik}\transpose [\psi_{\tau}(\widehat{\varepsilon}^{*}_{ik\tau})-\psi_{\tau}(\varepsilon^{*}_{ik\tau})]\Big\rvert \\
& \quad + \Big\lvert\frac{1}{nm^2}\sum_{i=1}^{n}\sum_{jk}^{m}\boldsymbol{x}_{ij}\boldsymbol{x}_{ik}\transpose [\psi_{\tau}(\widehat{\varepsilon}^{*}_{ij\tau})-\psi_{\tau}(\varepsilon^{*}_{ij\tau})]\Big\rvert \\
& \quad  + \Big\lvert\frac{1}{nm}\sum_{i=1}^{n}\sum_{jk}^{m}\boldsymbol{x}_{ij}\boldsymbol{x}_{ik}\transpose
w^{*}_{i\tau}\psi_{\tau}(\varepsilon^{*}_{ij\tau})\psi_{\tau}(\varepsilon^{*}_{ik\tau}) - \frac{1}{nm}\sum_{i=1}^{n}\sum_{jk}^{m}\boldsymbol{x}_{ij}\boldsymbol{x}_{ik}\transpose\E[w^{*}_{i\tau}\psi_{\tau}(\varepsilon^{*}_{ij\tau})\psi_{\tau}(\varepsilon^{*}_{ik\tau})]\Big\rvert.  \\
\end{split}
\end{align}
Finally, the convergence of the expression (\ref{second_d1_erfe}) results from the application of the relations (\ref{conv_diff_phi_erfe}) and (\ref{conv_poids_erfe} ) and from the application of Markov's Law of Large Numbers.
~~\\
~~\\
The second term of the variance-covariance matrix whose convergence we must show is:
\begin{align*}
    \begin{split}
        \widehat{\boldsymbol{D}}_0(\boldsymbol{\tau})
        &=\frac{1}{nm}(\boldsymbol{V}\otimes\widehat{\boldsymbol{X}^{*}})\transpose\boldsymbol{\Psi}_{\boldsymbol{\tau}}(\widehat{\boldsymbol{\varepsilon}_{\boldsymbol{\tau}}^{*}})
        \widehat{\boldsymbol{\varepsilon}_{\boldsymbol{\tau}}^{*}}\widehat{\boldsymbol{\varepsilon}_{\boldsymbol{\tau}}^{*}}\transpose
        \boldsymbol{\Psi}_{\boldsymbol{\tau}}(\widehat{\boldsymbol{\varepsilon}_{\boldsymbol{\tau}}^{*}}) (\boldsymbol{V}\otimes\widehat{\boldsymbol{X}^{*}}), \\
        &=\frac{1}{nm}\sum_{i=1}^{n}
         \begin{pmatrix}
          v_1^2\widehat{\boldsymbol{X}^{*}_i}\transpose\widehat{\boldsymbol{\Sigma}}^{*}_{i\tau_{1}\tau_{1}} \widehat{\boldsymbol{X}^{*}_i}
          & \cdots
          & \cdots &   v_1v_q\widehat{\boldsymbol{X}^{*}_i}\transpose\widehat{\boldsymbol{\Sigma}}^{*}_{i\tau_{1}\tau_{q}} \widehat{\boldsymbol{X}^{*}_i}  \\
          \vdots  & \vdots  & \ddots & \vdots \\
            v_1v_q\widehat{\boldsymbol{X}^{*}_i}\transpose\widehat{\boldsymbol{\Sigma}}^{*}_{i\tau_{q}\tau_{1}} \widehat{\boldsymbol{X}^{*}_i}  
          & \cdots
          & \cdots
          &   v_q^2\widehat{\boldsymbol{X}^{*}_i}\transpose\widehat{\boldsymbol{\Sigma}}^{*}_{i\tau_{q}\tau_{q}}\widehat{\boldsymbol{X}^{*}_i} 
         \end{pmatrix}
    \end{split}
\end{align*}
where 
\begin{equation*}
\widehat{\boldsymbol{\Sigma}}^{*}_{i\tau_{k}\tau_{j}}=\boldsymbol{\Psi}_{\tau_k}(\widehat{\boldsymbol{\varepsilon}}^{*}_{i\tau_k})\widehat{\boldsymbol{\varepsilon}}^{*}_{i\tau_k}\widehat{\boldsymbol{\varepsilon}}^{*}_{i\tau_j}\transpose
\boldsymbol{\Psi}_{\tau_j}(\widehat{\boldsymbol{\varepsilon}}^{*}_{i\tau_j}).
\end{equation*}
~~\\
Again, showing the convergence of the general term suffices:
\begin{equation}\label{terme_general_d0_erfe}
    \frac{1}{nm}\sum_{i=1}^{n}\widehat{\boldsymbol{X}^{*}_i}\transpose\widehat{\boldsymbol{\Sigma}}^{*}_{i\tau_{k}\tau_{j}}\widehat{\boldsymbol{X}^{*}_i}=\frac{1}{nm}
    \boldsymbol{\widehat{\boldsymbol{X}^{*}}}\transpose\boldsymbol{\Psi}_{\tau_k}(\widehat{\boldsymbol{\varepsilon}}^{*}_{\tau_k})\widehat{\boldsymbol{\varepsilon}}^{*}_{\tau_k}\widehat{\boldsymbol{\varepsilon}}^{*}_{\tau_j}\transpose
        \boldsymbol{\Psi}_{\tau_j}(\widehat{\boldsymbol{\varepsilon}}^{*}_{\tau_j})\boldsymbol{\widehat{\boldsymbol{X}^{*}}}
\end{equation}
~~\\
Note that,
\begin{equation*}
\widehat{\boldsymbol{\varepsilon}}^{*}_{\tau}=\boldsymbol{\varepsilon}^{*}_{\tau}-\boldsymbol{X}^{*}(\widehat{\boldsymbol{\beta}}_{\tau}-\boldsymbol{\beta}_{\tau}) + \Delta\boldsymbol{M}_{\boldsymbol{Z}}(\tau)[\boldsymbol{y}-\boldsymbol{X}\boldsymbol{\beta}_{\tau}],
\end{equation*}
where $\Delta\boldsymbol{M}_{\boldsymbol{Z}}(\tau)=\widehat{\boldsymbol{M}}_{\boldsymbol{Z}}(\tau)-\boldsymbol{M}_{\boldsymbol{Z}}(\tau)=\boldsymbol{P}_{\boldsymbol{Z}}(\tau)-\widehat{\boldsymbol{P}}_{\boldsymbol{Z}}(\tau)=\Delta\boldsymbol{P}_{\boldsymbol{Z}}(\tau).$ We have,
\begin{equation*}
\begin{split}
    \widehat{\boldsymbol{\varepsilon}_{\tau}}^{*}_{\tau_{k}}\widehat{\boldsymbol{\varepsilon}_{\tau}}^{*}_{\tau_{j}}\transpose 
    &=\boldsymbol{\varepsilon}^{*}_{\tau_{k}}\boldsymbol{\varepsilon}^{*}_{\tau_{j}}\transpose-\boldsymbol{\varepsilon}^{*}_{\tau_{k}}(\widehat{\boldsymbol{\beta}}_{\tau_{j}} - \boldsymbol{\beta}_{\tau_{j}})\transpose \boldsymbol{X}^{*}\transpose 
     + \boldsymbol{\varepsilon}^{*}_{\tau_{k}} [\boldsymbol{y}-\boldsymbol{X}\widehat{\boldsymbol{\beta}}_{\tau_{j}}]\transpose\Delta\boldsymbol{P}_{\boldsymbol{Z}}(\tau_{j})\transpose  \\
    &-\boldsymbol{X}^{*} (\widehat{\boldsymbol{\beta}}_{\tau_{k}} - \boldsymbol{\beta}_{\tau_{k}})\boldsymbol{\varepsilon}^{*}_{\tau_{j}}\transpose +
    \boldsymbol{X}^{*} (\widehat{\boldsymbol{\beta}}_{\tau_{k}} - \boldsymbol{\beta}_{\tau_{k}})(\widehat{\boldsymbol{\beta}}_{\tau_{j}} - \boldsymbol{\beta}_{\tau_{j}})\transpose\boldsymbol{X}^{*}\transpose \\
    & -\boldsymbol{X}^{*} (\widehat{\boldsymbol{\beta}}_{\tau_{k}} - \boldsymbol{\beta}_{\tau_{k}})[\boldsymbol{y}-\boldsymbol{X}\widehat{\boldsymbol{\beta}}_{\tau_{j}}]\transpose
        \Delta\boldsymbol{P}_{\boldsymbol{Z}}(\tau_{j})\transpose + \Delta\boldsymbol{P}_{\boldsymbol{Z}}(\tau_{k})[\boldsymbol{y}-\boldsymbol{X}\widehat{\boldsymbol{\beta}}_{\tau_{k}}]\boldsymbol{\varepsilon}^{*}_{\tau_{j}}\transpose  \\
    &-\Delta\boldsymbol{P}_{\boldsymbol{Z}}(\tau_{k})[\boldsymbol{y}-\boldsymbol{X}\widehat{\boldsymbol{\beta}}_{\tau_{k}}](\widehat{\boldsymbol{\beta}}_{\tau_{j}} - \boldsymbol{\beta}_{\tau_{j}})\transpose\boldsymbol{X}^{*}\transpose\\
    &+ \Delta\boldsymbol{P}_{\boldsymbol{Z}}(\tau_{k})[\boldsymbol{y}-\boldsymbol{X}\widehat{\boldsymbol{\beta}}_{\tau_{k}}][\boldsymbol{y}-\boldsymbol{X}
    \widehat{\boldsymbol{\beta}}_{\tau_{j}}]\transpose
        \Delta\boldsymbol{P}_{\boldsymbol{Z}}(\tau_{j})\transpose\\
\end{split}
\end{equation*}
Then, replacing $\boldsymbol{\widehat{\boldsymbol{X}^{*}}}=\widehat{\boldsymbol{M}}_{\boldsymbol{Z}}(\tau)\boldsymbol{X}=\boldsymbol{X}^{*} + \Delta\boldsymbol{M}_{\boldsymbol{Z}}(\tau)\boldsymbol{X},$ we obtain:

\begin{equation*}
\begin{split}
&{} \boldsymbol{\widehat{\boldsymbol{X}^{*}}}\transpose\boldsymbol{\Psi}_{\tau_k}(\widehat{\boldsymbol{\varepsilon}}^{*}_{\tau_k})\widehat{\boldsymbol{\varepsilon}}^{*}_{\tau_k}\widehat{\boldsymbol{\varepsilon}}^{*}_{\tau_j}\transpose
        \boldsymbol{\Psi}_{\tau_j}(\widehat{\boldsymbol{\varepsilon}}^{*}_{\tau_j})\boldsymbol{\widehat{\boldsymbol{X}^{*}}}= \\
& + \boldsymbol{X}^{*}\transpose\boldsymbol{\Psi}_{\tau_k}(\widehat{\boldsymbol{\varepsilon}}^{*}_{\tau_k})\boldsymbol{\varepsilon}^{*}_{\tau_{k}}\boldsymbol{\varepsilon}^{*}_{\tau_{j}}\transpose\boldsymbol{\Psi}_{\tau_j}(\widehat{\boldsymbol{\varepsilon}}^{*}_{\tau_j})
        \boldsymbol{X}^{*} - \boldsymbol{X}^{*}\transpose\boldsymbol{\Psi}_{\tau_k}(\widehat{\boldsymbol{\varepsilon}}^{*}_{\tau_k})\boldsymbol{\varepsilon}^{*}_{\tau_{k}}(\widehat{\boldsymbol{\beta}}_{\tau_{j}} - \boldsymbol{\beta}_{\tau_{j}})\transpose  \boldsymbol{X}^{*}\transpose\boldsymbol{\Psi}_{\tau_j}(\widehat{\boldsymbol{\varepsilon}}^{*}_{\tau_j})\boldsymbol{X}^{*}\\
& + \boldsymbol{X}^{*}\transpose\boldsymbol{\Psi}_{\tau_k}(\widehat{\boldsymbol{\varepsilon}}^{*}_{\tau_k})\boldsymbol{\varepsilon}^{*}_{\tau_{k}}[\boldsymbol{y}-\boldsymbol{X}\widehat{\boldsymbol{\beta}}_{\tau_{j}}]\transpose
        \Delta\boldsymbol{P}_{\boldsymbol{Z}}(\tau_{j})\transpose\boldsymbol{\Psi}_{\tau_j}(\widehat{\boldsymbol{\varepsilon}}^{*}_{\tau_j})\boldsymbol{X}^{*}\\
& - \boldsymbol{X}^{*}\transpose\boldsymbol{\Psi}_{\tau_k}(\widehat{\boldsymbol{\varepsilon}}^{*}_{\tau_k})\boldsymbol{X}^{*} (\widehat{\boldsymbol{\beta}}_{\tau_{k}} - \boldsymbol{\beta}_{\tau_{k}})\boldsymbol{\varepsilon}^{*}_{\tau_{j}}\transpose
    \boldsymbol{\Psi}_{\tau_j}(\widehat{\boldsymbol{\varepsilon}}^{*}_{\tau_j})\boldsymbol{X}^{*}\\
& + \boldsymbol{X}^{*}\transpose\boldsymbol{\Psi}_{\tau_k}(\widehat{\boldsymbol{\varepsilon}}^{*}_{\tau_k})\boldsymbol{X}^{*} (\widehat{\boldsymbol{\beta}}_{\tau_{k}} - \boldsymbol{\beta}_{\tau_{k}})(\widehat{\boldsymbol{\beta}}_{\tau_{j}} -         \boldsymbol{\beta}_{\tau_{j}})\transpose\boldsymbol{X}^{*}\transpose\boldsymbol{\Psi}_{\tau_j}(\widehat{\boldsymbol{\varepsilon}}^{*}_{\tau_j})\boldsymbol{X}^{*}\\
& - \boldsymbol{X}^{*}\transpose\boldsymbol{\Psi}_{\tau_k}(\widehat{\boldsymbol{\varepsilon}}^{*}_{\tau_k})\boldsymbol{X}^{*} (\widehat{\boldsymbol{\beta}}_{\tau_{k}} -     
    \boldsymbol{\beta}_{\tau_{k}})[\boldsymbol{y}-\boldsymbol{X}\widehat{\boldsymbol{\beta}}_{\tau_{j}}]\transpose\Delta\boldsymbol{P}_{\boldsymbol{Z}}(\tau_{j})\transpose\boldsymbol{\Psi}_{\tau_j}
    (\widehat{\boldsymbol{\varepsilon}}^{*}_{\tau_j})\boldsymbol{X}^{*}\\
& + \boldsymbol{X}^{*}\transpose\boldsymbol{\Psi}_{\tau_k}(\widehat{\boldsymbol{\varepsilon}}^{*}_{\tau_k})\Delta\boldsymbol{P}_{\boldsymbol{Z}}(\tau_{k})[\boldsymbol{y}-\boldsymbol{X}\widehat{\boldsymbol{\beta}}_{\tau_{k}}]
    \boldsymbol{\varepsilon}^{*}_{\tau_{j}}\transpose\boldsymbol{\Psi}_{\tau_j}(\widehat{\boldsymbol{\varepsilon}}^{*}_{\tau_j})\boldsymbol{X}^{*}\\
& - \boldsymbol{X}^{*}\transpose\boldsymbol{\Psi}_{\tau_k}(\widehat{\boldsymbol{\varepsilon}}^{*}_{\tau_k})\Delta\boldsymbol{P}_{\boldsymbol{Z}}(\tau_{k})[\boldsymbol{y}-\boldsymbol{X}\widehat{\boldsymbol{\beta}}_{\tau_{k}}]
    (\widehat{\boldsymbol{\beta}}_{\tau_{j}} - \boldsymbol{\beta}_{\tau_{j}})\transpose\boldsymbol{X}^{*}\transpose\boldsymbol{\Psi}_{\tau_j}(\widehat{\boldsymbol{\varepsilon}}^{*}_{\tau_j})\boldsymbol{X}^{*}\\
& + \boldsymbol{X}^{*}\transpose\boldsymbol{\Psi}_{\tau_k}(\widehat{\boldsymbol{\varepsilon}}^{*}_{\tau_k})\Delta\boldsymbol{P}_{\boldsymbol{Z}}(\tau_{k})[\boldsymbol{y}-\boldsymbol{X}\widehat{\boldsymbol{\beta}}_{\tau_{k}}]
    [\boldsymbol{y}-\boldsymbol{X}\widehat{\boldsymbol{\beta}}_{\tau_{j}}]\transpose\Delta\boldsymbol{P}_{\boldsymbol{Z}}(\tau_{j})\transpose\boldsymbol{\Psi}_{\tau_j}(\widehat{\boldsymbol{\varepsilon}}^{*}_{\tau_j})\boldsymbol{X}^{*}\\
& + \boldsymbol{X}^{*}\transpose\boldsymbol{\Psi}_{\tau_k}(\widehat{\boldsymbol{\varepsilon}}^{*}_{\tau_k})\boldsymbol{\varepsilon}^{*}_{\tau_{k}}\boldsymbol{\varepsilon}^{*}_{\tau_{j}}\transpose\boldsymbol{\Psi}_{\tau_j}(\widehat{\boldsymbol{\varepsilon}}^{*}_{\tau_j})
        \Delta\boldsymbol{P}_{\boldsymbol{Z}}(\tau_j)\boldsymbol{X} \\
& - \boldsymbol{X}^{*}\transpose\boldsymbol{\Psi}_{\tau_k}(\widehat{\boldsymbol{\varepsilon}}^{*}_{\tau_k})\boldsymbol{\varepsilon}^{*}_{\tau_{k}}(\widehat{\boldsymbol{\beta}}_{\tau_{j}} - \boldsymbol{\beta}_{\tau_{j}})\transpose   
       \boldsymbol{X}^{*}\transpose\boldsymbol{\Psi}_{\tau_j}(\widehat{\boldsymbol{\varepsilon}}^{*}_{\tau_j})\Delta\boldsymbol{P}_{\boldsymbol{Z}}(\tau_j)\boldsymbol{X}\\
& + \boldsymbol{X}^{*}\transpose\boldsymbol{\Psi}_{\tau_k}(\widehat{\boldsymbol{\varepsilon}}^{*}_{\tau_k})\boldsymbol{\varepsilon}^{*}_{\tau_{k}}[\boldsymbol{y}-\boldsymbol{X}\widehat{\boldsymbol{\beta}}_{\tau_{j}}]\transpose
        \Delta\boldsymbol{P}_{\boldsymbol{Z}}(\tau_{j})\transpose\boldsymbol{\Psi}_{\tau_j}(\widehat{\boldsymbol{\varepsilon}}^{*}_{\tau_j})\Delta\boldsymbol{P}_{\boldsymbol{Z}}(\tau_j)\boldsymbol{X}\\
& - \boldsymbol{X}^{*}\transpose\boldsymbol{\Psi}_{\tau_k}(\widehat{\boldsymbol{\varepsilon}}^{*}_{\tau_k})\boldsymbol{X}^{*} (\widehat{\boldsymbol{\beta}}_{\tau_{k}} - \boldsymbol{\beta}_{\tau_{k}})\boldsymbol{\varepsilon}^{*}_{\tau_{j}}\transpose
    \boldsymbol{\Psi}_{\tau_j}(\widehat{\boldsymbol{\varepsilon}}^{*}_{\tau_j})\Delta\boldsymbol{P}_{\boldsymbol{Z}}(\tau_j)\boldsymbol{X}\\
& + \boldsymbol{X}^{*}\transpose\boldsymbol{\Psi}_{\tau_k}(\widehat{\boldsymbol{\varepsilon}}^{*}_{\tau_k})\boldsymbol{X}^{*} (\widehat{\boldsymbol{\beta}}_{\tau_{k}} - \boldsymbol{\beta}_{\tau_{k}})(\widehat{\boldsymbol{\beta}}_{\tau_{j}} -         \boldsymbol{\beta}_{\tau_{j}})\transpose\boldsymbol{X}^{*}\transpose\boldsymbol{\Psi}_{\tau_j}(\widehat{\boldsymbol{\varepsilon}}^{*}_{\tau_j})\Delta\boldsymbol{P}_{\boldsymbol{Z}}(\tau_j)\boldsymbol{X}\\
& - \boldsymbol{X}^{*}\transpose\boldsymbol{\Psi}_{\tau_k}(\widehat{\boldsymbol{\varepsilon}}^{*}_{\tau_k})\boldsymbol{X}^{*} (\widehat{\boldsymbol{\beta}}_{\tau_{k}} -     
    \boldsymbol{\beta}_{\tau_{k}})[\boldsymbol{y}-\boldsymbol{X}\widehat{\boldsymbol{\beta}}_{\tau_{j}}]\transpose\Delta\boldsymbol{P}_{\boldsymbol{Z}}(\tau_{j})\transpose\boldsymbol{\Psi}_{\tau_j}
    (\widehat{\boldsymbol{\varepsilon}}^{*}_{\tau_j})\Delta\boldsymbol{P}_{\boldsymbol{Z}}(\tau_j)\boldsymbol{X}\\
& + \boldsymbol{X}^{*}\transpose\boldsymbol{\Psi}_{\tau_k}(\widehat{\boldsymbol{\varepsilon}}^{*}_{\tau_k})\Delta\boldsymbol{P}_{\boldsymbol{Z}}(\tau_{k})[\boldsymbol{y}-\boldsymbol{X}\widehat{\boldsymbol{\beta}}_{\tau_{k}}]
    \boldsymbol{\varepsilon}^{*}_{\tau_{j}}\transpose\boldsymbol{\Psi}_{\tau_j}(\widehat{\boldsymbol{\varepsilon}}^{*}_{\tau_j})\Delta\boldsymbol{P}_{\boldsymbol{Z}}(\tau_j)\boldsymbol{X}\\
& - \boldsymbol{X}^{*}\transpose\boldsymbol{\Psi}_{\tau_k}(\widehat{\boldsymbol{\varepsilon}}^{*}_{\tau_k})\Delta\boldsymbol{P}_{\boldsymbol{Z}}(\tau_{k})[\boldsymbol{y}-\boldsymbol{X}\widehat{\boldsymbol{\beta}}_{\tau_{k}}]
    (\widehat{\boldsymbol{\beta}}_{\tau_{j}} - \boldsymbol{\beta}_{\tau_{j}})\transpose\boldsymbol{X}^{*}\transpose\boldsymbol{\Psi}_{\tau_j}(\widehat{\boldsymbol{\varepsilon}}^{*}_{\tau_j})
    \Delta\boldsymbol{P}_{\boldsymbol{Z}}(\tau_j)\boldsymbol{X}\\
& + \boldsymbol{X}^{*}\transpose\boldsymbol{\Psi}_{\tau_k}(\widehat{\boldsymbol{\varepsilon}}^{*}_{\tau_k})\Delta\boldsymbol{P}_{\boldsymbol{Z}}(\tau_{k})[\boldsymbol{y}-\boldsymbol{X}\widehat{\boldsymbol{\beta}}_{\tau_{k}}]
    [\boldsymbol{y}-\boldsymbol{X}\widehat{\boldsymbol{\beta}}_{\tau_{j}}]\transpose\Delta\boldsymbol{P}_{\boldsymbol{Z}}(\tau_{j})\transpose\boldsymbol{\Psi}_{\tau_j}(\widehat{\boldsymbol{\varepsilon}}^{*}_{\tau_j})
    \Delta\boldsymbol{P}_{\boldsymbol{Z}}(\tau_j)\boldsymbol{X}\\
& + \boldsymbol{X}\transpose\Delta\boldsymbol{P}_{\boldsymbol{Z}}(\tau_k)\transpose\boldsymbol{\Psi}_{\tau_k}(\widehat{\boldsymbol{\varepsilon}}^{*}_{\tau_k})\boldsymbol{\varepsilon}^{*}_{\tau_{k}}\boldsymbol{\varepsilon}^{*}_{\tau_{j}}\transpose
        \boldsymbol{\Psi}_{\tau_j}(\widehat{\boldsymbol{\varepsilon}}^{*}_{\tau_j})\boldsymbol{X}^{*} \\
& - \boldsymbol{X}\transpose\Delta\boldsymbol{P}_{\boldsymbol{Z}}(\tau_k)\transpose\boldsymbol{\Psi}_{\tau_k}(\widehat{\boldsymbol{\varepsilon}}^{*}_{\tau_k})\boldsymbol{\varepsilon}^{*}_{\tau_{k}}(\widehat{\boldsymbol{\beta}}_{\tau_{j}} -   
        \boldsymbol{\beta}_{\tau_{j}})\transpose\boldsymbol{X}^{*}\transpose\boldsymbol{\Psi}_{\tau_j}(\widehat{\boldsymbol{\varepsilon}}^{*}_{\tau_j})\boldsymbol{X}^{*}\\
& + \boldsymbol{X}\transpose\Delta\boldsymbol{P}_{\boldsymbol{Z}}(\tau_k)\transpose\boldsymbol{\Psi}_{\tau_k}(\widehat{\boldsymbol{\varepsilon}}^{*}_{\tau_k})\boldsymbol{\varepsilon}^{*}_{\tau_{k}}[\boldsymbol{y}-\boldsymbol{X}
        \widehat{\boldsymbol{\beta}}_{\tau_{j}}]\transpose\Delta\boldsymbol{P}_{\boldsymbol{Z}}(\tau_{j})\transpose\boldsymbol{\Psi}_{\tau_j}(\widehat{\boldsymbol{\varepsilon}}^{*}_{\tau_j})\boldsymbol{X}^{*}\\
& - \boldsymbol{X}\transpose\Delta\boldsymbol{P}_{\boldsymbol{Z}}(\tau_k)\transpose\boldsymbol{\Psi}_{\tau_k}(\widehat{\boldsymbol{\varepsilon}}^{*}_{\tau_k})\boldsymbol{X}^{*} (\widehat{\boldsymbol{\beta}}_{\tau_{k}} -   
        \boldsymbol{\beta}_{\tau_{k}})\boldsymbol{\varepsilon}^{*}_{\tau_{j}}\transpose\boldsymbol{\Psi}_{\tau_j}(\widehat{\boldsymbol{\varepsilon}}^{*}_{\tau_j})\boldsymbol{X}^{*}\\
& + \boldsymbol{X}\transpose\Delta\boldsymbol{P}_{\boldsymbol{Z}}(\tau_k)\transpose\boldsymbol{\Psi}_{\tau_k}(\widehat{\boldsymbol{\varepsilon}}^{*}_{\tau_k})\boldsymbol{X}^{*} (\widehat{\boldsymbol{\beta}}_{\tau_{k}} -   
    \boldsymbol{\beta}_{\tau_{k}})(\widehat{\boldsymbol{\beta}}_{\tau_{j}} - \boldsymbol{\beta}_{\tau_{j}})\transpose\boldsymbol{X}^{*}\transpose\boldsymbol{\Psi}_{\tau_j}(\widehat{\boldsymbol{\varepsilon}}^{*}_{\tau_j})\boldsymbol{X}^{*}\\
&\hspace{5.12in}\llap{\text{(continued on next page)}}
\end{split}
\end{equation*}       
\clearpage
\begin{equation*}
\begin{split}
& \hspace{5.12in}\llap{\text{(continued from previous page)}}\\
& - \boldsymbol{X}\transpose\Delta\boldsymbol{P}_{\boldsymbol{Z}}(\tau_k)\transpose\boldsymbol{\Psi}_{\tau_k}(\widehat{\boldsymbol{\varepsilon}}^{*}_{\tau_k})\boldsymbol{X}^{*} (\widehat{\boldsymbol{\beta}}_{\tau_{k}} -
        \boldsymbol{\beta}_{\tau_{k}})[\boldsymbol{y}-\boldsymbol{X}\widehat{\boldsymbol{\beta}}_{\tau_{j}}]\transpose\Delta\boldsymbol{P}_{\boldsymbol{Z}}(\tau_{j})\transpose\boldsymbol{\Psi}_{\tau_j}
        (\widehat{\boldsymbol{\varepsilon}}^{*}_{\tau_j})\boldsymbol{X}^{*}\\
& + \boldsymbol{X}\transpose\Delta\boldsymbol{P}_{\boldsymbol{Z}}(\tau_k)\transpose\boldsymbol{\Psi}_{\tau_k}(\widehat{\boldsymbol{\varepsilon}}^{*}_{\tau_k})\Delta\boldsymbol{P}_{\boldsymbol{Z}}(\tau_{k})
[\boldsymbol{y}-\boldsymbol{X}\widehat{\boldsymbol{\beta}}_{\tau_{k}}]\boldsymbol{\varepsilon}^{*}_{\tau_{j}}\transpose\boldsymbol{\Psi}_{\tau_j}(\widehat{\boldsymbol{\varepsilon}}^{*}_{\tau_j})\boldsymbol{X}^{*}\\
& - \boldsymbol{X}\transpose\Delta\boldsymbol{P}_{\boldsymbol{Z}}(\tau_k)\transpose\boldsymbol{\Psi}_{\tau_k}(\widehat{\boldsymbol{\varepsilon}}^{*}_{\tau_k})\Delta\boldsymbol{P}_{\boldsymbol{Z}}(\tau_{k})
        [\boldsymbol{y}-\boldsymbol{X}\widehat{\boldsymbol{\beta}}_{\tau_{k}}](\widehat{\boldsymbol{\beta}}_{\tau_{j}} - \boldsymbol{\beta}_{\tau_{j}})\transpose\boldsymbol{X}^{*}\transpose\boldsymbol{\Psi}_{\tau_j}(\widehat{\boldsymbol{\varepsilon}}^{*}_{\tau_j})\boldsymbol{X}^{*}\\
& + \boldsymbol{X}\transpose\Delta\boldsymbol{P}_{\boldsymbol{Z}}(\tau_k)\transpose\boldsymbol{\Psi}_{\tau_k}(\widehat{\boldsymbol{\varepsilon}}^{*}_{\tau_k})\Delta\boldsymbol{P}_{\boldsymbol{Z}}(\tau_{k})[\boldsymbol{y}-
\boldsymbol{X}\widehat{\boldsymbol{\beta}}_{\tau_{k}}][\boldsymbol{y}-\boldsymbol{X}\widehat{\boldsymbol{\beta}}_{\tau_{j}}]\transpose\Delta\boldsymbol{P}_{\boldsymbol{Z}}(\tau_{j})\transpose
        \boldsymbol{\Psi}_{\tau_j}(\widehat{\boldsymbol{\varepsilon}}^{*}_{\tau_j})\boldsymbol{X}^{*}\\
& + \boldsymbol{X}\transpose\Delta\boldsymbol{P}_{\boldsymbol{Z}}(\tau_k)\transpose\boldsymbol{\Psi}_{\tau_k}(\widehat{\boldsymbol{\varepsilon}}^{*}_{\tau_k})\boldsymbol{\varepsilon}^{*}_{\tau_{k}}\boldsymbol{\varepsilon}^{*}_{\tau_{j}}\transpose
        \boldsymbol{\Psi}_{\tau_j}(\widehat{\boldsymbol{\varepsilon}}^{*}_{\tau_j})\Delta\boldsymbol{P}_{\boldsymbol{Z}}(\tau_j)\boldsymbol{X}\\
& - \boldsymbol{X}\transpose\Delta\boldsymbol{P}_{\boldsymbol{Z}}(\tau_k)\transpose\boldsymbol{\Psi}_{\tau_k}(\widehat{\boldsymbol{\varepsilon}}^{*}_{\tau_k})\boldsymbol{\varepsilon}^{*}_{\tau_{k}}(\widehat{\boldsymbol{\beta}}_{\tau_{j}} -   
        \boldsymbol{\beta}_{\tau_{j}})\transpose\boldsymbol{X}^{*}\transpose\boldsymbol{\Psi}_{\tau_j}(\widehat{\boldsymbol{\varepsilon}}^{*}_{\tau_j})\Delta\boldsymbol{P}_{\boldsymbol{Z}}(\tau_j)\boldsymbol{X}\\
& + \boldsymbol{X}\transpose\Delta\boldsymbol{P}_{\boldsymbol{Z}}(\tau_k)\transpose\boldsymbol{\Psi}_{\tau_k}(\widehat{\boldsymbol{\varepsilon}}^{*}_{\tau_k})\boldsymbol{\varepsilon}^{*}_{\tau_{k}}[\boldsymbol{y}-\boldsymbol{X}
        \widehat{\boldsymbol{\beta}}_{\tau_{j}}]\transpose\Delta\boldsymbol{P}_{\boldsymbol{Z}}(\tau_{j})\transpose\boldsymbol{\Psi}_{\tau_j}(\widehat{\boldsymbol{\varepsilon}}^{*}_{\tau_j})
        \Delta\boldsymbol{P}_{\boldsymbol{Z}}(\tau_j)\boldsymbol{X}\\
& - \boldsymbol{X}\transpose\Delta\boldsymbol{P}_{\boldsymbol{Z}}(\tau_k)\transpose\boldsymbol{\Psi}_{\tau_k}(\widehat{\boldsymbol{\varepsilon}}^{*}_{\tau_k})\boldsymbol{X}^{*} (\widehat{\boldsymbol{\beta}}_{\tau_{k}} -   
        \boldsymbol{\beta}_{\tau_{k}})\boldsymbol{\varepsilon}^{*}_{\tau_{j}}\transpose\boldsymbol{\Psi}_{\tau_j}(\widehat{\boldsymbol{\varepsilon}}^{*}_{\tau_j})\Delta\boldsymbol{P}_{\boldsymbol{Z}}(\tau_j)\boldsymbol{X}\\
& + \boldsymbol{X}\transpose\Delta\boldsymbol{P}_{\boldsymbol{Z}}(\tau_k)\transpose\boldsymbol{\Psi}_{\tau_k}(\widehat{\boldsymbol{\varepsilon}}^{*}_{\tau_k})\boldsymbol{X}^{*} (\widehat{\boldsymbol{\beta}}_{\tau_{k}} -   
    \boldsymbol{\beta}_{\tau_{k}})(\widehat{\boldsymbol{\beta}}_{\tau_{j}} - \boldsymbol{\beta}_{\tau_{j}})\transpose\boldsymbol{X}^{*}\transpose\boldsymbol{\Psi}_{\tau_j}(\widehat{\boldsymbol{\varepsilon}}^{*}_{\tau_j})\Delta\boldsymbol{P}_{\boldsymbol{Z}}(\tau_j)\boldsymbol{X}\\
& - \boldsymbol{X}\transpose\Delta\boldsymbol{P}_{\boldsymbol{Z}}(\tau_k)\transpose\boldsymbol{\Psi}_{\tau_k}(\widehat{\boldsymbol{\varepsilon}}^{*}_{\tau_k})\boldsymbol{X}^{*} (\widehat{\boldsymbol{\beta}}_{\tau_{k}} -
        \boldsymbol{\beta}_{\tau_{k}})[\boldsymbol{y}-\boldsymbol{X}\widehat{\boldsymbol{\beta}}_{\tau_{j}}]\transpose\Delta\boldsymbol{P}_{\boldsymbol{Z}}(\tau_{j})\transpose\boldsymbol{\Psi}_{\tau_j}
        (\widehat{\boldsymbol{\varepsilon}}^{*}_{\tau_j})\Delta\boldsymbol{P}_{\boldsymbol{Z}}(\tau_j)\boldsymbol{X}\\
& + \boldsymbol{X}\transpose\Delta\boldsymbol{P}_{\boldsymbol{Z}}(\tau_k)\transpose\boldsymbol{\Psi}_{\tau_k}(\widehat{\boldsymbol{\varepsilon}}^{*}_{\tau_k})\Delta\boldsymbol{P}_{\boldsymbol{Z}}(\tau_{k})[\boldsymbol{y}-\boldsymbol{X}
        \widehat{\boldsymbol{\beta}}_{\tau_{k}}]\boldsymbol{\varepsilon}^{*}_{\tau_{j}}\transpose\boldsymbol{\Psi}_{\tau_j}(\widehat{\boldsymbol{\varepsilon}}^{*}_{\tau_j})\Delta\boldsymbol{P}_{\boldsymbol{Z}}(\tau_j)\boldsymbol{X}\\
& - \boldsymbol{X}\transpose\Delta\boldsymbol{P}_{\boldsymbol{Z}}(\tau_k)\transpose\boldsymbol{\Psi}_{\tau_k}(\widehat{\boldsymbol{\varepsilon}}^{*}_{\tau_k})\Delta\boldsymbol{P}_{\boldsymbol{Z}}(\tau_{k})
        [\boldsymbol{y}-\boldsymbol{X}\widehat{\boldsymbol{\beta}}_{\tau_{k}}](\widehat{\boldsymbol{\beta}}_{\tau_{j}} - \boldsymbol{\beta}_{\tau_{j}})\transpose\boldsymbol{X}^{*}\transpose\boldsymbol{\Psi}_{\tau_j}(\widehat{\boldsymbol{\varepsilon}}^{*}_{\tau_j})\Delta\boldsymbol{P}_{\boldsymbol{Z}}(\tau_j)\boldsymbol{X}\\
& + \boldsymbol{X}\transpose\Delta\boldsymbol{P}_{\boldsymbol{Z}}(\tau_k)\transpose\boldsymbol{\Psi}_{\tau_k}(\widehat{\boldsymbol{\varepsilon}}^{*}_{\tau_k})\Delta\boldsymbol{P}_{\boldsymbol{Z}}(\tau_{k})[\boldsymbol{y}-\boldsymbol{X}
        \widehat{\boldsymbol{\beta}}_{\tau_{k}}][\boldsymbol{y}-\boldsymbol{X}\widehat{\boldsymbol{\beta}}_{\tau_{j}}]\transpose\Delta\boldsymbol{P}_{\boldsymbol{Z}}(\tau_{j})\transpose
        \boldsymbol{\Psi}_{\tau_j}(\widehat{\boldsymbol{\varepsilon}}^{*}_{\tau_j})\Delta\boldsymbol{P}_{\boldsymbol{Z}}(\tau_j)\boldsymbol{X}.\\
\end{split}
\end{equation*}
Now let's break down the following expression:
\begin{equation*}
\begin{split}
{}& \boldsymbol{X}^{*}\transpose\boldsymbol{\Psi}_{\tau_k}(\widehat{\boldsymbol{\varepsilon}}^{*}_{\tau_k})\boldsymbol{\varepsilon}^{*}_{\tau_{k}}\boldsymbol{\varepsilon}^{*}_{\tau_{j}}\transpose\boldsymbol{\Psi}_{\tau_j}(\widehat{\boldsymbol{\varepsilon}}^{*}_{\tau_j})
        \boldsymbol{X}^{*} = \boldsymbol{X}^{*}\transpose\boldsymbol{\Psi}_{\tau_k}(\boldsymbol{\varepsilon}^{*}_{\tau_{k}})\boldsymbol{\varepsilon}^{*}_{\tau_{k}}\boldsymbol{\varepsilon}^{*}_{\tau_{j}}\transpose\boldsymbol{\Psi}_{\tau_j}
        (\boldsymbol{\varepsilon}^{*}_{\tau_{j}})\boldsymbol{X}^{*}  \\ 
&  + \boldsymbol{X}^{*}\transpose\boldsymbol{\Psi}_{\tau_k}(\boldsymbol{\varepsilon}^{*}_{\tau_{k}})\boldsymbol{\varepsilon}^{*}_{\tau_{k}}\boldsymbol{\varepsilon}^{*}_{\tau_{j}}\transpose\Big[\boldsymbol{\Psi}_{\tau_j}(\widehat{\boldsymbol{\varepsilon}}^{*}_{\tau_j})
        -\boldsymbol{\Psi}_{\tau_j}(\boldsymbol{\varepsilon}^{*}_{\tau_{j}})\Big]\boldsymbol{X}^{*} \\
& + \boldsymbol{X}^{*} \transpose\Big[\boldsymbol{\Psi}_{\tau_k}(\widehat{\boldsymbol{\varepsilon}}^{*}_{\tau_k})
        -\boldsymbol{\Psi}_{\tau_k}(\boldsymbol{\varepsilon}^{*}_{\tau_{k}})\Big]
        \boldsymbol{\varepsilon}^{*}_{\tau_{k}}\boldsymbol{\varepsilon}^{*}_{\tau_{j}}\transpose\boldsymbol{\Psi}_{\tau_j}(\boldsymbol{\varepsilon}^{*}_{\tau_{j}})\boldsymbol{X}^{*}  \\
& + \boldsymbol{X}^{*}\transpose\Big[\boldsymbol{\Psi}_{\tau_k}(\widehat{\boldsymbol{\varepsilon}}^{*}_{\tau_k})-\boldsymbol{\Psi}_{\tau_k}(\boldsymbol{\varepsilon}^{*}_{\tau_{k}})\Big]
\boldsymbol{\varepsilon}^{*}_{\tau_{k}}\boldsymbol{\varepsilon}^{*}_{\tau_{j}}\transpose
\Big[\boldsymbol{\Psi}_{\tau_j}(\widehat{\boldsymbol{\varepsilon}}^{*}_{\tau_j})
        -\boldsymbol{\Psi}_{\tau_j}(\boldsymbol{\varepsilon}^{*}_{\tau_{j}})\Big]\boldsymbol{X}^{*} .
\end{split}
\end{equation*}
The final expression of (\ref{terme_general_d0_erfe}) is:
\begin{flalign*}
&{} \boldsymbol{\widehat{\boldsymbol{X}^{*}}}\transpose\boldsymbol{\Psi}_{\tau_k}(\widehat{\boldsymbol{\varepsilon}}^{*}_{\tau_k})\widehat{\boldsymbol{\varepsilon}}^{*}_{\tau_k}\widehat{\boldsymbol{\varepsilon}}^{*}_{\tau_j}\transpose
        \boldsymbol{\Psi}_{\tau_j}(\widehat{\boldsymbol{\varepsilon}}^{*}_{\tau_j})\boldsymbol{\widehat{\boldsymbol{X}^{*}}}= \\
& + \boldsymbol{X}^{*}\transpose\boldsymbol{\Psi}_{\tau_k}(\boldsymbol{\varepsilon}^{*}_{\tau_{k}})\boldsymbol{\varepsilon}^{*}_{\tau_{k}}\boldsymbol{\varepsilon}^{*}_{\tau_{j}}\transpose\boldsymbol{\Psi}_{\tau_j} 
        (\boldsymbol{\varepsilon}^{*}_{\tau_{j}})\boldsymbol{X}^{*} \tag{\mbox{e1}} \\        
&  + \boldsymbol{X}^{*}\transpose\boldsymbol{\Psi}_{\tau_k}(\boldsymbol{\varepsilon}^{*}_{\tau_{k}})\boldsymbol{\varepsilon}^{*}_{\tau_{k}}\boldsymbol{\varepsilon}^{*}_{\tau_{j}}\transpose\Big[\boldsymbol{\Psi}_{\tau_j}(\widehat{\boldsymbol{\varepsilon}}^{*}_{\tau_j})
        -\boldsymbol{\Psi}_{\tau_j}(\boldsymbol{\varepsilon}^{*}_{\tau_{j}})\Big]\boldsymbol{X}^{*} \tag{\mbox{e2}}\\
& + \boldsymbol{X}^{*} \transpose\Big[\boldsymbol{\Psi}_{\tau_k}(\widehat{\boldsymbol{\varepsilon}}^{*}_{\tau_k})-\boldsymbol{\Psi}_{\tau_k}(\boldsymbol{\varepsilon}^{*}_{\tau_{k}})\Big]
        \boldsymbol{\varepsilon}^{*}_{\tau_{k}}\boldsymbol{\varepsilon}^{*}_{\tau_{j}}\transpose\boldsymbol{\Psi}_{\tau_j}(\boldsymbol{\varepsilon}^{*}_{\tau_{j}})\boldsymbol{X}^{*}  \tag{\mbox{e3}}\\
& + \boldsymbol{X}^{*}\transpose\Big[\boldsymbol{\Psi}_{\tau_k}(\widehat{\boldsymbol{\varepsilon}}^{*}_{\tau_k})-\boldsymbol{\Psi}_{\tau_k}(\boldsymbol{\varepsilon}^{*}_{\tau_{k}})\Big]
        \boldsymbol{\varepsilon}^{*}_{\tau_{k}}\boldsymbol{\varepsilon}^{*}_{\tau_{j}}\transpose\Big[\boldsymbol{\Psi}_{\tau_j}(\widehat{\boldsymbol{\varepsilon}}^{*}_{\tau_j})-\boldsymbol{\Psi}_{\tau_j}(\boldsymbol{\varepsilon}^{*}_{\tau_{j}})\Big]\boldsymbol{X}^{*}  \tag{\mbox{e4}}\\
& - \boldsymbol{X}^{*}\transpose\boldsymbol{\Psi}_{\tau_k}(\widehat{\boldsymbol{\varepsilon}}^{*}_{\tau_k})\boldsymbol{\varepsilon}^{*}_{\tau_{k}}(\widehat{\boldsymbol{\beta}}_{\tau_{j}} - \boldsymbol{\beta}_{\tau_{j}})\transpose        
        \boldsymbol{X}^{*}\transpose\boldsymbol{\Psi}_{\tau_j}(\widehat{\boldsymbol{\varepsilon}}^{*}_{\tau_j})\boldsymbol{X}^{*} \tag{\mbox{e5}}\\
& + \boldsymbol{X}^{*}\transpose\boldsymbol{\Psi}_{\tau_k}(\widehat{\boldsymbol{\varepsilon}}^{*}_{\tau_k})\boldsymbol{\varepsilon}^{*}_{\tau_{k}}[\boldsymbol{y}-\boldsymbol{X}\widehat{\boldsymbol{\beta}}_{\tau_{j}}]\transpose
        \Delta\boldsymbol{P}_{\boldsymbol{Z}}(\tau_{j})\transpose\boldsymbol{\Psi}_{\tau_j}(\widehat{\boldsymbol{\varepsilon}}^{*}_{\tau_j})\boldsymbol{X}^{*} \tag{\mbox{e6}}\\
& - \boldsymbol{X}^{*}\transpose\boldsymbol{\Psi}_{\tau_k}(\widehat{\boldsymbol{\varepsilon}}^{*}_{\tau_k})\boldsymbol{X}^{*} (\widehat{\boldsymbol{\beta}}_{\tau_{k}} - \boldsymbol{\beta}_{\tau_{k}})\boldsymbol{\varepsilon}^{*}_{\tau_{j}}\transpose
    \boldsymbol{\Psi}_{\tau_j}(\widehat{\boldsymbol{\varepsilon}}^{*}_{\tau_j})\boldsymbol{X}^{*} \tag{\mbox{e7}}\\
& + \boldsymbol{X}^{*}\transpose\boldsymbol{\Psi}_{\tau_k}(\widehat{\boldsymbol{\varepsilon}}^{*}_{\tau_k})\boldsymbol{X}^{*} (\widehat{\boldsymbol{\beta}}_{\tau_{k}} - \boldsymbol{\beta}_{\tau_{k}})(\widehat{\boldsymbol{\beta}}_{\tau_{j}} -         \boldsymbol{\beta}_{\tau_{j}})\transpose\boldsymbol{X}^{*}\transpose\boldsymbol{\Psi}_{\tau_j}(\widehat{\boldsymbol{\varepsilon}}^{*}_{\tau_j})\boldsymbol{X}^{*} \tag{\mbox{e8}}\\
& - \boldsymbol{X}^{*}\transpose\boldsymbol{\Psi}_{\tau_k}(\widehat{\boldsymbol{\varepsilon}}^{*}_{\tau_k})\boldsymbol{X}^{*} (\widehat{\boldsymbol{\beta}}_{\tau_{k}} -     
    \boldsymbol{\beta}_{\tau_{k}})[\boldsymbol{y}-\boldsymbol{X}\widehat{\boldsymbol{\beta}}_{\tau_{j}}]\transpose\Delta\boldsymbol{P}_{\boldsymbol{Z}}(\tau_{j})\transpose\boldsymbol{\Psi}_{\tau_j}
    (\widehat{\boldsymbol{\varepsilon}}^{*}_{\tau_j})\boldsymbol{X}^{*} \tag{\mbox{e9}}\\
& + \boldsymbol{X}^{*}\transpose\boldsymbol{\Psi}_{\tau_k}(\widehat{\boldsymbol{\varepsilon}}^{*}_{\tau_k})\Delta\boldsymbol{P}_{\boldsymbol{Z}}(\tau_{k})[\boldsymbol{y}-\boldsymbol{X}\widehat{\boldsymbol{\beta}}_{\tau_{k}}]
    \boldsymbol{\varepsilon}^{*}_{\tau_{j}}\transpose\boldsymbol{\Psi}_{\tau_j}(\widehat{\boldsymbol{\varepsilon}}^{*}_{\tau_j})\boldsymbol{X}^{*} \tag{\mbox{e10}}\\
& - \boldsymbol{X}^{*}\transpose\boldsymbol{\Psi}_{\tau_k}(\widehat{\boldsymbol{\varepsilon}}^{*}_{\tau_k})\Delta\boldsymbol{P}_{\boldsymbol{Z}}(\tau_{k})[\boldsymbol{y}-\boldsymbol{X}\widehat{\boldsymbol{\beta}}_{\tau_{k}}]
    (\widehat{\boldsymbol{\beta}}_{\tau_{j}} - \boldsymbol{\beta}_{\tau_{j}})\transpose\boldsymbol{X}^{*}\transpose\boldsymbol{\Psi}_{\tau_j}(\widehat{\boldsymbol{\varepsilon}}^{*}_{\tau_j})\boldsymbol{X}^{*} \tag{\mbox{e11}}\\
& + \boldsymbol{X}^{*}\transpose\boldsymbol{\Psi}_{\tau_k}(\widehat{\boldsymbol{\varepsilon}}^{*}_{\tau_k})\Delta\boldsymbol{P}_{\boldsymbol{Z}}(\tau_{k})[\boldsymbol{y}-\boldsymbol{X}\widehat{\boldsymbol{\beta}}_{\tau_{k}}]
    [\boldsymbol{y}-\boldsymbol{X}\widehat{\boldsymbol{\beta}}_{\tau_{j}}]\transpose\Delta\boldsymbol{P}_{\boldsymbol{Z}}(\tau_{j})\transpose\boldsymbol{\Psi}_{\tau_j}(\widehat{\boldsymbol{\varepsilon}}^{*}_{\tau_j})\boldsymbol{X}^{*} \tag{\mbox{e12}}\\
& + \boldsymbol{X}^{*}\transpose\boldsymbol{\Psi}_{\tau_k}(\widehat{\boldsymbol{\varepsilon}}^{*}_{\tau_k})\boldsymbol{\varepsilon}^{*}_{\tau_{k}}\boldsymbol{\varepsilon}^{*}_{\tau_{j}}\transpose\boldsymbol{\Psi}_{\tau_j}(\widehat{\boldsymbol{\varepsilon}}^{*}_{\tau_j})
        \Delta\boldsymbol{P}_{\boldsymbol{Z}}(\tau_j)\boldsymbol{X} \tag{\mbox{e13}}\\
& - \boldsymbol{X}^{*}\transpose\boldsymbol{\Psi}_{\tau_k}(\widehat{\boldsymbol{\varepsilon}}^{*}_{\tau_k})\boldsymbol{\varepsilon}^{*}_{\tau_{k}}(\widehat{\boldsymbol{\beta}}_{\tau_{j}} - \boldsymbol{\beta}_{\tau_{j}})\transpose   
       \boldsymbol{X}^{*}\transpose\boldsymbol{\Psi}_{\tau_j}(\widehat{\boldsymbol{\varepsilon}}^{*}_{\tau_j})\Delta\boldsymbol{P}_{\boldsymbol{Z}}(\tau_j)\boldsymbol{X}\tag{\mbox{e14}}\\
& + \boldsymbol{X}^{*}\transpose\boldsymbol{\Psi}_{\tau_k}(\widehat{\boldsymbol{\varepsilon}}^{*}_{\tau_k})\boldsymbol{\varepsilon}^{*}_{\tau_{k}}[\boldsymbol{y}-\boldsymbol{X}\widehat{\boldsymbol{\beta}}_{\tau_{j}}]\transpose
        \Delta\boldsymbol{P}_{\boldsymbol{Z}}(\tau_{j})\transpose\boldsymbol{\Psi}_{\tau_j}(\widehat{\boldsymbol{\varepsilon}}^{*}_{\tau_j})\Delta\boldsymbol{P}_{\boldsymbol{Z}}(\tau_j)\boldsymbol{X}\tag{\mbox{e15}}\\
& - \boldsymbol{X}^{*}\transpose\boldsymbol{\Psi}_{\tau_k}(\widehat{\boldsymbol{\varepsilon}}^{*}_{\tau_k})\boldsymbol{X}^{*} (\widehat{\boldsymbol{\beta}}_{\tau_{k}} - \boldsymbol{\beta}_{\tau_{k}})\boldsymbol{\varepsilon}^{*}_{\tau_{j}}\transpose
    \boldsymbol{\Psi}_{\tau_j}(\widehat{\boldsymbol{\varepsilon}}^{*}_{\tau_j})\Delta\boldsymbol{P}_{\boldsymbol{Z}}(\tau_j)\boldsymbol{X}\tag{\mbox{e16}}\\
& + \boldsymbol{X}^{*}\transpose\boldsymbol{\Psi}_{\tau_k}(\widehat{\boldsymbol{\varepsilon}}^{*}_{\tau_k})\boldsymbol{X}^{*} (\widehat{\boldsymbol{\beta}}_{\tau_{k}} - \boldsymbol{\beta}_{\tau_{k}})(\widehat{\boldsymbol{\beta}}_{\tau_{j}} -         \boldsymbol{\beta}_{\tau_{j}})\transpose\boldsymbol{X}^{*}\transpose\boldsymbol{\Psi}_{\tau_j}(\widehat{\boldsymbol{\varepsilon}}^{*}_{\tau_j})\Delta\boldsymbol{P}_{\boldsymbol{Z}}(\tau_j)\boldsymbol{X}\tag{\mbox{e17}}\\
& - \boldsymbol{X}^{*}\transpose\boldsymbol{\Psi}_{\tau_k}(\widehat{\boldsymbol{\varepsilon}}^{*}_{\tau_k})\boldsymbol{X}^{*} (\widehat{\boldsymbol{\beta}}_{\tau_{k}} -     
    \boldsymbol{\beta}_{\tau_{k}})[\boldsymbol{y}-\boldsymbol{X}\widehat{\boldsymbol{\beta}}_{\tau_{j}}]\transpose\Delta\boldsymbol{P}_{\boldsymbol{Z}}(\tau_{j})\transpose\boldsymbol{\Psi}_{\tau_j}
    (\widehat{\boldsymbol{\varepsilon}}^{*}_{\tau_j})\Delta\boldsymbol{P}_{\boldsymbol{Z}}(\tau_j)\boldsymbol{X} \tag{\mbox{e18}}\\
& + \boldsymbol{X}^{*}\transpose\boldsymbol{\Psi}_{\tau_k}(\widehat{\boldsymbol{\varepsilon}}^{*}_{\tau_k})\Delta\boldsymbol{P}_{\boldsymbol{Z}}(\tau_{k})[\boldsymbol{y}-\boldsymbol{X}\widehat{\boldsymbol{\beta}}_{\tau_{k}}]
    \boldsymbol{\varepsilon}^{*}_{\tau_{j}}\transpose\boldsymbol{\Psi}_{\tau_j}(\widehat{\boldsymbol{\varepsilon}}^{*}_{\tau_j})\Delta\boldsymbol{P}_{\boldsymbol{Z}}(\tau_j)\boldsymbol{X} \tag{\mbox{e19}}\\
& - \boldsymbol{X}^{*}\transpose\boldsymbol{\Psi}_{\tau_k}(\widehat{\boldsymbol{\varepsilon}}^{*}_{\tau_k})\Delta\boldsymbol{P}_{\boldsymbol{Z}}(\tau_{k})[\boldsymbol{y}-\boldsymbol{X}\widehat{\boldsymbol{\beta}}_{\tau_{k}}]
    (\widehat{\boldsymbol{\beta}}_{\tau_{j}} - \boldsymbol{\beta}_{\tau_{j}})\transpose\boldsymbol{X}^{*}\transpose\boldsymbol{\Psi}_{\tau_j}(\widehat{\boldsymbol{\varepsilon}}^{*}_{\tau_j})
    \Delta\boldsymbol{P}_{\boldsymbol{Z}}(\tau_j)\boldsymbol{X} \tag{\mbox{e20}}\\
& + \boldsymbol{X}^{*}\transpose\boldsymbol{\Psi}_{\tau_k}(\widehat{\boldsymbol{\varepsilon}}^{*}_{\tau_k})\Delta\boldsymbol{P}_{\boldsymbol{Z}}(\tau_{k})[\boldsymbol{y}-\boldsymbol{X}\widehat{\boldsymbol{\beta}}_{\tau_{k}}]
    [\boldsymbol{y}-\boldsymbol{X}\widehat{\boldsymbol{\beta}}_{\tau_{j}}]\transpose\Delta\boldsymbol{P}_{\boldsymbol{Z}}(\tau_{j})\transpose\boldsymbol{\Psi}_{\tau_j}(\widehat{\boldsymbol{\varepsilon}}^{*}_{\tau_j})
    \Delta\boldsymbol{P}_{\boldsymbol{Z}}(\tau_j)\boldsymbol{X} \tag{\mbox{e21}}\\
&\hspace{5.12in}\llap{\text{(continued on next page)}}
\end{flalign*}      
\clearpage
\begin{flalign*}
& \hspace{5.12in}\llap{\text{(continued from previous page)}}\\
& + \boldsymbol{X}\transpose\Delta\boldsymbol{P}_{\boldsymbol{Z}}(\tau_k)\transpose\boldsymbol{\Psi}_{\tau_k}(\widehat{\boldsymbol{\varepsilon}}^{*}_{\tau_k})\boldsymbol{\varepsilon}^{*}_{\tau_{k}}
\boldsymbol{\varepsilon}^{*}_{\tau_{j}}\transpose \boldsymbol{\Psi}_{\tau_j}(\widehat{\boldsymbol{\varepsilon}}^{*}_{\tau_j})\boldsymbol{X}^{*} \tag{\mbox{e22}}\\
& - \boldsymbol{X}\transpose\Delta\boldsymbol{P}_{\boldsymbol{Z}}(\tau_k)\transpose\boldsymbol{\Psi}_{\tau_k}(\widehat{\boldsymbol{\varepsilon}}^{*}_{\tau_k})\boldsymbol{\varepsilon}^{*}_{\tau_{k}}
(\widehat{\boldsymbol{\beta}}_{\tau_{j}} - \boldsymbol{\beta}_{\tau_{j}})\transpose\boldsymbol{X}^{*}\transpose\boldsymbol{\Psi}_{\tau_j}(\widehat{\boldsymbol{\varepsilon}}^{*}_{\tau_j})\boldsymbol{X}^{*}\tag{\mbox{e23}}\\
& + \boldsymbol{X}\transpose\Delta\boldsymbol{P}_{\boldsymbol{Z}}(\tau_k)\transpose\boldsymbol{\Psi}_{\tau_k}(\widehat{\boldsymbol{\varepsilon}}^{*}_{\tau_k})\boldsymbol{\varepsilon}^{*}_{\tau_{k}}[\boldsymbol{y}-\boldsymbol{X}
        \widehat{\boldsymbol{\beta}}_{\tau_{j}}]\transpose\Delta\boldsymbol{P}_{\boldsymbol{Z}}(\tau_{j})\transpose\boldsymbol{\Psi}_{\tau_j}(\widehat{\boldsymbol{\varepsilon}}^{*}_{\tau_j})\boldsymbol{X}^{*}\tag{\mbox{e24}}\\
& - \boldsymbol{X}\transpose\Delta\boldsymbol{P}_{\boldsymbol{Z}}(\tau_k)\transpose\boldsymbol{\Psi}_{\tau_k}(\widehat{\boldsymbol{\varepsilon}}^{*}_{\tau_k})\boldsymbol{X}^{*} (\widehat{\boldsymbol{\beta}}_{\tau_{k}} -   
        \boldsymbol{\beta}_{\tau_{k}})\boldsymbol{\varepsilon}^{*}_{\tau_{j}}\transpose\boldsymbol{\Psi}_{\tau_j}(\widehat{\boldsymbol{\varepsilon}}^{*}_{\tau_j})\boldsymbol{X}^{*}\tag{\mbox{e25}}\\
& + \boldsymbol{X}\transpose\Delta\boldsymbol{P}_{\boldsymbol{Z}}(\tau_k)\transpose\boldsymbol{\Psi}_{\tau_k}(\widehat{\boldsymbol{\varepsilon}}^{*}_{\tau_k})\boldsymbol{X}^{*} (\widehat{\boldsymbol{\beta}}_{\tau_{k}} -   
    \boldsymbol{\beta}_{\tau_{k}})(\widehat{\boldsymbol{\beta}}_{\tau_{j}} - \boldsymbol{\beta}_{\tau_{j}})\transpose\boldsymbol{X}^{*}\transpose\boldsymbol{\Psi}_{\tau_j}(\widehat{\boldsymbol{\varepsilon}}^{*}_{\tau_j})\boldsymbol{X}^{*}\tag{\mbox{e26}}\\
& - \boldsymbol{X}\transpose\Delta\boldsymbol{P}_{\boldsymbol{Z}}(\tau_k)\transpose\boldsymbol{\Psi}_{\tau_k}(\widehat{\boldsymbol{\varepsilon}}^{*}_{\tau_k})\boldsymbol{X}^{*} (\widehat{\boldsymbol{\beta}}_{\tau_{k}} -
        \boldsymbol{\beta}_{\tau_{k}})[\boldsymbol{y}-\boldsymbol{X}\widehat{\boldsymbol{\beta}}_{\tau_{j}}]\transpose\Delta\boldsymbol{P}_{\boldsymbol{Z}}(\tau_{j})\transpose\boldsymbol{\Psi}_{\tau_j}
        (\widehat{\boldsymbol{\varepsilon}}^{*}_{\tau_j})\boldsymbol{X}^{*}\tag{\mbox{e27}}\\
& + \boldsymbol{X}\transpose\Delta\boldsymbol{P}_{\boldsymbol{Z}}(\tau_k)\transpose\boldsymbol{\Psi}_{\tau_k}(\widehat{\boldsymbol{\varepsilon}}^{*}_{\tau_k})\Delta\boldsymbol{P}_{\boldsymbol{Z}}(\tau_{k})[\boldsymbol{y}-\boldsymbol{X}
        \widehat{\boldsymbol{\beta}}_{\tau_{k}}]\boldsymbol{\varepsilon}^{*}_{\tau_{j}}\transpose\boldsymbol{\Psi}_{\tau_j}(\widehat{\boldsymbol{\varepsilon}}^{*}_{\tau_j})\boldsymbol{X}^{*}\tag{\mbox{e28}}\\
& - \boldsymbol{X}\transpose\Delta\boldsymbol{P}_{\boldsymbol{Z}}(\tau_k)\transpose\boldsymbol{\Psi}_{\tau_k}(\widehat{\boldsymbol{\varepsilon}}^{*}_{\tau_k})\Delta\boldsymbol{P}_{\boldsymbol{Z}}(\tau_{k})
        [\boldsymbol{y}-\boldsymbol{X}\widehat{\boldsymbol{\beta}}_{\tau_{k}}](\widehat{\boldsymbol{\beta}}_{\tau_{j}} - \boldsymbol{\beta}_{\tau_{j}})\transpose\boldsymbol{X}^{*}\transpose\boldsymbol{\Psi}_{\tau_j}(\widehat{\boldsymbol{\varepsilon}}^{*}_{\tau_j})\boldsymbol{X}^{*}\tag{\mbox{e29}}\\
& + \boldsymbol{X}\transpose\Delta\boldsymbol{P}_{\boldsymbol{Z}}(\tau_k)\transpose\boldsymbol{\Psi}_{\tau_k}(\widehat{\boldsymbol{\varepsilon}}^{*}_{\tau_k})\Delta\boldsymbol{P}_{\boldsymbol{Z}}(\tau_{k})[\boldsymbol{y}-\boldsymbol{X}
        \widehat{\boldsymbol{\beta}}_{\tau_{k}}][\boldsymbol{y}-\boldsymbol{X}\widehat{\boldsymbol{\beta}}_{\tau_{j}}]\transpose\Delta\boldsymbol{P}_{\boldsymbol{Z}}(\tau_{j})\transpose
        \boldsymbol{\Psi}_{\tau_j}(\widehat{\boldsymbol{\varepsilon}}^{*}_{\tau_j})\boldsymbol{X}^{*}\tag{\mbox{e30}}\\
& + \boldsymbol{X}\transpose\Delta\boldsymbol{P}_{\boldsymbol{Z}}(\tau_k)\transpose\boldsymbol{\Psi}_{\tau_k}(\widehat{\boldsymbol{\varepsilon}}^{*}_{\tau_k})\boldsymbol{\varepsilon}^{*}_{\tau_{k}}\boldsymbol{\varepsilon}^{*}_{\tau_{j}}\transpose
        \boldsymbol{\Psi}_{\tau_j}(\widehat{\boldsymbol{\varepsilon}}^{*}_{\tau_j})\Delta\boldsymbol{P}_{\boldsymbol{Z}}(\tau_j)\boldsymbol{X}\tag{\mbox{e31}}\\
& - \boldsymbol{X}\transpose\Delta\boldsymbol{P}_{\boldsymbol{Z}}(\tau_k)\transpose\boldsymbol{\Psi}_{\tau_k}(\widehat{\boldsymbol{\varepsilon}}^{*}_{\tau_k})\boldsymbol{\varepsilon}^{*}_{\tau_{k}}(\widehat{\boldsymbol{\beta}}_{\tau_{j}} -   
        \boldsymbol{\beta}_{\tau_{j}})\transpose\boldsymbol{X}^{*}\transpose\boldsymbol{\Psi}_{\tau_j}(\widehat{\boldsymbol{\varepsilon}}^{*}_{\tau_j})\Delta\boldsymbol{P}_{\boldsymbol{Z}}(\tau_j)\boldsymbol{X}\tag{\mbox{e32}}\\
& + \boldsymbol{X}\transpose\Delta\boldsymbol{P}_{\boldsymbol{Z}}(\tau_k)\transpose\boldsymbol{\Psi}_{\tau_k}(\widehat{\boldsymbol{\varepsilon}}^{*}_{\tau_k})\boldsymbol{\varepsilon}^{*}_{\tau_{k}}[\boldsymbol{y}-\boldsymbol{X}
        \widehat{\boldsymbol{\beta}}_{\tau_{j}}]\transpose\Delta\boldsymbol{P}_{\boldsymbol{Z}}(\tau_{j})\transpose\boldsymbol{\Psi}_{\tau_j}(\widehat{\boldsymbol{\varepsilon}}^{*}_{\tau_j})
        \Delta\boldsymbol{P}_{\boldsymbol{Z}}(\tau_j)\boldsymbol{X}\tag{\mbox{e33}}\\
& - \boldsymbol{X}\transpose\Delta\boldsymbol{P}_{\boldsymbol{Z}}(\tau_k)\transpose\boldsymbol{\Psi}_{\tau_k}(\widehat{\boldsymbol{\varepsilon}}^{*}_{\tau_k})\boldsymbol{X}^{*} (\widehat{\boldsymbol{\beta}}_{\tau_{k}} -   
        \boldsymbol{\beta}_{\tau_{k}})\boldsymbol{\varepsilon}^{*}_{\tau_{j}}\transpose\boldsymbol{\Psi}_{\tau_j}(\widehat{\boldsymbol{\varepsilon}}^{*}_{\tau_j})\Delta\boldsymbol{P}_{\boldsymbol{Z}}(\tau_j)
        \boldsymbol{X}\tag{\mbox{e34}}\\
& \hspace{5.12in}\llap{\text{(continued on next page)}}
\end{flalign*}      
\clearpage
\begin{flalign*}
& \hspace{5.12in}\llap{\text{(continued from previous page)}}\\
& + \boldsymbol{X}\transpose\Delta\boldsymbol{P}_{\boldsymbol{Z}}(\tau_k)\transpose\boldsymbol{\Psi}_{\tau_k}(\widehat{\boldsymbol{\varepsilon}}^{*}_{\tau_k})\boldsymbol{X}^{*} (\widehat{\boldsymbol{\beta}}_{\tau_{k}} -   
    \boldsymbol{\beta}_{\tau_{k}})(\widehat{\boldsymbol{\beta}}_{\tau_{j}} - \boldsymbol{\beta}_{\tau_{j}})\transpose\boldsymbol{X}^{*}\transpose\boldsymbol{\Psi}_{\tau_j}(\widehat{\boldsymbol{\varepsilon}}^{*}_{\tau_j})\Delta\boldsymbol{P}_{\boldsymbol{Z}}(\tau_j)\boldsymbol{X}\tag{\mbox{e35}}\\
& - \boldsymbol{X}\transpose\Delta\boldsymbol{P}_{\boldsymbol{Z}}(\tau_k)\transpose\boldsymbol{\Psi}_{\tau_k}(\widehat{\boldsymbol{\varepsilon}}^{*}_{\tau_k})\boldsymbol{X}^{*} (\widehat{\boldsymbol{\beta}}_{\tau_{k}} -
        \boldsymbol{\beta}_{\tau_{k}})[\boldsymbol{y}-\boldsymbol{X}\widehat{\boldsymbol{\beta}}_{\tau_{j}}]\transpose\Delta\boldsymbol{P}_{\boldsymbol{Z}}(\tau_{j})\transpose\boldsymbol{\Psi}_{\tau_j}
        (\widehat{\boldsymbol{\varepsilon}}^{*}_{\tau_j})\Delta\boldsymbol{P}_{\boldsymbol{Z}}(\tau_j)\boldsymbol{X}\tag{\mbox{e36}}\\
& + \boldsymbol{X}\transpose\Delta\boldsymbol{P}_{\boldsymbol{Z}}(\tau_k)\transpose\boldsymbol{\Psi}_{\tau_k}(\widehat{\boldsymbol{\varepsilon}}^{*}_{\tau_k})\Delta\boldsymbol{P}_{\boldsymbol{Z}}(\tau_{k})[\boldsymbol{y}-\boldsymbol{X}
        \widehat{\boldsymbol{\beta}}_{\tau_{k}}]\boldsymbol{\varepsilon}^{*}_{\tau_{j}}\transpose\boldsymbol{\Psi}_{\tau_j}(\widehat{\boldsymbol{\varepsilon}}^{*}_{\tau_j})\Delta\boldsymbol{P}_{\boldsymbol{Z}}(\tau_j)\boldsymbol{X}\tag{\mbox{e37}}\\
& - \boldsymbol{X}\transpose\Delta\boldsymbol{P}_{\boldsymbol{Z}}(\tau_k)\transpose\boldsymbol{\Psi}_{\tau_k}(\widehat{\boldsymbol{\varepsilon}}^{*}_{\tau_k})\Delta\boldsymbol{P}_{\boldsymbol{Z}}(\tau_{k})
        [\boldsymbol{y}-\boldsymbol{X}\widehat{\boldsymbol{\beta}}_{\tau_{k}}](\widehat{\boldsymbol{\beta}}_{\tau_{j}} - \boldsymbol{\beta}_{\tau_{j}})\transpose\boldsymbol{X}^{*}\transpose\boldsymbol{\Psi}_{\tau_j}(\widehat{\boldsymbol{\varepsilon}}^{*}_{\tau_j})\Delta\boldsymbol{P}_{\boldsymbol{Z}}(\tau_j)\boldsymbol{X}\tag{\mbox{e38}}\\
& + \boldsymbol{X}\transpose\Delta\boldsymbol{P}_{\boldsymbol{Z}}(\tau_k)\transpose\boldsymbol{\Psi}_{\tau_k}(\widehat{\boldsymbol{\varepsilon}}^{*}_{\tau_k})\Delta\boldsymbol{P}_{\boldsymbol{Z}}(\tau_{k})
[\boldsymbol{y}-\boldsymbol{X}
        \widehat{\boldsymbol{\beta}}_{\tau_{k}}][\boldsymbol{y}-\boldsymbol{X}\widehat{\boldsymbol{\beta}}_{\tau_{j}}]\transpose\Delta\boldsymbol{P}_{\boldsymbol{Z}}(\tau_{j})\transpose
        \boldsymbol{\Psi}_{\tau_j}(\widehat{\boldsymbol{\varepsilon}}^{*}_{\tau_j})\Delta\boldsymbol{P}_{\boldsymbol{Z}}(\tau_j)\boldsymbol{X}.\tag{\mbox{e39}}\\
\end{flalign*} 
The demonstration is based on showing the convergence of each of these terms (\text{e2})-(\text{e39}). We have three types of expression: those that are function of $\Delta\boldsymbol{P}_{\boldsymbol{Z}},$ those that are function of $\Big[\boldsymbol{\Psi}_{\tau_k}(\widehat{\boldsymbol{\varepsilon}}^{*}_{\tau_k})-\boldsymbol{\Psi}_{\tau_k}(\boldsymbol{\varepsilon}^{*}_{\tau_{k}})\Big]$ and those that are function of $(\widehat{\boldsymbol{\beta}}_{\tau_{j}} - \boldsymbol{\beta}_{\tau_{j}}).$ Those that are a function of $\Delta\boldsymbol{P}_{\boldsymbol{Z}}$ are shown by following the approach of (\ref{conv_poids_erfe} )  and those that are function of $\Big[\boldsymbol{\Psi}_{\tau_k}(\widehat{\boldsymbol{\varepsilon}}^{*}_{\tau_k})-\boldsymbol{\Psi}_{\tau_k}(\boldsymbol{\varepsilon}^{*}_{\tau_{k}})\Big]$ according to the approach (\ref{conv_diff_phi_erfe}).
The convergence technique of those that are a function of $(\widehat{\boldsymbol{\beta}}_{\tau_{j}} - \boldsymbol{\beta}_{\tau_{j}})$ is identical to the convergence procedure used for (\text{e7}) and (\text{e8}). Thus, in the following, we will show the convergence of (\text{e7}) and (\text{e8}).
~~\\
~~\\
Using the relation, $\Vect(ABC)=(C\transpose \otimes A)\Vect(B),$ equation (\text{e7}) is transformed as follows:
\begin{multline*}
    \Vect\Big(\frac{1}{nm}\sum_{i=1}^{n}\boldsymbol{X}^{*}_i\transpose \boldsymbol{\Psi}_{\tau_k}(\widehat{\boldsymbol{\varepsilon}}^{*}_{i\tau_k})\boldsymbol{X}^{*}_i (\widehat{\boldsymbol{\beta}}_{\tau_k}-\boldsymbol{\beta}_{\tau_k})\boldsymbol{\varepsilon}^{*}_{i\tau_j}\transpose\boldsymbol{\Psi}_{\tau_j}(
    \widehat{\boldsymbol{\varepsilon}}^{*}_{i\tau_j})
    \boldsymbol{X}^{*}_i\Big) \\ 
=\frac{1}{nm}\sum_{i=1}^{n} \boldsymbol{X}^{*}_i\transpose\boldsymbol{\Psi}_{\tau_k}(\widehat{\boldsymbol{\varepsilon}}^{*}_{i\tau_k})
\boldsymbol{\varepsilon}^{*}_{i\tau_j} \otimes\boldsymbol{X}^{*}_i\transpose \boldsymbol{\Psi}_{\tau_j}(\widehat{\boldsymbol{\varepsilon}}_{i\tau_j})\boldsymbol{X}^{*}_i 
\Vect(\widehat{\boldsymbol{\beta}}_{\tau_k}-\boldsymbol{\beta}_{\tau_k}).
\end{multline*}
Applying \textbf{Lemma \ref{hans1}}, we have
\begin{multline*}
\E\norm{\boldsymbol{X}^{*}_i\transpose\boldsymbol{\Psi}_{\tau_k}(\widehat{\boldsymbol{\varepsilon}}^{*}_{i\tau_k})
\boldsymbol{\varepsilon}^{*}_{i\tau_j} \otimes\boldsymbol{X}^{*}_i\transpose \boldsymbol{\Psi}_{\tau_j}(\widehat{\boldsymbol{\varepsilon}}_{i\tau_j})\boldsymbol{X}^{*}_i }^{1+\nu} \\ 
\leq \Big(\E\norm{\boldsymbol{X}^{*}_i\transpose\boldsymbol{\Psi}_{\tau_k}(\widehat{\boldsymbol{\varepsilon}}^{*}_{i\tau_k})
\boldsymbol{\varepsilon}^{*}_{i\tau_j} }^{2+2\nu} \E\norm{\boldsymbol{X}^{*}_i\transpose \boldsymbol{\Psi}_{\tau_j}(\widehat{\boldsymbol{\varepsilon}}_{i\tau_j})\boldsymbol{X}^{*}_i }^{2+2\nu} \Big)^{1/2}.
\end{multline*}
Repeated application of Minkowski and Holder inequalities shows that 
\begin{equation*}
\begin{split}
\E\norm{ \boldsymbol{X}^{*}_i\transpose\boldsymbol{\Psi}_{\tau_k}(\widehat{\boldsymbol{\varepsilon}}^{*}_{i\tau_k}) \boldsymbol{\varepsilon}^{*}_{i\tau_j} }^{2+2\nu} 
&=\E\bigg[\sum_{k=1}^{p}\Big\lvert\sum_{j=1}^{m}x_{ij}^{*k}\psi_{\tau_k}(\widehat{\varepsilon}^{*}_{ij\tau_k})\varepsilon^{*}_{ij\tau_j}\Big\rvert^{2}\bigg]^{1+\nu}\\
& \leq \bigg[\sum_{k=1}^{p}\bigg( \E\Big\lvert\sum_{j=1}^{m}x_{ij}^{*k}\psi_{\tau_k}(\widehat{\varepsilon}^{*}_{ij\tau_k})\varepsilon^{*}_{ij\tau_j}
\Big\rvert^{2+2\nu}\bigg)^{\frac{1}{1+\nu}}
\bigg]^{1+\nu}\\
& \leq \bigg\lbrace  \sum_{k=1}^{p}\bigg[\sum_{j=1}^{m} \Big(\E\lvert x_{ij}^{*k}\psi_{\tau_k}(\widehat{\varepsilon}^{*}_{ij\tau_k})\varepsilon^{*}_{ij\tau_j} \rvert^{2+2\nu}\Big)^{\frac{1}{2+2\nu}} \bigg]^{2+2\nu} \bigg\rbrace^{1+\nu} \\
& \leq (p\Delta)^{1+\nu}(Mm)^{2+\nu}.
\end{split}
\end{equation*}
The last inequality is obtained by applying the assumptions $\E\lvert\psi_{\tau}(\widehat{\varepsilon}_{ij\tau}) \rvert^{4+\nu} < \Delta \mbox{ and } \E\lvert \varepsilon_{ij\tau}\rvert^{4 + \nu} < \Delta.$ Similarly
\begin{equation*}
\begin{split}
\E\norm{\boldsymbol{X}^{*}_i\transpose \boldsymbol{\Psi}_{\tau_j}(\widehat{\boldsymbol{\varepsilon}}_{i\tau_j})\boldsymbol{X}^{*}_i }^{2+2\nu}  
& \leq \E\norm{\boldsymbol{X}^{*}_i\transpose \boldsymbol{\Psi}_{\tau_j}^{1/2}(\widehat{\boldsymbol{\varepsilon}}_{i\tau_j}) }^{4+4\nu} \\
& \leq \bigg[\sum_{k=1}^{p}\sum_{j=1}^{m}\Big(\E\lvert(x_{ij}^{*k}\psi_{\tau_j}^{1/2}(\widehat{\varepsilon}^{*}_{ij\tau_j}))^2\rvert^{2+2\nu}\Big)^{\frac{1}{2+2\nu}}\bigg]^{2+2\nu}\\
& \leq (pm)^{2+\nu}\Delta.
\end{split}
\end{equation*} 
Then, by  the Markov LLN it follows that
\begin{equation*}
    \Vect\Big(\frac{1}{nm}\sum_{i=1}^{n}\boldsymbol{X}^{*}_i\transpose \boldsymbol{\Psi}_{\tau_k}(\widehat{\boldsymbol{\varepsilon}}^{*}_{i\tau_k})\boldsymbol{X}^{*}_i (\widehat{\boldsymbol{\beta}}_{\tau_k}-\boldsymbol{\beta}_{\tau_k})\boldsymbol{\varepsilon}^{*}_{i\tau_j}\transpose\boldsymbol{\Psi}_{\tau_j}(\widehat{\boldsymbol{\varepsilon}}^{*}_{i\tau_j})
    \boldsymbol{X}^{*}_i\Big) = \frac{1}{\sqrt{nm}}\bigO_p(1)\bigO_p(1).
\end{equation*}
Considering that $\Vect(\widehat{\boldsymbol{\beta}}_{\tau}-\boldsymbol{\beta}_{\tau})=\bigO_p((nm)^{-1/2})$ and that ~~\\
$\E\norm{\boldsymbol{X}^{*}_i\transpose\boldsymbol{\Psi}_{\tau_k}(\widehat{\boldsymbol{\varepsilon}}^{*}_{i\tau_k})\boldsymbol{X}^{*}_i }^{2+2\nu} < (pm)^{2+\nu}\Delta,$ term (\text{e8}) is
\begin{equation*}
\begin{split}
{}&\frac{1}{nm}\Vect\Big(\sum_{i=1}^{n} \boldsymbol{X}^{*}_i\transpose\boldsymbol{\Psi}_{\tau_k}(\widehat{\boldsymbol{\varepsilon}}^{*}_{i\tau_k})\boldsymbol{X}^{*}_i (\widehat{\boldsymbol{\beta}}_{\tau_{k}} - \boldsymbol{\beta}_{\tau_{k}})(\widehat{\boldsymbol{\beta}}_{\tau_{j}} - \boldsymbol{\beta}_{\tau_{j}})\transpose\boldsymbol{X}^{*}_i\transpose\boldsymbol{\Psi}_{\tau_j}(\widehat{\boldsymbol{\varepsilon}}^{*}_{i\tau_j})\boldsymbol{X}^{*}_i\Big)\\
&= \frac{1}{nm}\sum_{i=1}^{n}\boldsymbol{X}^{*}_i\transpose\boldsymbol{\Psi}_{\tau_j}(\widehat{\boldsymbol{\varepsilon}}^{*}_{i\tau_j})\boldsymbol{X}^{*}_i \otimes\boldsymbol{X}^{*}_i\transpose\boldsymbol{\Psi}_{\tau_k}(\widehat{\boldsymbol{\varepsilon}}^{*}_{i\tau_k})\boldsymbol{X}^{*}_i 
\Vect[(\widehat{\boldsymbol{\beta}}_{\tau}-\boldsymbol{\beta}_{\tau})(\widehat{\boldsymbol{\beta}}_{\tau_{k}} - \boldsymbol{\beta}_{\tau_{k}})\transpose]\\
&= \frac{1}{nm}\bigO_p(1)\bigO_p(1).
\end{split}
\end{equation*}
We have just shown that 
\begin{equation*}
    \frac{1}{nm}\sum_{i=1}^{n}\widehat{\boldsymbol{X}^{*}_i}\transpose\widehat{\boldsymbol{\Sigma}}^{*}_{i\tau_{k}\tau_{j}}\widehat{\boldsymbol{X}^{*}_i} -  \frac{1}{nm}\sum_{i=1}^{n}
    \boldsymbol{X}^{*}_i\transpose\boldsymbol{\Psi}_{\tau_k}(\boldsymbol{\varepsilon}^{*}_{i\tau_{k}})\boldsymbol{\varepsilon}^{*}_{i\tau_{k}}
    \boldsymbol{\varepsilon}^{*}_{i\tau_{j}}\transpose\boldsymbol{\Psi}_{\tau_j} 
        (\boldsymbol{\varepsilon}^{*}_{i\tau_{j}})\boldsymbol{X}^{*}_i
\xrightarrow{p} \boldsymbol{0}.
\end{equation*}
Application of the Markov LLN also yields
\begin{equation*}
    \frac{1}{nm}\sum_{i=1}^{n}\boldsymbol{X}^{*}_i\transpose\boldsymbol{\Psi}_{\tau_k}(\boldsymbol{\varepsilon}^{*}_{i\tau_{k}})\boldsymbol{\varepsilon}^{*}_{i\tau_{k}}\boldsymbol{\varepsilon}^{*}_{i\tau_{j}}\transpose\boldsymbol{\Psi}_{\tau_j} 
        (\boldsymbol{\varepsilon}^{*}_{i\tau_{j}})\boldsymbol{X}^{*}_i -  
    \frac{1}{nm}\sum_{i=1}^{n}\boldsymbol{X}^{*}_i\transpose \boldsymbol{\Sigma}_{i\tau_{k}\tau_{j}}^{*} \boldsymbol{X}^{*}_i \xrightarrow{p} \boldsymbol{0}.
\end{equation*}
Thus, application of the triangular inequality shows that $\widehat{\boldsymbol{D}}_{0}(\boldsymbol{\tau})\xrightarrow{p}\boldsymbol{D}_{0}(\boldsymbol{\tau}).$
\end{proof}
\begin{proof}[\textbf{Proof of Corollary \ref{cor1_erfe}}]$ $
~~\\
Proof of \textbf{Corollary} \ref{cor1_erfe} follows immediately from the proof of \textbf{Theorem} \ref{theo3_erfe}.
\end{proof}


\section{Additional simulation results}

\begin{center}
\begin{figure}[H]
\includegraphics[width=0.8\linewidth]{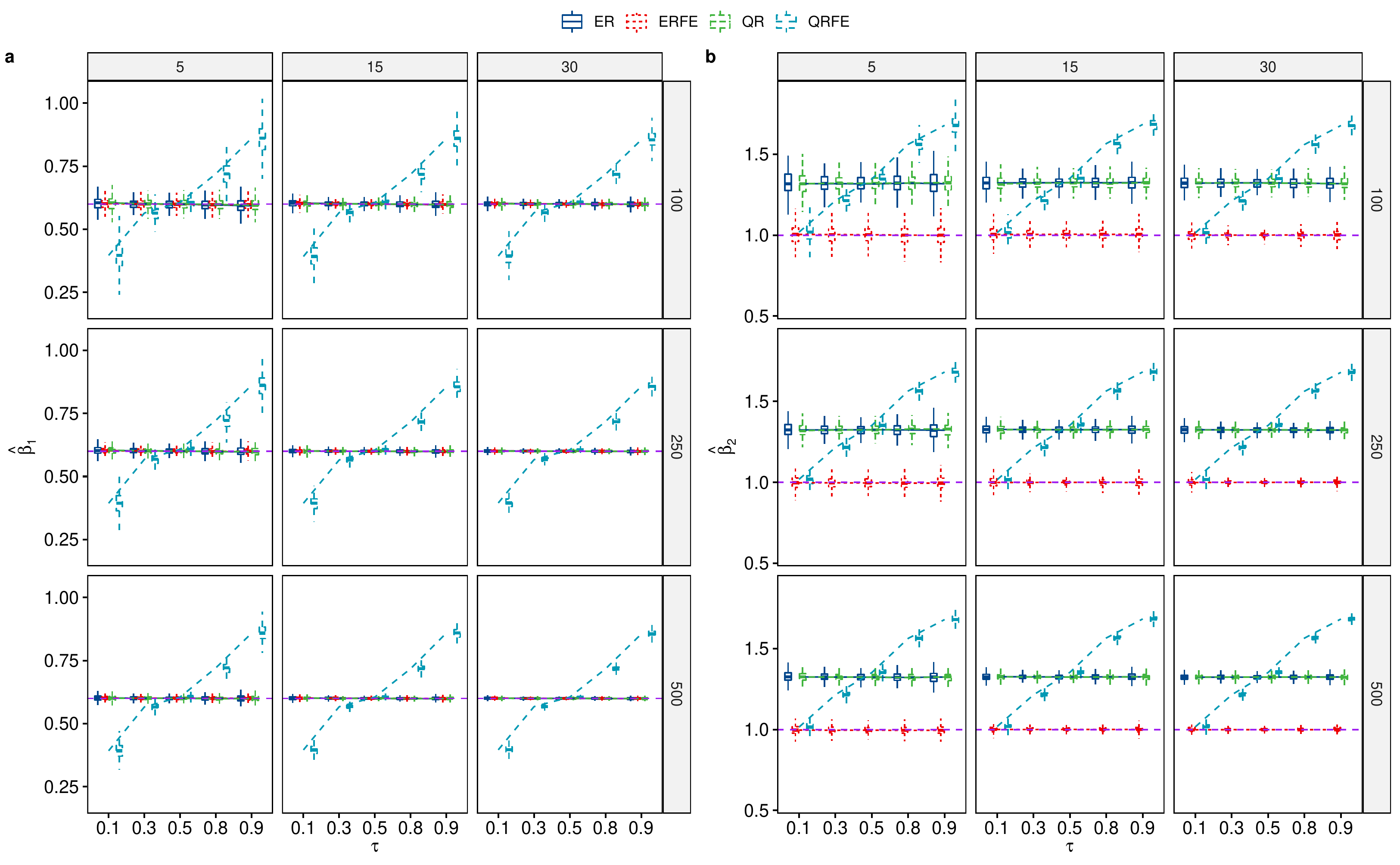}
 \caption{Distribution of the coefficient estimates of the parameter $\beta_1$ (Figure \ref{fig:gstud1_erfe}\textbf{a}) and the parameter $\beta_2$ (Figure \ref{fig:gstud1_erfe}\textbf{b}) represented as boxplot according to the sample size $n\in(100,  \ 250,  \ 500),$ the repeated measurements $m=(5,\ 15,\ 30),$ the asymmetric points $\tau\in (0.1,  \ 0.3,  \  0.5, \  0.8,\ 0.9)$ and the error term $\varepsilon\sim\mathcal{T}(3)$ in the location-shift scenario.}\label{fig:gstud1_erfe}
\end{figure}
\end{center}

\begin{center}
\begin{figure}[H]
\includegraphics[width=0.8\linewidth]{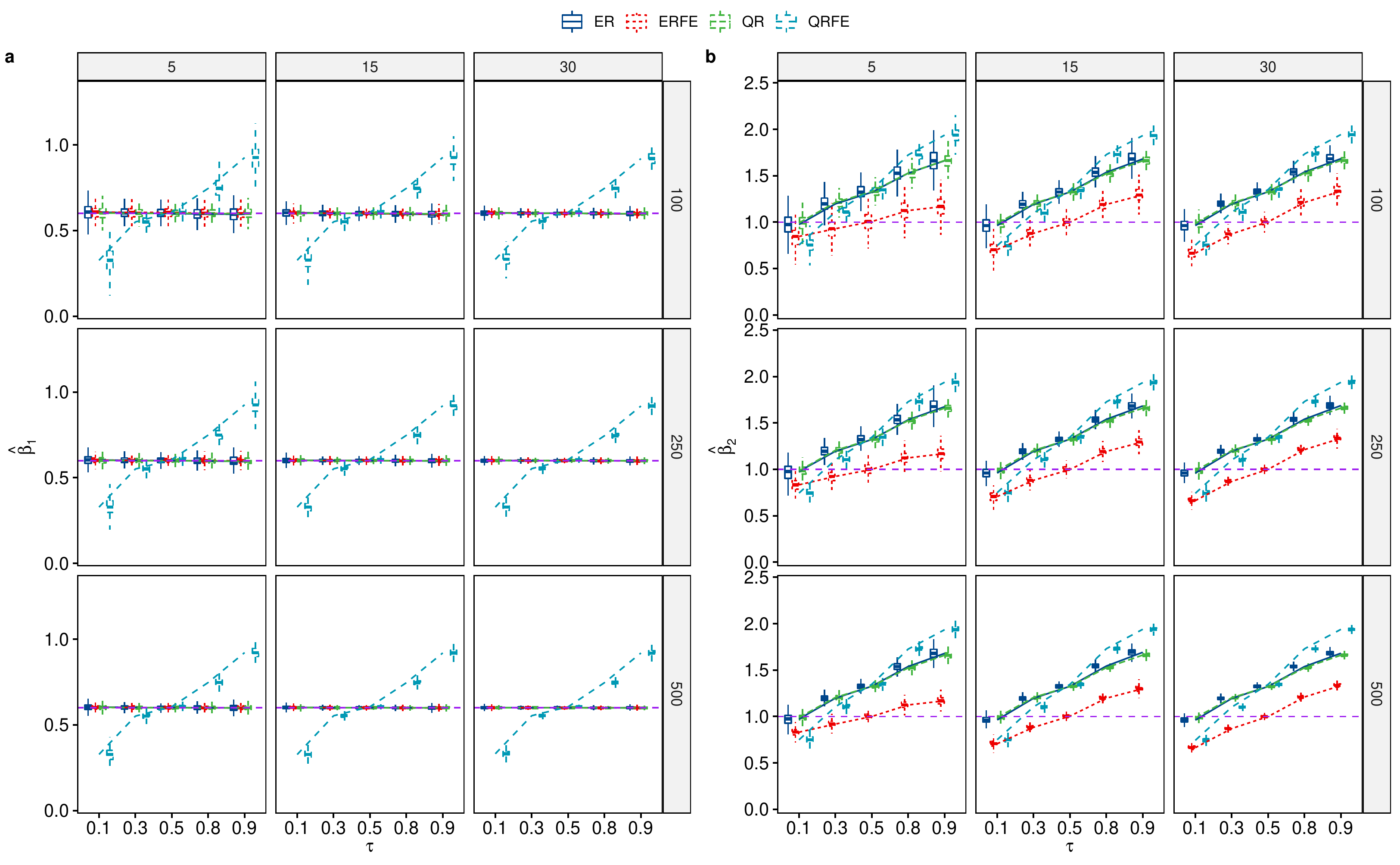}
 \caption{Distribution of the coefficient estimates of the parameter $\beta_1$ (Figure \ref{fig:gstud2_erfe}\textbf{a}) and the parameter $\beta_2$ (Figure \ref{fig:gstud2_erfe}\textbf{b}) represented as boxplot according to the sample size $n\in(100,  \ 250,  \ 500),$ the repeated measurements $m=(5,\ 15,\ 30),$ the asymmetric points $\tau\in (0.1,  \ 0.3,  \  0.5, \  0.8,\ 0.9)$ and the error term $\varepsilon\sim\mathcal{T}(3)$ in the location-scale-shift scenario.}\label{fig:gstud2_erfe}
\end{figure}
\end{center} 

\begin{center}
\begin{figure}[H]
\includegraphics[width=0.8\linewidth]{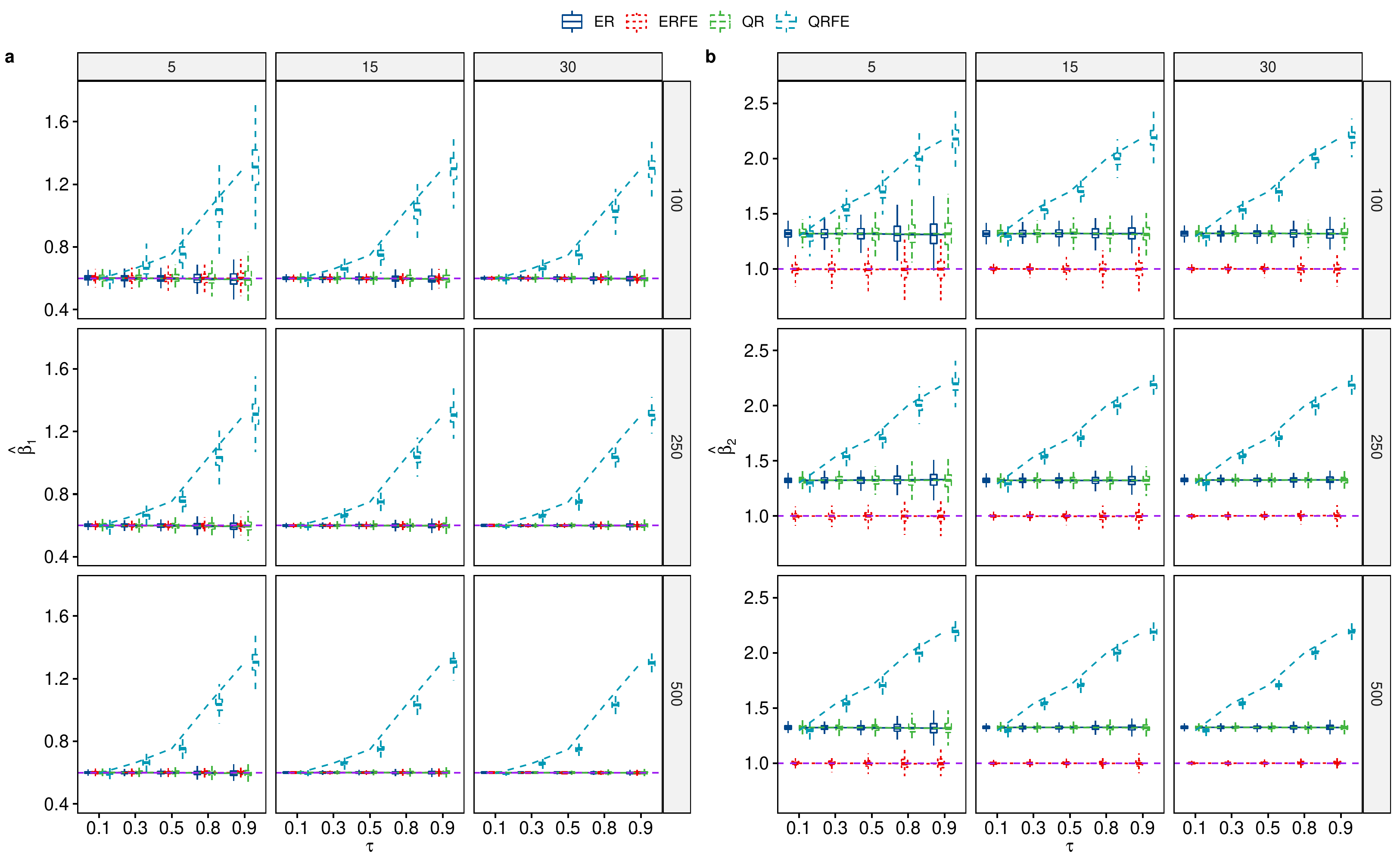}
 \caption{Distribution of the coefficient estimates of the parameter $\beta_1$ (Figure \ref{fig:gchi1_erfe}\textbf{a}) and the parameter $\beta_2$ (Figure \ref{fig:gchi1_erfe}\textbf{b}) represented as boxplot according to the sample size $n\in(100,  \ 250,  \ 500),$ the repeated measurements $m=(5,\ 15,\ 30),$ the asymmetric points $\tau\in (0.1,  \ 0.3,  \  0.5, \  0.8,\ 0.9)$ and the error term $\varepsilon\sim\chi_2(3)$ in the location-shift scenario.}\label{fig:gchi1_erfe}
\end{figure}
\end{center}

\begin{center}
\begin{figure}[H]
\includegraphics[width=0.8\linewidth]{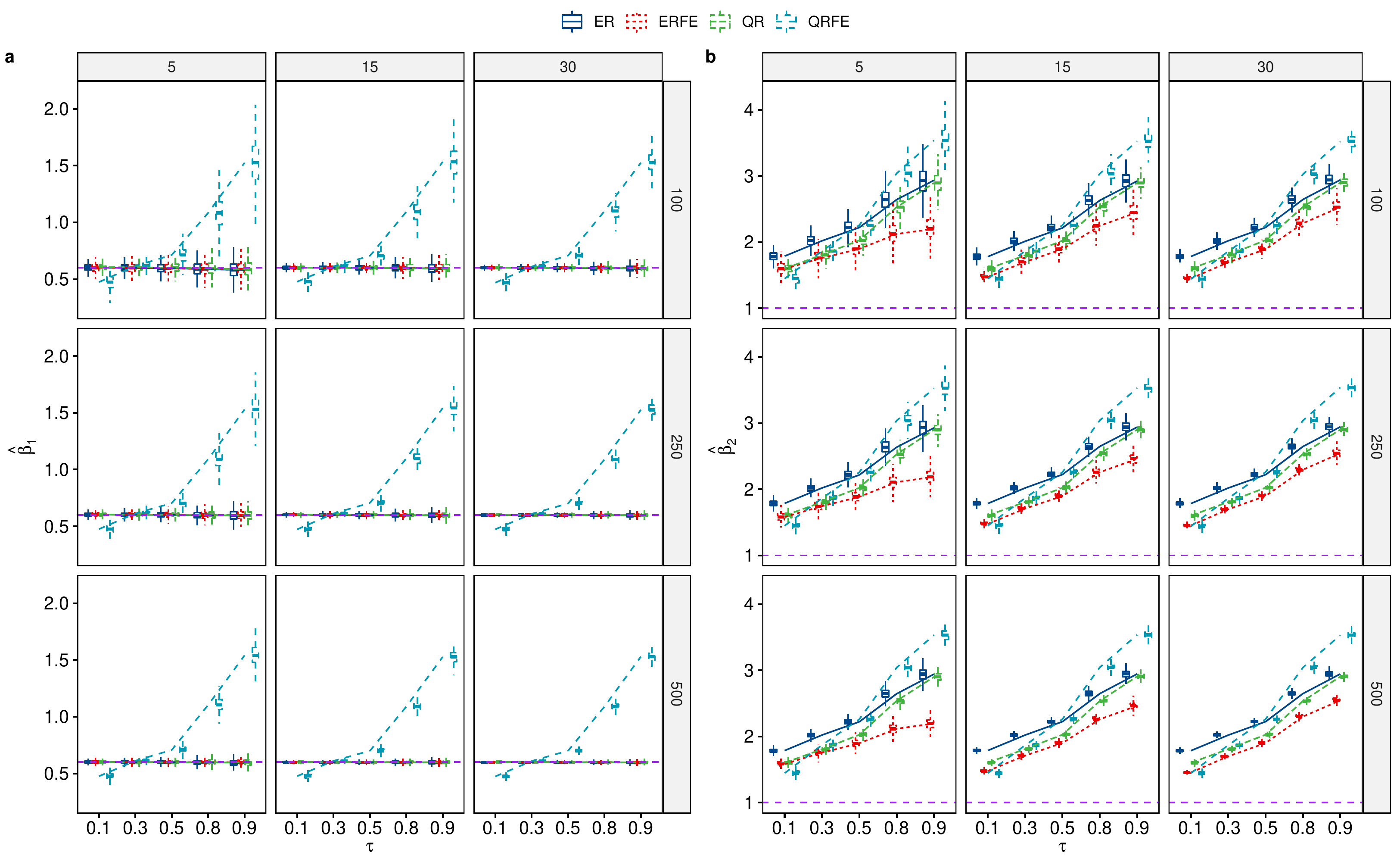}
 \caption{Distribution of the coefficient estimates of the parameter $\beta_1$ (Figure \ref{fig:gchi2_erfe}\textbf{a}) and the parameter $\beta_2$ (Figure \ref{fig:gchi2_erfe}\textbf{b}) represented as boxplot according to the sample size $n\in(100,  \ 250,  \ 500),$ the repeated measurements $m=(5,\ 15,\ 30),$ the asymmetric points $\tau\in (0.1,  \ 0.3,  \  0.5, \  0.8,\ 0.9)$ and the error term $\varepsilon\sim\chi_2(3)$ in the location-scale-shift scenario.}\label{fig:gchi2_erfe}
\end{figure}
\end{center} 

\begin{center}
\begin{figure}[hbt!]
\includegraphics[width=0.75\linewidth]{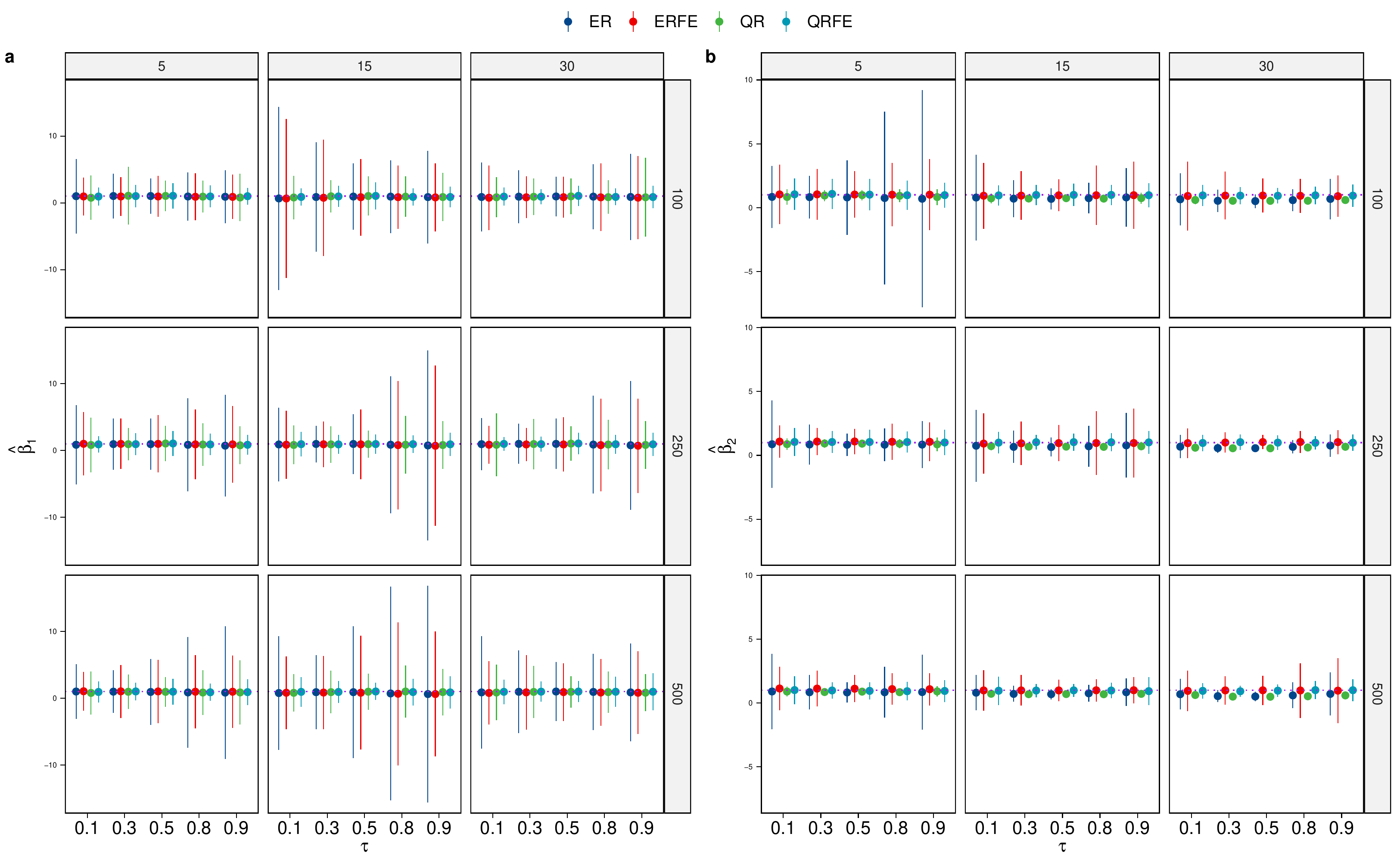}
    \caption{Distribution of the ratio $\frac{\SE}{\SD}$ for the parameter estimate $\widehat{\beta}_1$ (Figure \ref{fig:SdSe_stud_homo}\textbf{a}) and the parameter estimate $\widehat{\beta}_2$ (Figure \ref{fig:SdSe_stud_homo}\textbf{b}) represented as an error plot with respect to the sample size $n\in(100,  \ 250,  \ 500),$ the repeated measurements $m=(5,\ 15,\ 30),$ the asymmetric points $\tau\in (0.1,  \ 0.3,  \  0.5, \  0.8,\ 0.9)$ and the error term $\varepsilon\sim\mathcal{T}(3)$ in the location-shift scenario.} \label{fig:SdSe_stud_homo}
\end{figure}
\end{center}

\begin{center}
\begin{figure}[hbt!]
\includegraphics[width=0.75\linewidth]{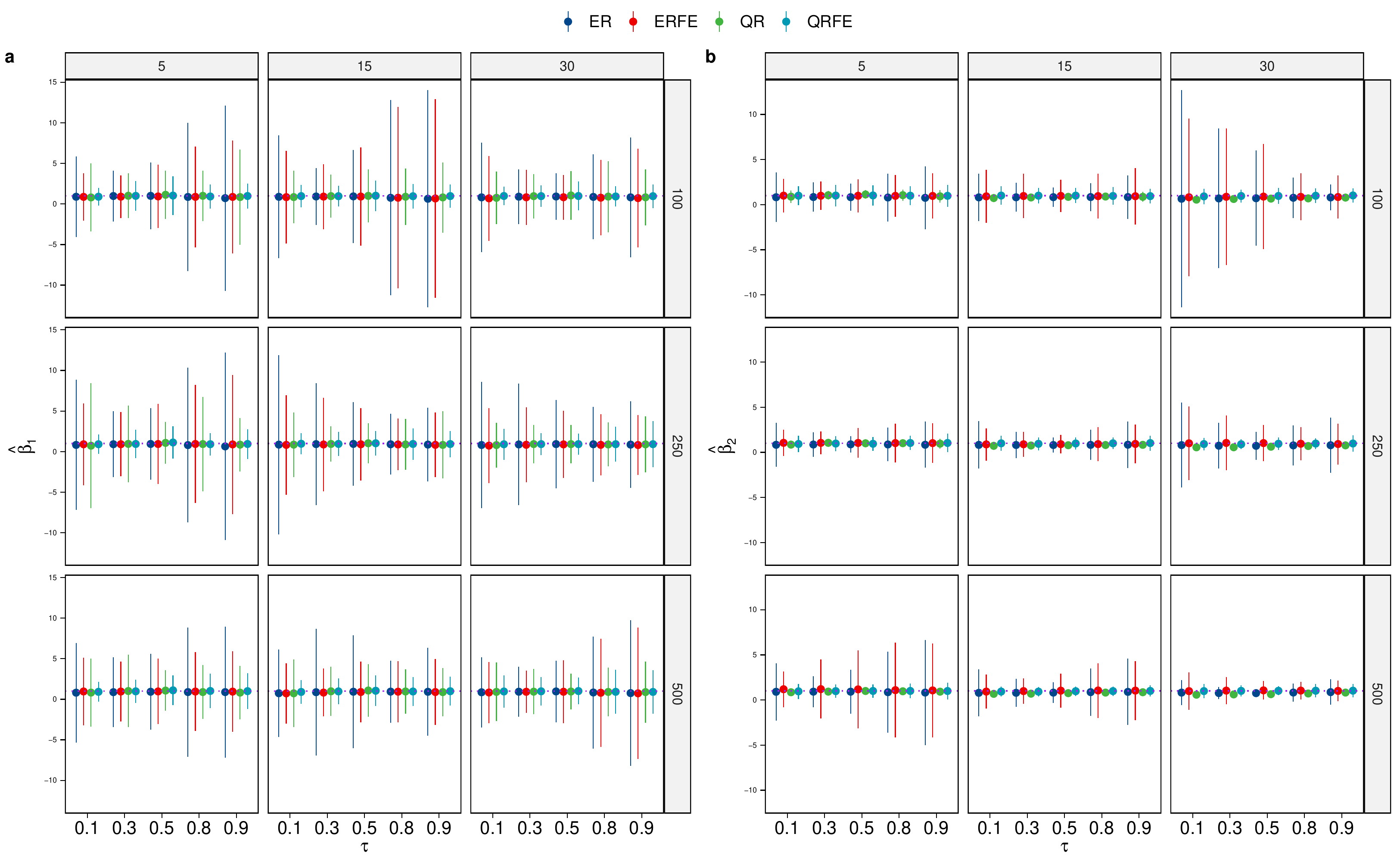}
    \caption{Distribution of the ratio $\frac{\SE}{\SD}$ for the parameter estimate $\widehat{\beta}_1$ (Figure \ref{fig:SdSe_stud_hetero}\textbf{a}) and the parameter estimate $\widehat{\beta}_2$ (Figure \ref{fig:SdSe_stud_hetero}\textbf{b}) represented as an error plot with respect to the sample size $n\in(100,  \ 250,  \ 500),$ the repeated measurements $m=(5,\ 15,\ 30),$ the asymmetric points $\tau\in (0.1,  \ 0.3,  \  0.5, \  0.8,\ 0.9)$ and the error term $\varepsilon\sim\mathcal{T}(3)$ in the location-scale-shift scenario.} \label{fig:SdSe_stud_hetero}
\end{figure}
\end{center}

\begin{center}
\begin{figure}[hbt!]
\includegraphics[width=0.75\linewidth]{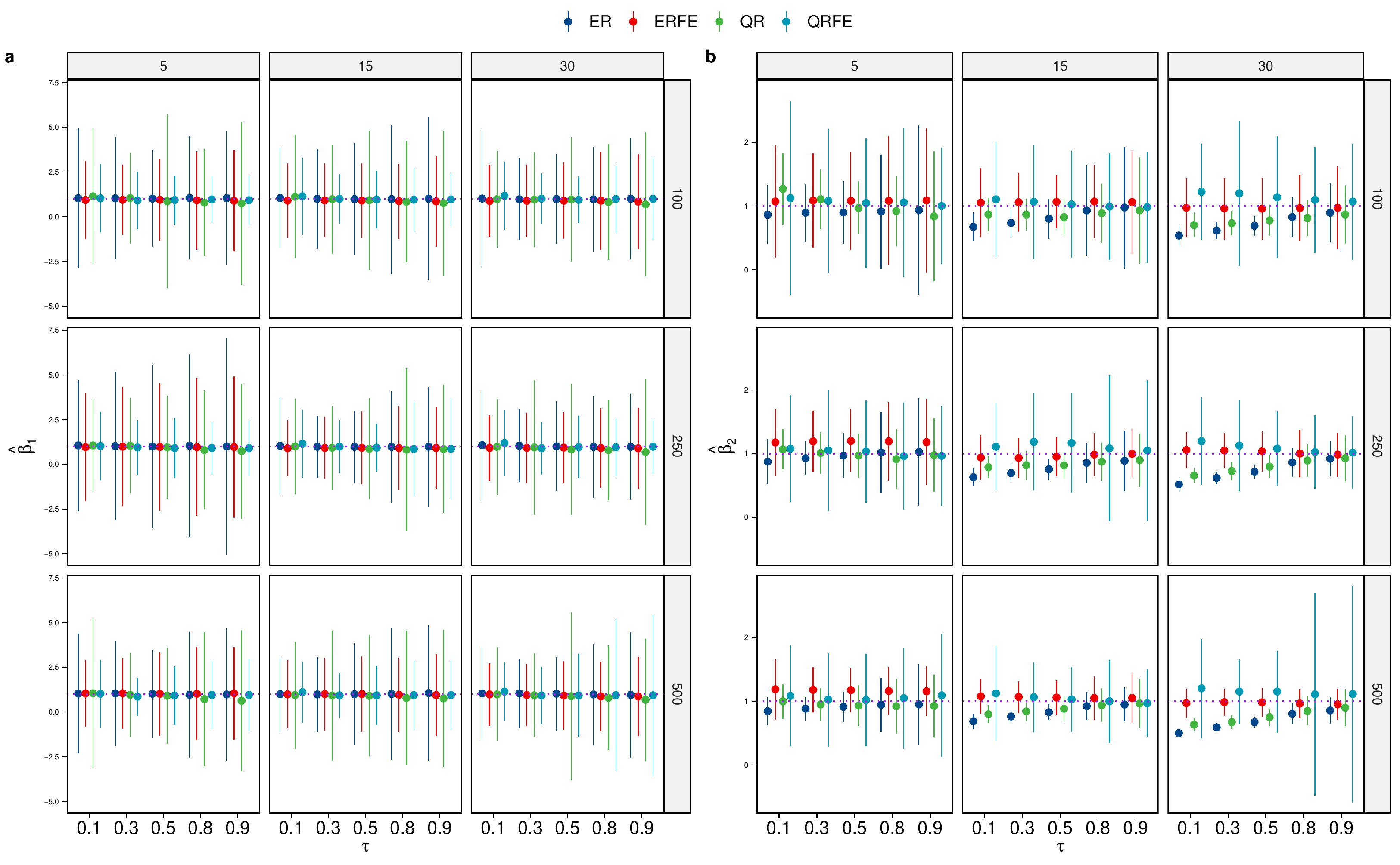}
    \caption{Distribution of the ratio $\frac{\SE}{\SD}$ for the parameter estimate $\widehat{\beta}_1$ (Figure \ref{fig:SdSe_chi2_homo}\textbf{a}) and the parameter estimate $\widehat{\beta}_2$ (Figure \ref{fig:SdSe_chi2_homo}\textbf{b}) represented as an error plot with respect to the sample size $n\in(100,  \ 250,  \ 500),$ the repeated measurements $m=(5,\ 15,\ 30),$ the asymmetric points $\tau\in (0.1,  \ 0.3,  \  0.5, \  0.8,\ 0.9)$ and the error term $\varepsilon\sim\chi_2(3)$ in the location-shift scenario.} \label{fig:SdSe_chi2_homo}
\end{figure}
\end{center}

\begin{center}
\begin{figure}[hbt!]
\includegraphics[width=0.75\linewidth]{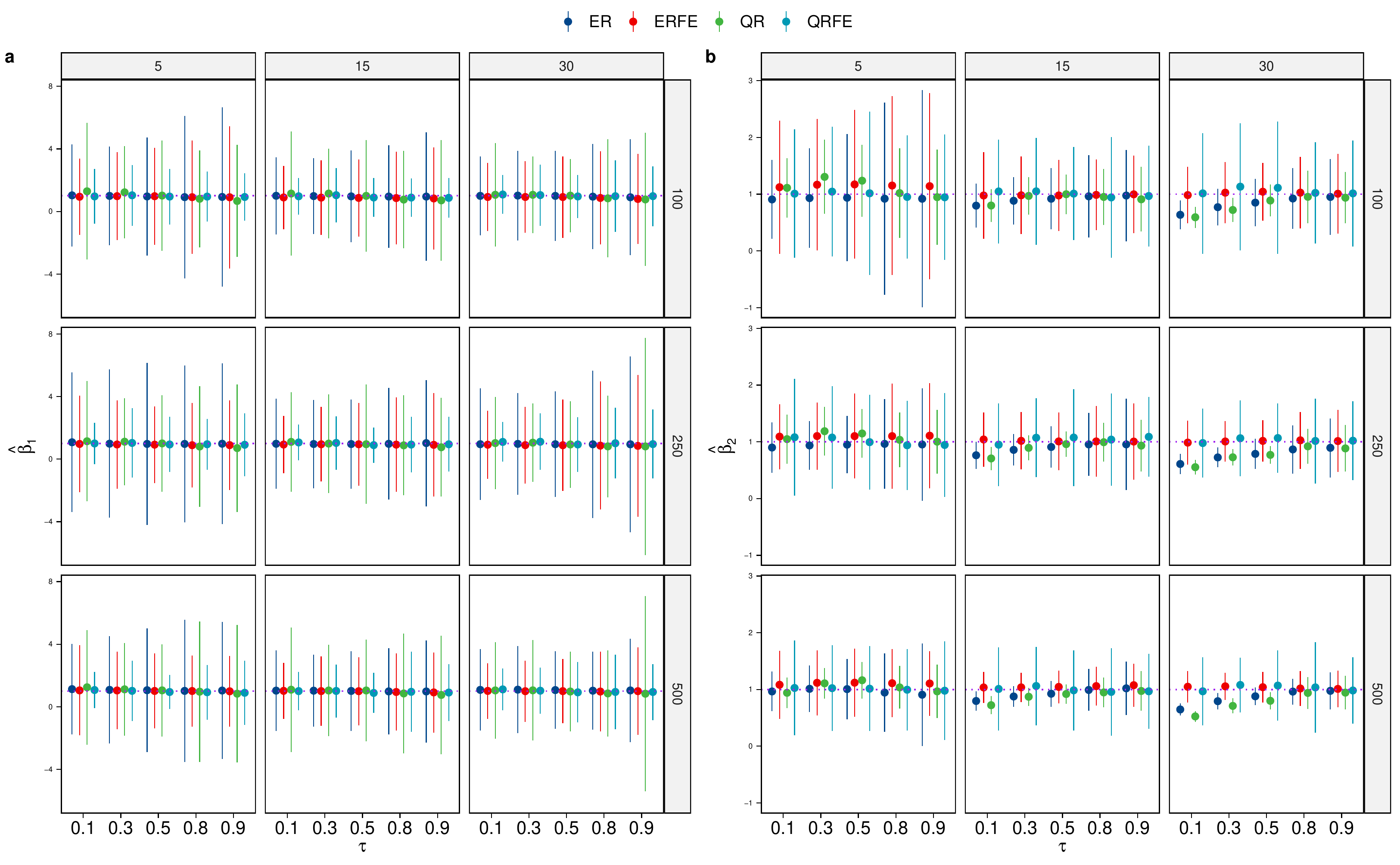}
    \caption{Distribution of the ratio $\frac{\SE}{\SD}$ for the parameter estimate $\widehat{\beta}_1$ (Figure \ref{fig:SdSe_chi2_hetero}\textbf{a}) and the parameter estimate $\widehat{\beta}_2$ (Figure \ref{fig:SdSe_chi2_hetero}\textbf{b}) represented as an error plot with respect to the sample size $n\in(100,  \ 250,  \ 500),$ the repeated measurements $m=(5,\ 15,\ 30),$ the asymmetric points $\tau\in (0.1,  \ 0.3,  \  0.5, \  0.8,\ 0.9)$ and the error term $\varepsilon\sim\chi_2(3)$ in the location-scale-shift scenario.} \label{fig:SdSe_chi2_hetero}
\end{figure}
\end{center}

\clearpage
\bibliographystyle{apalike}
\bibliography{bib_fix_effect.bib}